\documentclass[]{JFM-FLM-arxiv}

\usepackage{graphicx}
\usepackage{xcolor,soul}
\usepackage{mathtools}
\usepackage{newtxtext}
\usepackage{newtxmath}
\usepackage{natbib}
\usepackage{hyperref}
\usepackage{placeins}
\usepackage{cases}
\usepackage{overpic}
\usepackage{siunitx}
\usepackage{textcase}
\usepackage{lineno}
\usepackage[normalem]{ulem}
\usepackage{bm}
\sethlcolor{yellow!50}

\hypersetup{
    colorlinks = true,
    urlcolor   = blue,
    citecolor  = black,
}

\newcommand{\RomanNumeralCaps}[1]

 \lefttitle{ }
 \righttitle{ }
\newcommand{\Wi}{{Wi}}
\newcommand{\cref}{c_\mathrm{ref}}
\newcommand{\C}{\mathsfbi{C}}
\newcommand{\Cc}{\mathsfi{C}}
\newcommand{\trCb}{\mathrm{tr}(\bar{\C})}
\newcommand{\T}{\mathsfbi{T}}
\newcommand{\Tc}{\mathsfi{T}}
\newcommand{\dd}{\mathrm{d}}
\title{
Localising polymers promotes the centre-mode elastic instability
}

\author{Shailendra Kumar Yadav\aff{1}
 \and Jason R. Picardo\aff{1} }

\affiliation{\aff{1}Department of Chemical Engineering, Indian Institute of Technology Bombay, Mumbai 400076, India}

\corresau{Jason R. Picardo, \email{picardo@iitb.ac.in}}

\begin{document}

\maketitle

\begin{abstract}
The centre-mode elastic instability, prevalent in rectilinear flows of dilute polymer solutions, allows dynamic states to emerge even in the absence of inertia. Here, we show that this instability can be significantly enhanced when the base flow has a nonuniform spatial-distribution of polymers. 
In such a flow, the polymeric stress not only depends on the conformation tensor (determined here by the Oldroyd-B equation) but also varies proportionally with the polymer concentration field, which satisfies a scalar transport equation. Focusing on the inertialess Stokes limit, we first consider the simple setting of periodic Kolmogorov flow.
A linear stability analysis shows that a nonuniform polymer concentration either promotes or suppresses the centre-mode instability, depending on whether the polymers are localised near or away from the maximum of the base-velocity profile.
We then focus on a base flow with a distinct polymer-laden stream overlying the base-velocity maximum, and sandwiched between polymer-free streams. We find that the critical Weissenberg number (product of the elastic relaxation time and the typical strain-rate) at the onset of instability undergoes a sudden decrease as the total polymer loading is increased. An energy analysis attributes this destabilisation to elastic feedback forces that arise from concentration gradients near the interfaces between the polymer-laden and polymer-free streams. 
The insights from Kolmogorov flow are shown to carry over to channel flow, where localising polymers about the centreline strongly promotes the centre-mode instability. 
% These results are expected to be relevant in applications where polymers are added to trigger instabilities and enhance mixing.

% the neutral stability curve of the nonuniform system develops a double-lobe form, which results in a sudden decrease in the
% while it is suppressed when polymers are localised away from the base-velocity maximum. 
\end{abstract}


% \footnotesizeskip
% \section{Introduction}
% \label{secch3:Intro}

\section{Introduction} \label{sec:Intro}

Dilute polymer solutions can exhibit hydrodynamic instabilities even in the inertialess limit, owing to the nonlinear coupling between the flow and viscoelastic stress. While it initially seemed that such elastic instabilities required curvilinear streamlines and their associated hoop-stress~\citep{shaqfeh1996_ARFM,McKinley96}, it is now clear that rectilinear base flows can also be linearly unstable~\citep{khalid_prl,BeneitezPRF2023,Lewy_Kerswell_2025,Dutta_et_al}.
Post instability, the flow can transition to elastic turbulence~\citep{steinberg_2021_ARFM}, a state marked by chaotic fluctuations and a consequent increase in drag and mixing~\citep{groisman2000elastic,groisman2001efficient,T_Bhergelae_PRE,bonn2011,pan_et_al_2013,Qin_Arratia_2017,Shnapp_2022_PRF,Li-Steinberg-23}. 

The default assumption in analyses of viscoelastic instabilities is a constant, spatially-uniform, polymer concentration, i.e., the polymers are assumed to have been premixed and uniformly distributed in the solvent, before the solution is admitted into the flow-setting of interest. For microfluidic applications, such premixing not only represents an additional processing step but also surrenders a potential means for manipulating the instability. Consider, for example, a polymer-laden stream that is introduced into a microchannel via a T-junction, so that it flows alongside polymer-free streams. Would such a configuration with spatially-localised polymers be more unstable than its uniformly-distributed counterpart?

A few studies have examined flows with a nonuniform polymer concentration. Collins and coworkers simulated the vertical mixing of polymers in a high-Reynolds-number turbulent shear flow, wherein polymers were initially localised within a horizontal layer \citep{VAITHIANATHAN_COLLINS_jfm_2007}. More recently, Yamani et al. \citep{Yamani21,Yamani23} have experimentally analysed the elasto-inertial turbulent flow that develops when a jet of polymer solution is introduced into a bath of Newtonian solvent. Here, we examine how an inertialess elastic instability is modified by a spatially-varying polymer concentration.
Specifically, we focus on the centre-mode instability.
% in the simplified scenario of two-dimensional viscoelastic Kolmogorov flow.

The centre-mode instability was first discovered in the elasto-inertial regime of pipe~\citep{garg_prl,chaudhary_pipe_jfm_2021} and channel flows~\citep{khalid_jfm}, and then shown to continue down to the inertialess limit, though only for channel flows~\citep{khalid_prl,Buza_Page_Kerswell_2022,Khalid_FENEP_JFM_2025}. Its name derives from the fact that the corresponding unstable normal mode propagates along the flow direction with a speed close to the maximum base-flow velocity, located at the channel's centreline; the spatial structure of the mode is also focused about the centreline. This instability provides a subcritical route, via nonlinear travelling waves called arrowheads or narwhals~\citep{Buza_Page_Kerswell_2022,Buza_Beneitez_Page_Kerswell_2022,Morozov22}, to inertialess elastic turbulence (chaotic flow) in three-dimensional channel flows~\citep{Morozov24,morozov2025blessings}. 
A different route to elastic turbulence in wall-bounded rectilinear flows may be provided by the polymer-diffusive instability, which is a wall-mode that arises from the diffusivity of polymer molecules~\citep{BeneitezPRF2023,Lewy}; however, this instability is suppressed when the anisotropy of polymer diffusion is taken into account~\citep{Pandey25}. 
Mechanisms for flow transition in the absence of a linear instability have also been proposed: non-modal linear-growth~\citep{jovanovic2010,jovanovic2011,Shnapp_2022_PRF}, and hoop-stress-driven nonlinear growth of finite-amplitude perturbations~\citep{morozov_prl2005}. Multiple instabilities can compete in a given flow, e.g., the hoop-stress and polymer-diffusive instabilities in elastic Taylor-Couette flow~\citep{Surya_Shankar_ProceedingsA2025}, and the hoop-stress and centre-mode instabilities in elastic Dean flow~\citep{tej_Dean_2025_arxiv}. Clearly, there is a rich variety of viscoelastic scenarios in which one could investigate the effect of a nonuniform polymer concentration. 
We focus, in this work, on the inertialess centre-mode. 

The centre-mode instability requires the base velocity profile to have a maximum in the transverse direction, and hence it is absent in Couette flow~\citep{Kerswell_page,Yadav_et_al_PRF2024}. A particularly simple flow that satisfies this requirement is periodic Kolmogorov flow. It consists of flow in a periodic box, driven by an applied unidirectional body force with transverse sinusoidal variations. The corresponding unidirectional steady base velocity field varies with the body force and exhibits a maximum at each peak of the force. Just as in channel flow, Kolmogorov flow can be destabilised by elastic stresses even in the inertialess limit \citep{BOFFETTA_et_al_JFM_2005}. The centre-mode character of this Kolmogorov-flow elastic instability has recently been established by \citet{Kerswell_page} and \citet{Lewy_Kerswell_2025}. The emergent travelling waves in Kolmogorov flow~\citep{Berti_Boffetta_PRE_2010,Lewy_Kerswell_2025,Kerswell2026TW} are very similar to their counterparts in channel flow, and they are therefore also called narwhals or arrowheads. As in channel flow, the interaction of these structures leads to elastic turbulence~\citep{Berti_PRE_2008,Berti_Boffetta_PRE_2010,Lewy_Kerswell_2025,Thomases25,Kerswell26building-blocks}.
% these waves are very similar in structure to the narwhals or arrow-heads \citep{Dubief_et_al_PRF_2022} that emerge via the centre-mode instability in inertialess channel flow~\citep{Buza_Beneitez_Page_Kerswell_2022,Morozov22,Morozov24}.
 
% this instability leads to elastic turbulence in Kolmogorov flow via bursting~\citep{Lewy_Kerswell_2025} as well as period-doubling~\citep{Thomases25}.

The relative simplicity of Kolmogorov flow makes it ideal for a first exploration of the consequences on stability of a nonuniform spatial distribution of polymers. Specifically, we consider a strip of polymer-rich solution, aligned with the base-flow direction, with polymer-depleted or polymer-free solvent streams on either side. We place the strip either at the location of maximum velocity or of maximum shear (the problem is formulated in \S~\ref{sec:prob-form}, along with a discussion of the base state and its linear stability analysis). 
On comparing the stability of this non-uniform flow with its uniform counterpart, where the same mass of polymers is premixed throughout the domain, we find that localising polymers about the maximum of the velocity (shear) destabilises (stabilises) the flow (\S~\ref{sec:neutral}). The extent of destabilisation increases strongly with the total polymer loading---the corresponding neutral curves change qualitatively, from single to double-lobed loops, as more polymer is added to the flow (\S~\ref{sec:beta-effect}).
% This makes sense, given the central role in the centre-mode instability of a maximum in the base velocity profile. 
Using a disturbance kinetic energy analysis, we trace this behaviour to 
% (i) the sensitivity of the centre-mode to the local concentration of polymers near the velocity maximum, (ii) the flattening of the base-state velocity profile about its maximum due to the high shear-viscosity of the polymer strip, and (iii) 
 elastic feedback forces arising from gradients in the polymer concentration (\S~\ref{sec:energy}).
% The second point is demonstrated by analyzing a uniform polymer solution with a modified base-velocity profile, obtained by suitably modifying the forcing function.
These insights from Kolmogorov flow suggest that it may be possible to promote the centre-mode instability in channel flow by localising polymers about the centreline. We show that this is indeed the case (\S~\ref{sec:channel}).
% though a systematic analysis of channel flow will be reported elsewhere. 
Our results naturally lead one to question how a nonuniform polymer distribution may affect other viscoelastic instabilities; such avenues for future work are discussed, following a summary of our key results, in the concluding section (\S~\ref{sec:conclusions}).

%==========================================================================================

\section{Problem formulation}
\label{sec:prob-form} 

Our goal is to understand how the hydrodynamic stability of a nonuniform polymer solution changes relative to one in which polymers are premixed and uniformly distributed. Specifically, we consider situations in which the polymer concentration varies in the transverse direction, so that we have a strip of polymer solution aligned with the streamwise direction and sandwiched between Newtonian fluid streams (see Fig.~\ref{fig:base}(a) below).  In the unidirectional base-flow, such a distribution of polymers will not be altered by advection, since the concentration gradients are perpendicular to the flow streamlines. The diffusion of polymers, however, will cause the concentrated strip to spread out in the cross-stream, transverse direction. This evolution of the concentration profile will be very slow owing to the small diffusivity of long polymer molecules \citep[$\sim 10^{-12}$ \si{m^{2}.s^{-1}},][]{kl1989}. 
% and, in the limit of vanishing diffusivity, 
% /so that the concentration profile will diffuse extremely slowly;
One may thus freeze the concentration profile and perform a quasi-static stability analysis. Alternatively, one may exclude polymer diffusion from the governing equations so that an exact steady-state base flow obtains. We follow the latter approach here (though, as discussed below, we introduce weak diffusion in the stability calculation as a numerical technique to aid in tracing the stability boundary). 

Note that polymer diffusion is relevant even in uniform polymer solutions because it acts to diffusively smear out large-gradients in the the conformation tensor (and therefore in the polymer stress)~\citep{kl1989}. In simulations, very high spatial resolutions are necessitated by the small value of the diffusivity. In fact, this value is often enhanced in order to achieve numerical stability at low to moderate resolutions~\citep{sb95,gv19,aop21}; an alternative is to drop explicit diffusion from the equations and use a high-order shock-capturing advection scheme~\citep{VAITHIANATHAN_jnnfm2006,pa17,Yerasi2024}. Physically, polymer diffusion has been shown recently to have a nontrivial, destabilizing role in wall-bounded viscoelastic flows, giving rise to the polymer diffusive instability ~\citep{BeneitezPRF2023,Lewy,Couchman,BENEITEZ_jnnfm_2025,Surya_Shankar_ProceedingsA2025}. In the context of unbounded Kolmogorov flow, polymer diffusion has no qualitative effect other than weakly stabilizing the centre-mode~\citep{Thomases25,Lewy_Kerswell_2025}.  

\subsection{Governing equations}\label{subsec:gov-eqns}

We consider the 
% inertialess two-dimensional 
flow of a dilute polymer solution modeled as an Oldroyd-B fluid, with elastic relaxation time $\lambda$. The dissolved polymer molecules may be distributed nonuniformly in the solvent, and so the polymer concentration field $c$ can vary in space. Normalizing $c$ with a constant reference value $\cref$ yields the normalised polymer concentration field $\theta = c/\cref$. A uniform solution at the reference concentration has a polymer contribution to the viscosity $\mu_p$, which along with the solvent contribution $\mu_s$ yields the zero-shear viscosity of the uniform solution $\mu$. In a nonuniform solution, the total viscosity varies in proportion to the variations in $\theta$.

A viscoelastic version of Squire's theorem, which implies that two-dimensional disturbances are the most unstable, has been established by \citet{bistagnino} for the Oldroyd-B model with a uniform polymer concentration. On repeating the derivation in \citet{bistagnino} while allowing for concentration variations, we find that the Squires theorem remains valid for the present situation wherein the base-state concentration varies in the transverse direction alone.
We therefore limit our attention to two dimensions.

For a dilute solution with a spatially-nonuniform distribution of polymers, whose ensemble averaged dynamics are described by the Oldroyd-B model,
% and in the limit of vanishing inertida, 
we have 
the following nondimensional governing equations for the velocity $\bm{u}$, the pressure $p$, the conformation tensor $\C$, the polymeric elastic stress $\T$, and the polymer concentration $\theta$~\citep{VAITHIANATHAN_jnnfm2006,VAITHIANATHAN_COLLINS_jfm_2007}:
\begin{gather} 
  \bm{\nabla} \cdot \bm{u} = 0 \label{eq:cont}\\
  \Rey \left({\partial_t \bm u}+ \bm{u} \cdot \bm{\nabla u}\right)=-\bm{\nabla} p + \beta\bm{\nabla}{^2} \bm{u} + \bm{\nabla} \cdot \T + F(y) \bm{e_x} \label{eq:navier-stokes}\\
  % \end{gather}
  % \begin{gather}
% Re\left({\frac{\partial  \bm{u}}{\partial t} + \bm{u} \cdot \bm{\nabla}\bm{u}}\right) = -\bm{\nabla} p + \beta\bm{\nabla}{^2} \bm{u} + \bm{\nabla} \cdot \bm{\tau} + F_x(y) \bm{e_x}.
%-\bm{\nabla} p + \beta\bm{\nabla}{^2} \bm{u} + \bm{\nabla} \cdot \bm{\tau} + F_x(y) = 0.
{\partial_t \C} + \bm{u} \cdot \bm{\nabla}\C  - (\bm{\nabla} \bm{u}){^T} \cdot \C -\C \cdot (\bm{\nabla \bm{u}}) =   -\frac{\C - \mathsfbi{I}}{Wi}, \label{eq:oldroydB}\\
 \T =  \theta(y)\frac{1-\beta}{Wi}(\C - \mathsfbi{I}) \label{eq:stress},\\
 %   \mathcal{D} \bm{\nabla}^2 \C
 	{ \partial_t \theta} + \bm{u}\cdot\bm{\nabla} \theta =0 
     \label{eq:theta}.
\end{gather}
where $\mathsfbi{I}$ is the second-order identity tensor, $\beta =  {\mu_s}/{\mu}$, $\Rey = L V \rho/\mu$ is the Reynolds number, and $\Wi = \lambda V/L$ is the Weissenberg number, with $L$ being the transverse length scale of the domain
% the domain (which is limited to one forcing wavelength)} 
and $V$ being the magnitude of the maximum velocity of the unidirectional steady Newtonian flow.  We have used $L$, $V$, and $L/V$ as length, velocity, and time scales to nondimensionalize the equations. 

% The parameter $\beta$ varies with the magnitude of $\cref$: a uniform solution will be a Newtonian fluid if $\beta = 1$ and a upper-convected Maxwell fluid (no solvent contribution to the viscosity) if $\beta = 0$. Moreover, for a given value of the parameter $\beta$, a region of the fluid with a local concentration $ \theta \cref$ will have a local polymer-contribution to the viscosity of $\theta \mu_p $.

The sinusoidal Kolmogorov forcing function in the momentum equation \eqref{eq:navier-stokes} is $F(y) = -\cos(y)$. For simplicity and to better mimic channel flow, we restrict the domain to one forcing wavelength; hence the domain has periodic boundaries in the normal direction at $y = 0,$ $2 \pi$ while it is unbounded in the streamwise ($x$) direction.
% set $F_x = \cos(y)$, 
% so that only one pair of forward and backward flowing streams are present in the base flow [see Fig.~\ref{fig:base}(b) below]. 
The centre-mode instability in this flow, for the case of a uniform polymer solution, was analysed recently by \citet{Thomases25}. (The case of multiple interacting stream-pairs has been studied by \citet{Lewy_Kerswell_2025}, who show that the flow is most unstable when the domain size is twice the forcing wavelength.) Note that when channel flow is considered in \S~\ref{sec:channel} we set $F(y) = 2$ and impose no-slip and impermeable walls at $y = -1,1$.
% \hl{Hence, the nondimensional sinusoidal Kolmogorov forcing function is $F(y) = \cos(y)$, with the dimensionless coordinate $y$ varying from $0$ to $2 \pi$. }
% (this forcing is used throughout, except in \S~\hl{?} where a non-sinusoidal forcing is used to gain insight into the role of the modified base-state velocity profile in nonuniform flows).

% $Re = \frac{U_m L}{2\pi \mu}$, $\mathcal {D} = \frac{2\pi D}{U_m L}$.

If $\theta$ is set to unity, then equations \eqref{eq:cont} - \eqref{eq:stress} become those of the Oldroyd-B model for a uniform solution with a constant polymer concentration $\cref$. 
In a nonuniform solution with spatio-temporal variations of $\theta$, the local polymer concentration is  $\theta \cref$.
% and the local polymer-contribution to the viscosity is $\theta \mu_p $. The latter relation follows from the fact that $(1-\beta) = \mu_p/\mu$ is proportional to $\cref$ in dilute solutions. enters the problem via the parameter $\beta$ varies proportionally with $\cref$ and therefore represents the total polymer loading. 
As detailed in \S~\ref{sec:base}, we consider base steady states (denoted by an overbar) where either $\bar\theta = 1$ (uniform solutions) or $\bar\theta$ varies in $y$ (nonuniform solutions) such that its transverse average is unity:
% (thus the value of the transverse-integral is maintained):
% . The $\theta$-transport equation \eqref{eq:theta} will then preserve this integral so that 
\begin{align}\label{eq:int-thetabar}
    \frac{1}{2\pi}\int_0^{2\pi}\bar\theta(y) dy &= 1 \;\;\;\mathrm{(Kolmogorov\;flow),\;} \nonumber\\
    \mathrm{or}\;\; \frac{1}{2}\int_{-1}^{1}\bar\theta(y) dy &= 1\;\;\;
    \mathrm{(channel\;flow)} 
\end{align}
Owing to this condition, 
% (which of course is preserved by \eqref{eq:theta}), 
the total polymer loading is the same for uniform and nonuniform cases. This total polymer loading
% , which is $\cref L$ per unit longitudinal length, 
is varied by changing the parameter $\beta$, since 
% $(1-\beta)$ determines 
the polymer contribution to the solution's zero-shear viscosity, whether spatially varying or uniform, is $\theta(1-\beta)$. The polymer-free Newtonian limit corresponds to $\beta = 1$, while solutions with increasing polymer-loading are realised as $\beta$ is decreased toward zero.
% an upper-convected Maxwell fluid (no solvent contribution to the viscosity) is approached as $\beta \rightarrow 0$.
% with the same value of $\beta$. 
This setup allows us to study how the stability of the flow is affected when a given amount of polymer is nonuniformly distributed. 
Moreover, by varying $\beta$, we can examine how the effect of a nonuniform polymer distribution varies with the total polymer loading. 

Our focus in this study is on the purely-elastic centre-mode instability, prevalent in the limit of vanishingly small inertia. We therefore perform calculations for $\Rey = 0$, except for the energy analysis of \S~\ref{sec:energy} where we also consider a small value of $\Rey = 0.01$. (The $\Rey \rightarrow 0$ limit is regular and so our conclusions remain valid for finite small values of $\Rey$ )
% In a nonuniform solution, the concentration field $\theta$ is advected by the flow:
% \begin{gather} \label{eq:theta} 
% 	\frac{ \partial \theta} {\partial t} + \bm{u}\cdot\bm{\nabla} \theta =0.
%     % \mathcal {D} \bm{\nabla}^2 \theta(y)
% \end{gather} 

% The viscoelastic fluid is governed by the Oldroyd-B model \citep{larson1992instabilities, zhang_zaki_jfm, chaudhary_pipe_jfm_2021, khalid_jfm, khalid_prl, wan_sun_zhang_jfm_2021,dong_zhang_2022, wan_dong_zhang_jfm_2022}, which predicts shear-rate-independent viscosity and normal stress, making it suitable for modeling the shear rheology of dilute polymer solutions.

% \begin{figure}
% \centering
% 	\includegraphics[width=.33\linewidth]  	{fig-SI/base_state_trace_B0pt8_W1_three_ht.eps} 
%             \hspace{0.05\linewidth}
% 	\includegraphics[width=.33\linewidth]  	{fig-SI/base_state_trace_B0pt5_W1_three_ht.eps} 
% 	\caption{Panels (a) \& (b) show base-state trace; data are shown for $Wi = 1$, and $\beta  = 0.8$ \& 0.5.} %The area under the curve,  for each strip represnts the  total polymer present the system and it remains same for all polymer strips. }
% \end{figure}

\subsection{Base state}\label{sec:base}
The governing equations admit a base steady flow in the $x$ direction that varies in the transverse y direction, i.e, $\bar{\bm{ u}} = \bar{U}(y) \bm{e_x}$. The conformation tensor 
\begin{align} 
	\bar{\C}= 
		\begin{bmatrix}
           		1+ {2{Wi}^2\, (\bar{U}^{\prime})^2} & Wi\, \bar{U}^{\prime}\\
           	 Wi\, \bar{U}^{\prime} & 1 
      \end{bmatrix}.
\end{align}
where 
% the overbar denotes base state quantities and 
the prime denotes a derivative of a base-state field in the y-direction. The base velocity $\bar{U}$ is determined by 
% the simplified momentum equation
% \begin{equation}
%     \beta \frac{\dd^2 \bar{U}}{\dd y^2} +\frac{1-\beta}{Wi} \frac{\dd}{\dd y} (\bar{\theta}\, \bar{C}_{xy}) +F(y) = 0, \label{eq:baseU}
% \end{equation}
\begin{equation}
    \beta \, \dd^2_y \bar{U} +\frac{1-\beta}{Wi}\dd_y(\bar{\theta}\, \bar{C}_{xy}) +F(y) = 0, \label{eq:baseU}
\end{equation}
where $\dd_y  = {\dd}/{\dd y}$ and  $\dd^2_y  = {\dd^2}/{\dd y^2}$. Given a base-state concentration profile $\bar{\theta}(y)$, we numerically solve \eqref{eq:baseU}, subject to periodic boundary conditions, using the finite difference solver \textit{bvp4c} of MATLAB \citep{MATLAB:R2025a} which implements a fourth-order accurate collocation formula with adaptive mesh refinement \citep{shampine2003solving}.

We consider the following family of smoothened pulse profiles for the base concentration:
\begin{equation} \label{eq:pulse}
	% \theta(y) =\frac{c1}{\left[\frac{1}{2} - \frac{1}{\pi}\, tan^{-1} \left(\frac{ - \pi/4 - \tfrac{w_d}{2}}{\delta}\right) + \theta_{\min}\right] } *\left[\frac{1}{2} - \frac{1}{\pi}\, tan^{-1} \left(\frac{(y-\pi)^2 - \pi/4 - \tfrac{w_d}{2}}{\delta}\right) + \theta_{\min}\right],
    \bar\theta(y) ={\bar\theta_\mathrm{amp}}\,\frac{\phi(y)}{\phi(\pi) } \; \; \mathrm{where}\;\;\phi(y)={\frac{1}{2} - \frac{1}{\pi}\, \tan^{-1} \left(\frac{(y-\pi)^2 - \pi/4 - {w_d}/{2}}{\delta}\right) + \bar\theta_{\min}},
\end{equation} 
The parameters $w_d$ and $\delta$ control the width and the steepness of the pulse, while $\bar\theta_{\min}$ sets the minimum polymer number density in the base-state distribution. We fix $\bar\theta_{\min} = 10^{-3}$ and $w_d = 0.1$ (other values of $w_d$ are considered briefly in Appendix~\ref{app:wd}), and vary $\delta$ to realise a range of base-state polymer distributions, from uniform (large $\delta$) to a near-square pulse (small $\delta$). The amplitude of the pulse profile is controlled by $\theta_\mathrm{amp}$, which is adjusted as $\delta$ is varied to satisfy the integral constraint \eqref{eq:int-thetabar}.
% ensure that $\int_0^{2\pi}\bar\theta(y) dy = 1$ for Kolmogorov flow (and $\int_{-1}^{1}\bar\theta(y) dy = 1$ for channel flow).

\begin{figure}
	\includegraphics[width=.33\linewidth]{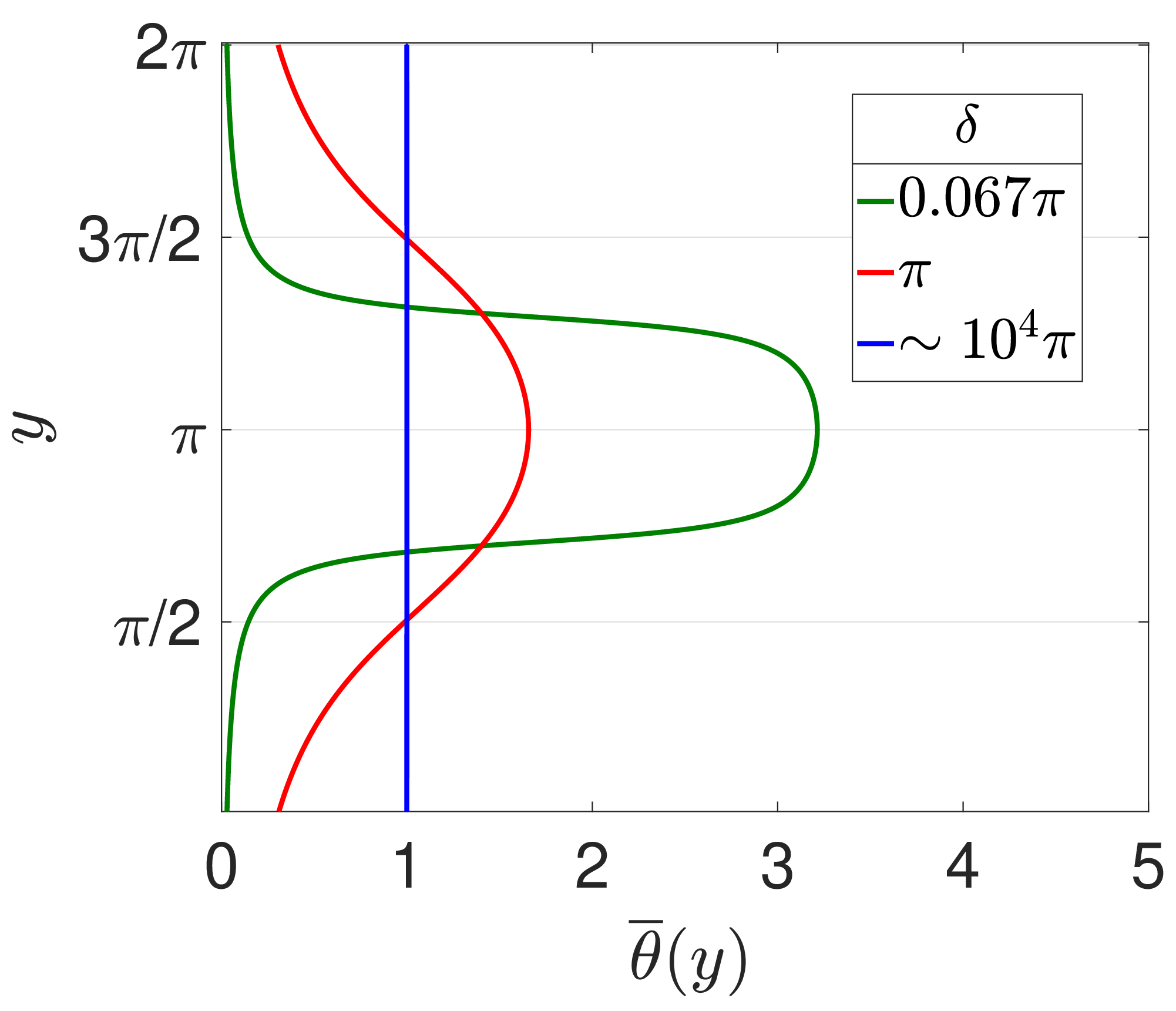} 
    \put (-20,25){\footnotesize (a)}
	\includegraphics[width=.33\linewidth]{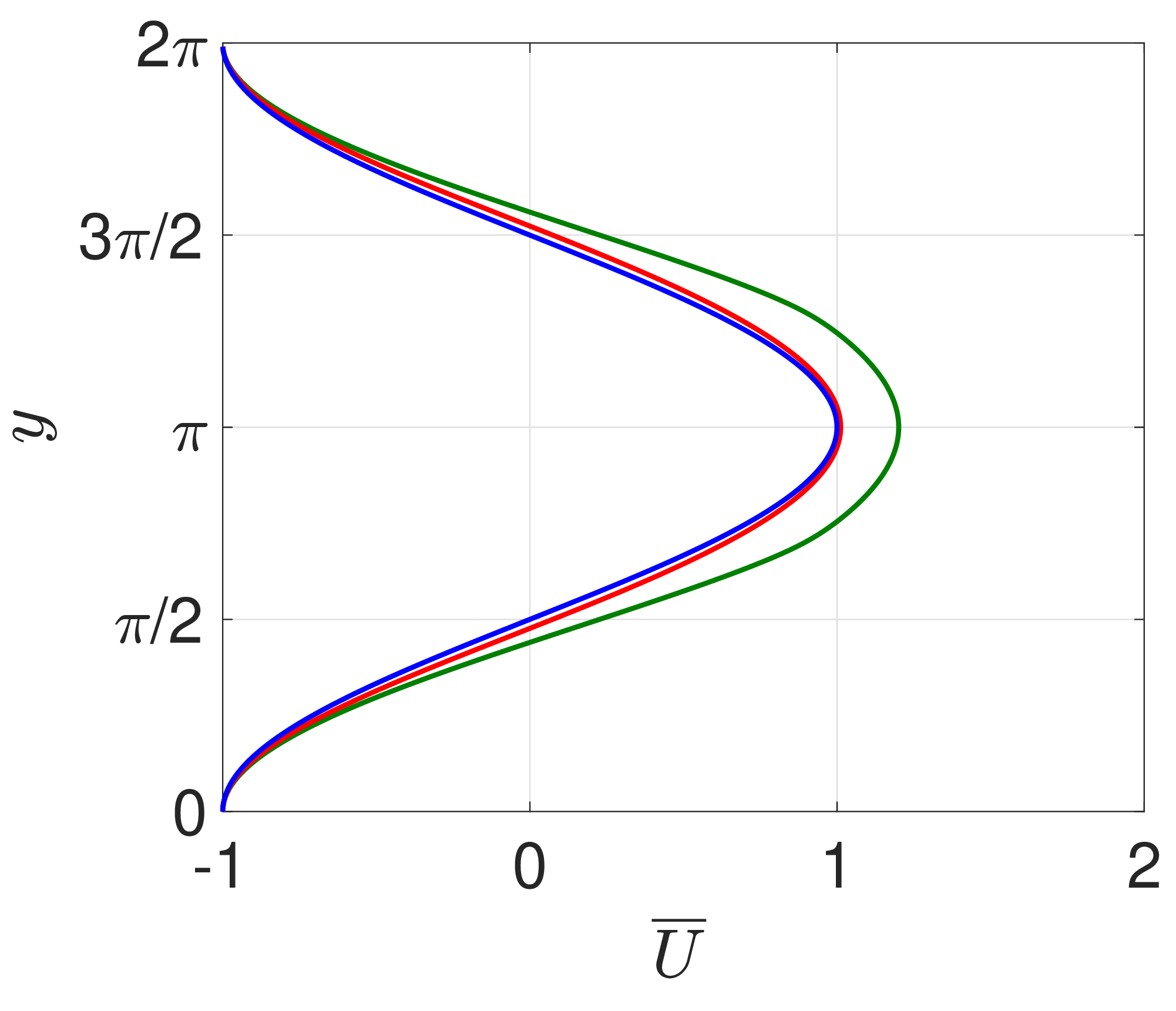} 
     % \put (-45,25){\footnotesize (b) \scriptsize $\beta = 0.8$}
    \put (-20,25){\footnotesize (b) }
	\includegraphics[width=.33\linewidth]{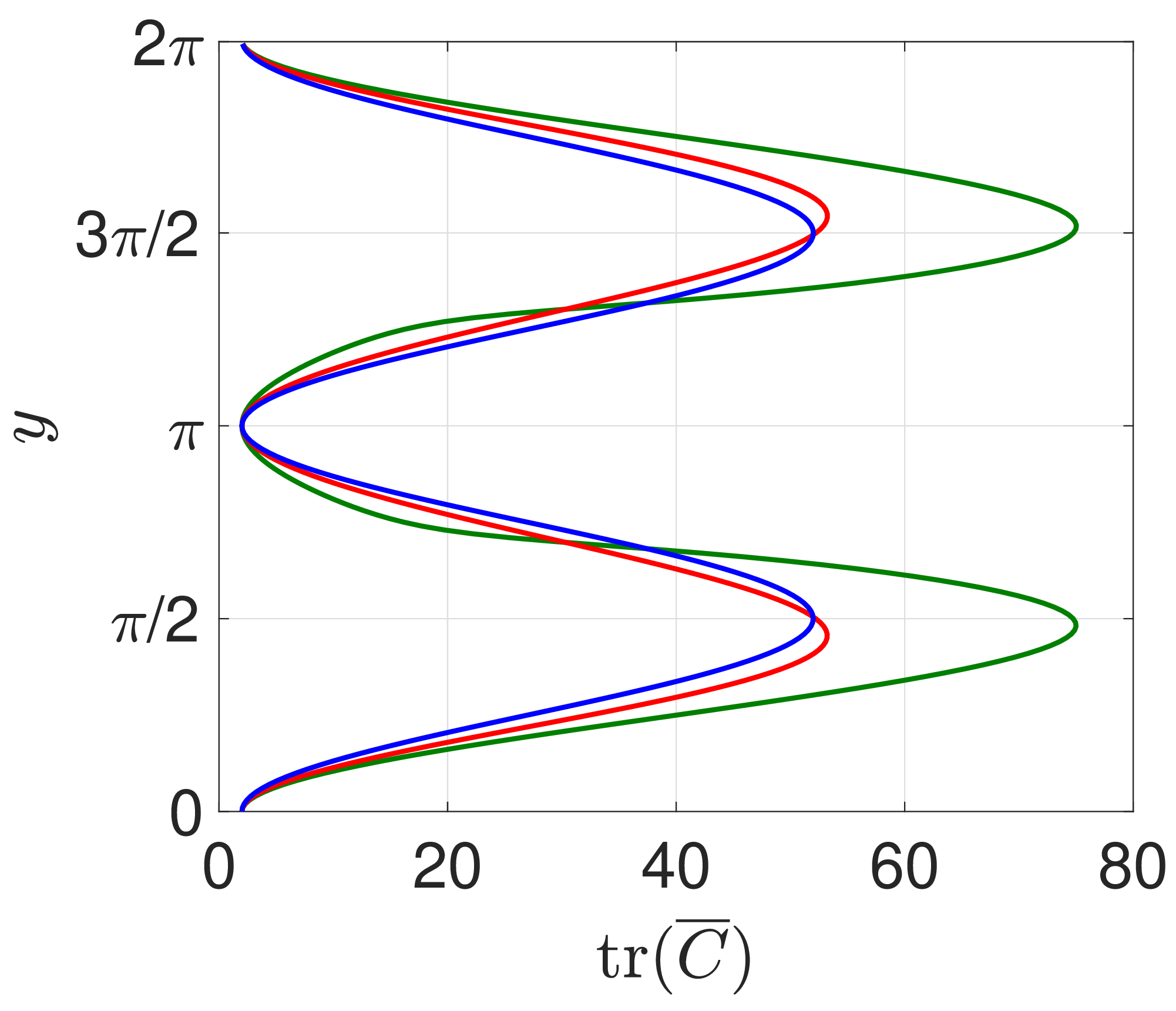} 
    % \put (-45,25){\footnotesize (c) \scriptsize $\beta = 0.5$}
    \put (-20,25){\footnotesize (c) }    
	% \includegraphics[width=.33\linewidth]{fig/base_velocity_B0pt5_three_ht.eps} 
 %    % \put (-45,25){\footnotesize (c) \scriptsize $\beta = 0.5$}
 %    \put (-20,25){\footnotesize (c) }
	% \includegraphics[width=.33\linewidth]{fig/base_vel_B0pt8_three_ht_all_max_at1.eps} 
 %    \put (-20,30){(b)}
	% \includegraphics[width=.33\linewidth]{fig/base_vel_B0pt5_three_ht_all_max_at1.eps} 
 %    \put (-20,30){(c)}
	\caption{Base states of Kolmogorov flow with nonuniform polymer concentrations. (a) Concentration profiles for different extents of localisation ($\delta$) near the velocity maximum; (b) base-state velocity; (c) base-state polymer squared extension. Here $\Wi = 5$ and $\beta = 0.8$; results for $\beta = 0.5$ are presented in the \href{https://bighome.iitb.ac.in/index.php/s/g6wRiyEY8H3SNyR}{supplementary material}. 
    } %The area under the curve,  for each strip represnts the  total polymer present the system and it remains same for all polymer strips. }
		\label{fig:base}
\end{figure}

The profile of $\bar\theta$ in \eqref{eq:pulse} localises polymers about the location of the maximum velocity, $y = \pi$, as illustrated in Fig.~\ref{fig:base}(a) below. 
By adding a constant shift of $\pi/2$ to $y$, we also study the
situation where polymers are localised near the location of maximum shear $y = \pi/2$ (more precisely, the location where the shear is maximum in a flow without polymers or with a uniform concentration of polymers); the corresponding base state is illustrated in the \href{https://bighome.iitb.ac.in/index.php/s/g6wRiyEY8H3SNyR}{supplementary material}.

Figure~\ref{fig:base}(a) shows three distributions for $\delta =$ $ \sim10^4\pi$, $\pi$, and $0.067 \pi$, all with $w_d = 10^{-1}$; the first is uniform, while the second and third are increasingly nonuniform profiles with the polymers localised about the velocity maximum at $y = \pi$. In particular, the third profile is a smoothened pulse which mimics a two-dimensional (2D) jet of polymer solution sandwiched between near-pure solvent streams. 
% The reflection symmetry of the uniform problem in the $x$ direction
We choose, without loss of generality, to localise the polymers about the maximum of the velocity in the positive $x$ direction, located at $y = \pi$. 

The velocity profiles $\bar{U}(y)$ corresponding to the three concentration distributions in Fig.~\ref{fig:base}(a) are depicted in Fig.~\ref{fig:base}(b), for $\beta = 0.8$. Localising the polymers about the velocity-maximum increases the extent of variation of the velocity. Here, we fix the maximum velocity in the negative direction at $-1$ (equations \eqref{eq:cont}-\eqref{eq:theta} are invariant to the addition of a constant vector to $\bm{u}$) and so the localisation of polymers increases the maximum positive velocity $\bar{U}_m$.
The trace of the base-state conformation tensor $\trCb$, i.e., the polymer squared-extension in the base flow, is presented in Fig.~\ref{fig:base}(c). Clearly, the increase in the velocity gradients, seen in {Fig.~\ref{fig:base}(b)}, produces more stretching in the nonuniform flow.  
These quantitative differences aside, no significant qualitative change occurs in the base profiles of the velocity and conformation tensor when the polymer is localised. And yet, we shall see in \S~\ref{sec:beta-effect} that the stability properties of the nonuniform case differ dramatically from the uniform case (the difference first becomes prominent when $\beta$ is decreased to $0.8$, which is the value used in Fig.~\ref{fig:base}). This observation suggests that the primary cause of the strong destabilizing effect of a nonuniform concentration is \textit{not} the modulation of the base state velocity or polymer extension fields. Rather, as will be shown in \S~\ref{sec:energy}, the strong destabilisation is driven by additional elastic feedback forces arising from gradients in the polymer concentration.

\subsection{Linear stability analysis}\label{sec:lsa}

We perform a modal linear stability analysis of the base state
by
% The  procedure used to demonstrate this result [ref] does not extend to flows with concentration variations and so 3D disturbances could be more unstable than 2D ones. Nevertheless, for simplicity, we only consider 2D disturbances in this work. Our results should therefore be viewed as conservative predictions of the destabilizing effect of concentration variations.// We apply normal mode disturbances...
% We therefore apply two dimensional
applying normal mode disturbances:
\begin{equation} \label{eq:normal-mode}
  {\psi}(x, y, t) = \bar{\psi}(y) + \hat{{\psi}}(x,y,t)=\bar{\psi}(y) + \tilde{\psi}(y){e}^{ik(x-ct)},
\end{equation}
where $\psi$ represents a field variable ($\bm{u}, \C, p, \theta$), and $k$ and $c$ are the real wavenumber and complex wave speed ($c = c_r + i c_i$) of the mode. Modes with $c_i > 0$ grow in time and render the base flow temporally unstable. 

On expanding all field variables as per \eqref{eq:normal-mode} and linearising the governing equations \eqref{eq:cont}-\eqref{eq:theta}, we obtain the following perturbation equations in component form:
% Show eqn without diffusion here; after showing spectra and talking about validation, example spectra, shift-reflect symmetry and its breaking; and continuous spectra ballon and new line of continuous spectra from -1 to 1. say that we add diffusion only to track the neutral curve and that we have verified the results with calculations without diffusion - so diffusion added to just th linearised equations should be view as a numerical technique.
%Cont. Eqn.
\begin{equation} \label{eq:tilde-cont}
  ik\tilde{u}_x +  \dd_y\tilde{u}_y = 0,
\end{equation}
%x-mom eqn
\begin{equation} \label{eq:tilde-vx}
	\begin{aligned}
Re\left(ik(\bar{U} -c) \tilde{u}_x + {\bar{U} }^{\prime}\tilde{u}_y \right)- \beta\left({\dd}^2_y - {k}^2\right)\tilde{u}_x + ik\tilde{p}
-\frac{1-\beta}{Wi}ik {\bar{\theta}}\tilde {{\Cc}}_{xx} \\
- \frac{1-\beta}{Wi} \left( {\bar{\theta}} \dd_y + \bar{\theta}^{\prime}\right) \tilde {{\Cc}}_{xy} 
- \frac{1-\beta}{Wi} \left( ik({\bar{{\Cc}}_{xx}} -1) + \bar{\Cc}_{xy}^{\prime} + \bar{\Cc}_{xy} \dd_y
\right) \tilde{\theta}= 0,
	\end{aligned}
\end{equation}
%z-mom eqn

\begin{equation} \label{eq:tilde-vy}
	\begin{aligned}
		Re ik(\bar{U} -c) \tilde{u}_y  - \beta\left({\dd}^2_y - {k}^2\right)\tilde{u}_y + \dd_y \tilde{p}
		-\frac{1-\beta}{Wi}ik {\bar{\theta}}\tilde {{\Cc}}_{xy} \\
		- \frac{1-\beta}{Wi} \left( {\bar{\theta}} \dd_y + \bar{\theta}^{\prime} \right) \tilde {{\Cc}}_{yy} 
		- \frac{1-\beta}{Wi}
        % \left( 
        ik{\bar{{\Cc}}_{xy}}  
        % + ({ \bar{\Cc}_{yy}} -1 ) d_y
		% \right) 
        \tilde{\theta} = 0,
	\end{aligned}
\end{equation}

%Cxx-stress eqn
\begin{equation} \label{eq:tilde-Cxy}
	\begin{aligned}
		-2\left(ik \bar{\Cc}_{xx} + \bar{\Cc}_{xy} \dd_y \right)\tilde{u}_x +\bar{\Cc}^{\prime}_{xx} \tilde{u}_y  + \left( \frac{1}{Wi} + ik(\bar{U} -c) \right)\tilde {{\Cc}}_{xx} -2 \, \bar{U}^\prime \tilde {{\Cc}}_{xy} = 0,
	\end{aligned}
\end{equation}

%Cxy-stress eqn
\begin{equation} \label{eq:tilde-Cxx}
	\begin{aligned}
		-\left(ik \bar{\Cc}_{xy} + \bar{\Cc}_{yy} \dd_y \right)\tilde{u}_x +\left( \bar{\Cc}^{\prime}_{xy} -\bar{\Cc}_{xy} \dd_y -ik \bar{\Cc}_{xx}  \right) \tilde{u}_y  \\
        + \left( \frac{1}{Wi} + ik(\bar{U} -c) \right)\tilde {{\Cc}}_{xy} - \bar{U}^\prime \tilde {{\Cc}}_{yy} = 0, 
	\end{aligned}
\end{equation}

%Cyy-stress eqn
\begin{equation} \label{eq:tilde-Cyy}
	\begin{aligned}
		-2 \left(ik \bar{\Cc}_{xy} + \bar{\Cc}_{yy} \dd_y \right)\tilde{u}_y  + \left( \frac{1}{Wi} + ik(\bar{U} -c) \right)\tilde {{\Cc}}_{yy} = 0,
	\end{aligned}
\end{equation}

%\theta- eqn
\begin{equation} \label{eq:theta-tilde}
	\begin{aligned}
		\bar{\theta}^{\prime}\tilde{u}_y  +  ik(\bar{U} -c) \tilde {\theta} = 0.
	\end{aligned}
\end{equation}
% where a prime denotes a $y$-derivative of a base-state field (e.g., $\bar{U}^{\prime} = d\bar{U}/dy$).

Equations\,(\ref{eq:tilde-cont}-\ref{eq:theta-tilde}), along with periodic boundary conditions at $y = 0$ and $2\pi$, define an eigenvalue problem for 
$c$, which is solved numerically using the Fourier spectral collocation method \citep{boyd, weideman}.
We use the MATLAB solver, \textit{eig}, to compute the eigenvalues. To eliminate non-physical, spurious eigenvalues, we compare the eigenspectra obtained using different numbers of collocation points $N$ \citep{SchmidHenningsonBook}; we find that $N \approx 250$ is sufficient to accurately capture the genuine eigenvalues.
% , with a relative error of less than $10^{-5}$ (see the online supplementary material). 
We have verified our computations by comparing the results for a uniform concentration with those of \cite{Lewy_Kerswell_2025} (see the \href{https://bighome.iitb.ac.in/index.php/s/g6wRiyEY8H3SNyR}{supplementary material}).

\begin{figure}
	        \centering
	\includegraphics[width=.32\linewidth]  	{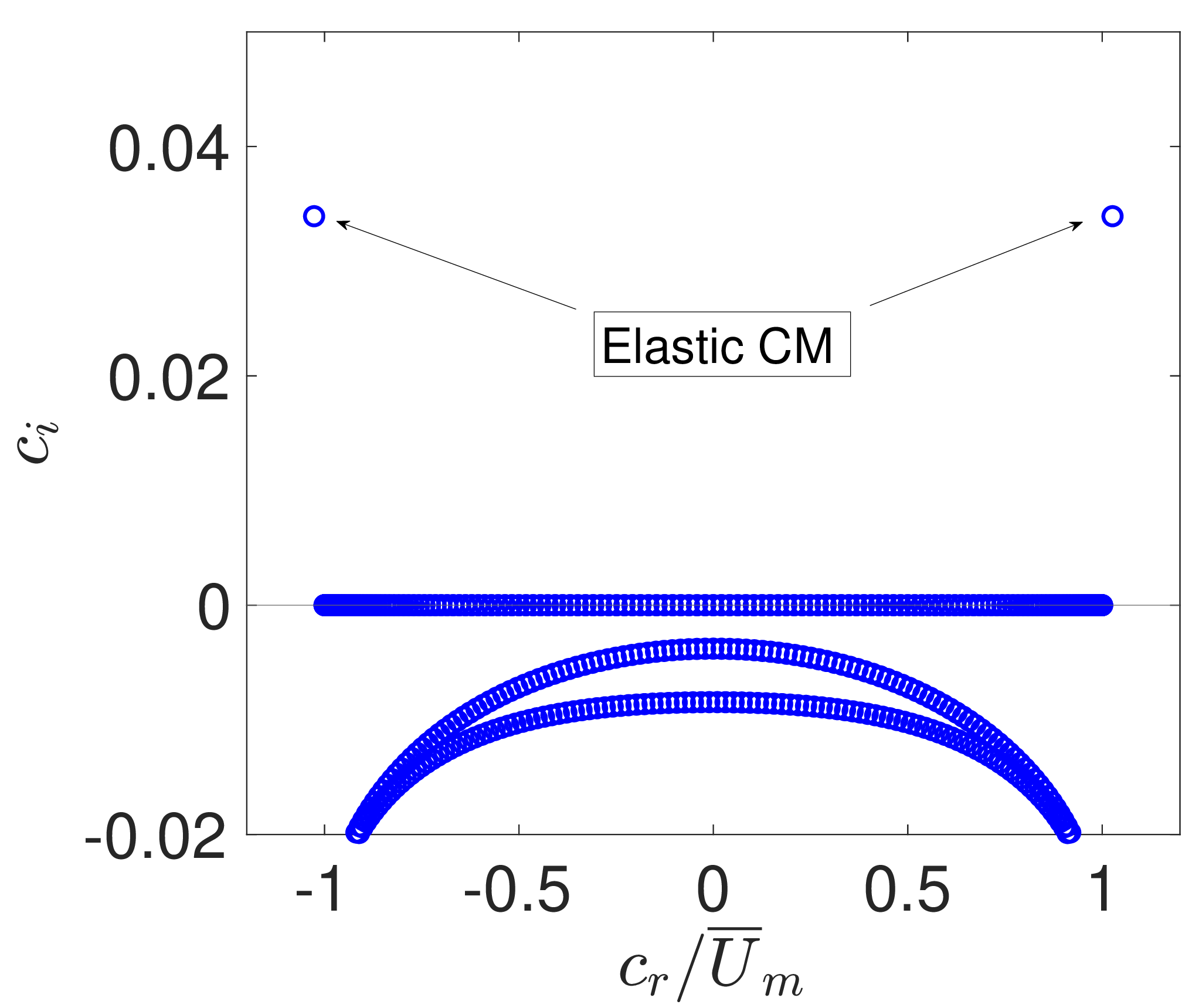}
    \put (-95,93){\footnotesize (a)}
	\includegraphics[width=.32\linewidth]  	{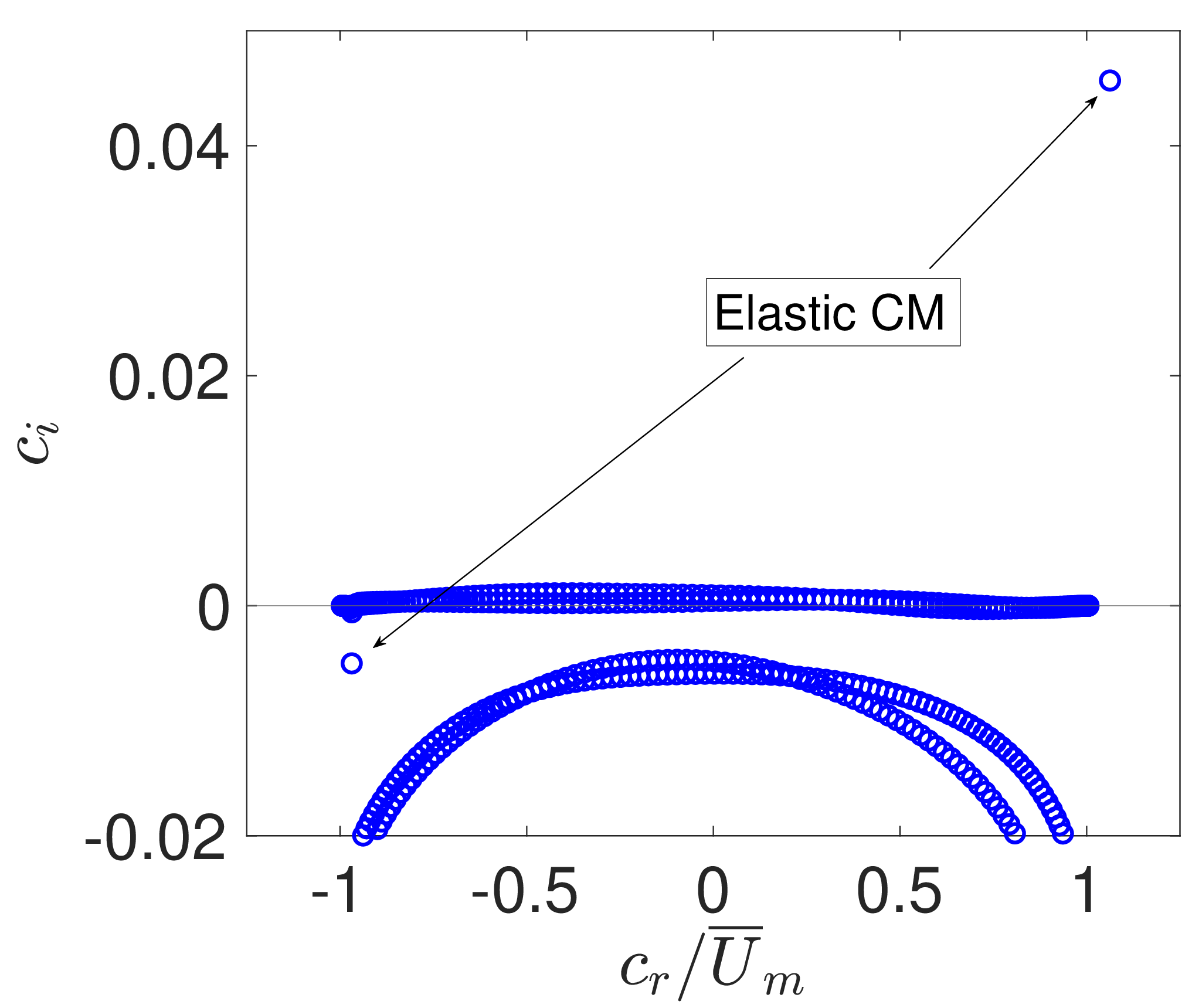}
    \put (-95,93){\footnotesize (b)}
    \includegraphics[width=.315\linewidth]  	{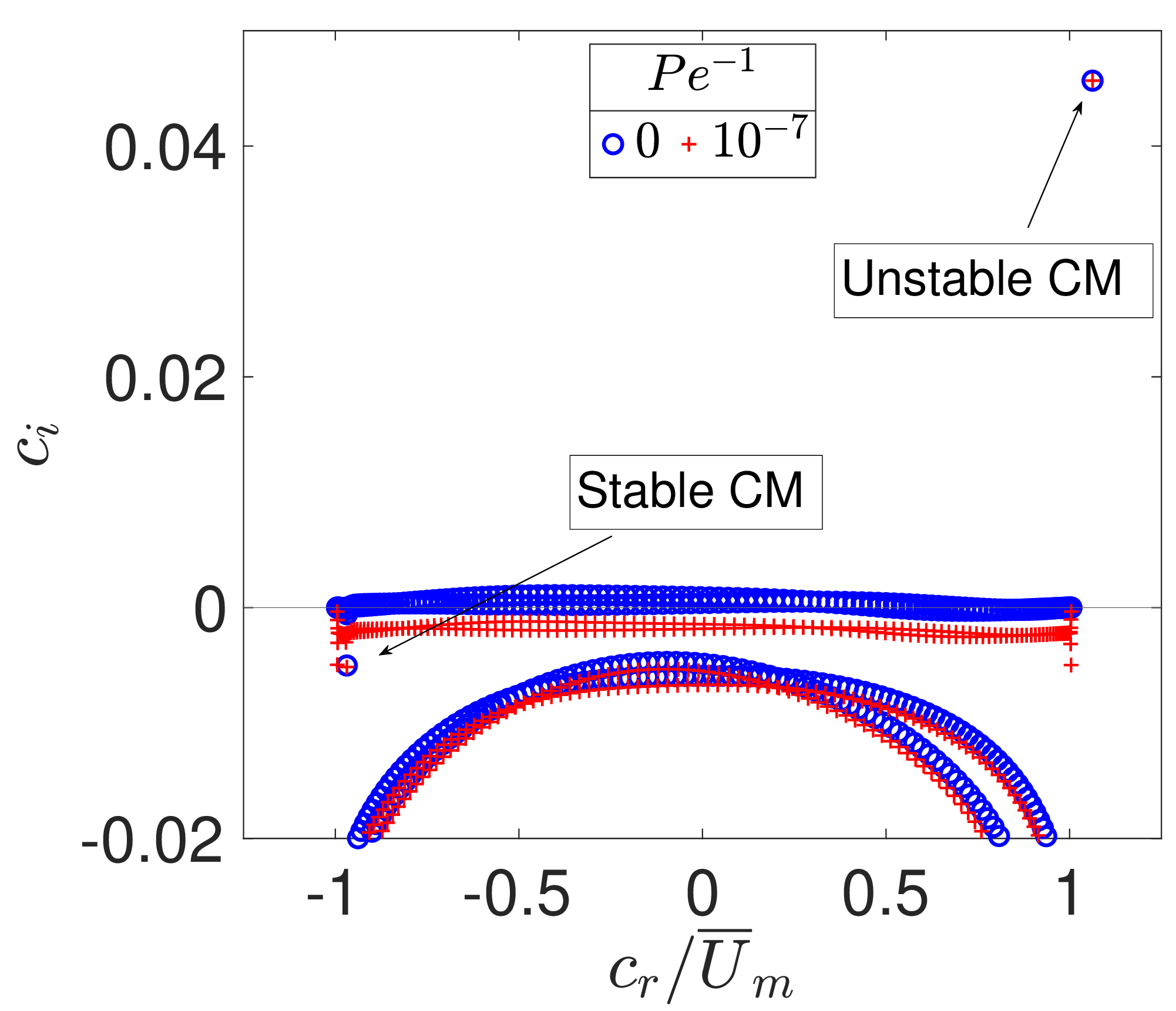}
    \put (-94,93){\footnotesize (c)}

    \caption{Eigenspectra for Kolmogorov flow of an Oldroyd-B fluid. (a) Uniform base-state concentration with $Re = 0$,  $\Wi =162$, $k = 0.2$, $\beta = 0.95$; the elastic centre-mode eigenvalues are $ \pm1.0265 + 0.03393\,\mathrm{i}$. (b) nonuniform  base-state concentration ($\bar{\theta}$ corresponds to $\delta = 0.417\pi$) with the same parameters as panel (a); the symmetry of the forward and backward propagating centre-mode eigenvalues is lost and they are now $1.06767 + 0.004567\,\mathrm{i}$ and $-0.973405 - 0.005013\,\mathrm{i}$
    % , and corresponding $c_r/\bar{U}_m$ are $1.06257$, and $-0.968755$, respectively. 
    (c) Comparison of the eigenspectra for the case in panel (b) without and with polymer diffusion in the linearised equations (see the legend).
    }
\label{fig:spectra}
\end{figure}

The typical eigenspectra for the case of a uniform base-state concentration  is shown in Fig.~\ref{fig:spectra}(a). Here, two centre-mode eigenvalues corresponding to forward and backward propagating modes are unstable; these modes have the same stability characteristics because of the shift-reflect symmetry of the uniform-concentration problem~\citep{Lewy_Kerswell_2025}. 
% We also see a stable balloon of eigenvalues which are the numerical approximation of the continuous spectrum. 
% Compared to previous results ~\citep{Lewy_Kerswell_2025}, 
The band of neutral eigenvalues, with $c_r$ ranging from $-\bar{U}_m$ to $+\bar{U}_m$, are numerical approximations of modes of the continuous spectrum that are introduced by the advection equation \eqref{eq:theta-tilde} for $\tilde {\theta}$: modes with $\tilde {\bm u} = 0$ and with $\tilde \theta$ zero everywhere except at $y^*$ will propagate with speed $c_r = \overline{U}(y^*)$. These modes are present regardless of whether the base-state concentration $\bar\theta$ is uniform or nonuniform and do not alter the stability characteristics of the system.
% Even when $\bar\theta$ is uniform are present the uniform system ...

On localising polymers around a positive velocity maximum, i.e., a forward flowing layer of fluid, the shift-reflect symmetry is broken and the two centre-mode eigenvalues no longer have the same imaginary parts. Fig.~\ref{fig:spectra}(b) presents the spectra for the same parameters as Fig.~\ref{fig:spectra}(a) but with a nonuniform base-concentration 
(corresponding to $\delta = 0.417 \pi$);
% in Fig.~\ref{fig:base}(a))
now the backward propagating centre-mode is stable while the forward propagating centre-mode is unstable (of course the reverse is true when polymers are localised about the negative velocity maximum). The nonuniform case (Fig.~\ref{fig:spectra}(b)) is more unstable than the uniform one (Fig.~\ref{fig:spectra}(a)), since $c_i$ of the unstable mode is larger in the former. 

In the following sections, we systematically study the effect of polymer localisation on the flow's stability by examining the change in the neutral stability curve in the $\Wi - k$ plane. To construct this curve, we fix $k$ and vary $\Wi$ until the eigenvalue with the largest $c_i$ becomes marginally unstable ($c_i = 0$). Repeating this process for different $k$ yields the $Wi-k$ neutral curve. This procedure is hampered by the presence of the neutral modes corresponding to the continuous spectrum of the $\tilde \theta$ advection equation. So, we stabilise these modes by introducing weak diffusion through the addition of $\Pen^{-1} \nabla^2 \tilde \theta$ to \eqref{eq:theta-tilde}, where $\Pen = VL/\mathcal{D}$ and the diffusivity $\mathcal{D}$ is such that $\Pen^{-1} \ll 1$; we also add diffusion to the perturbed conformation tensor equations \eqref{eq:tilde-Cxy} - \eqref{eq:tilde-Cyy} via $\Pen^{-1} \nabla^2 \tilde \C$ (see the \href{https://bighome.iitb.ac.in/index.php/s/g6wRiyEY8H3SNyR}{supplementary material}). The added diffusion stabilises the continuous-spectrum modes while leaving the centre-mode nearly unaltered ~\citep{Thomases25}, as evidenced by Fig.~\ref{fig:spectra}(c) which compares the spectra computed without diffusion to that computed using $\Pen^{-1} = 10^{-7}$. Hence, the addition of weak-diffusion to the perturbed equations alone (the base state is computed without diffusion) serves as a numerical strategy for facilitating the construction of neutral stability curves. We have cross-checked the neutral curves thus constructed by examining the spectra computed without diffusion at values of ($\Wi,k$) just inside and outside the curves.

\section{Localising polymers near the velocity-maximum is destabilizing}\label{sec:neutral}

\begin{figure}
       \centering
       \includegraphics[width=.4\linewidth]{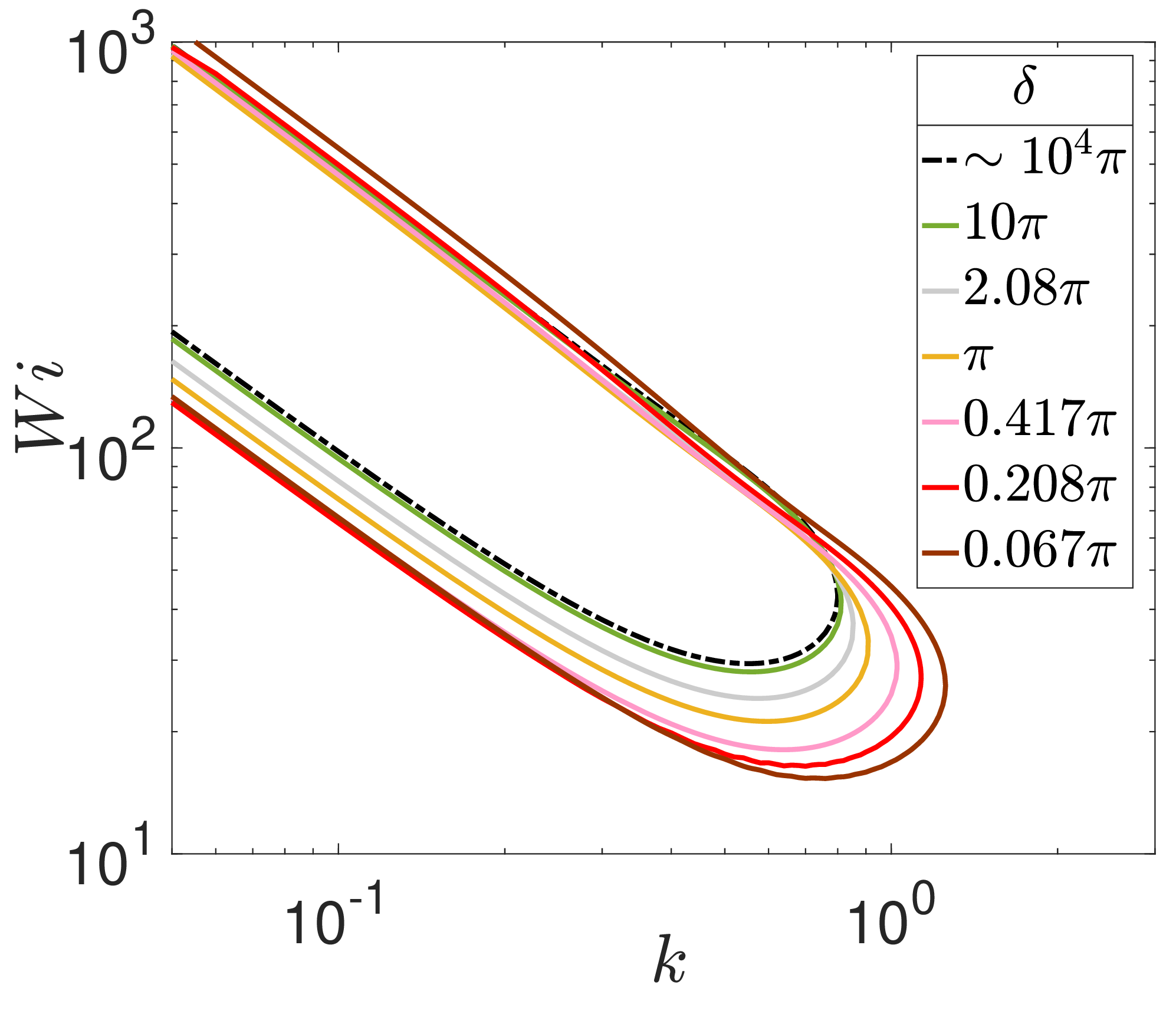}
       \put (-127,25){\footnotesize (a)}
       \hspace{1 em}
      \includegraphics[width=.4\linewidth]{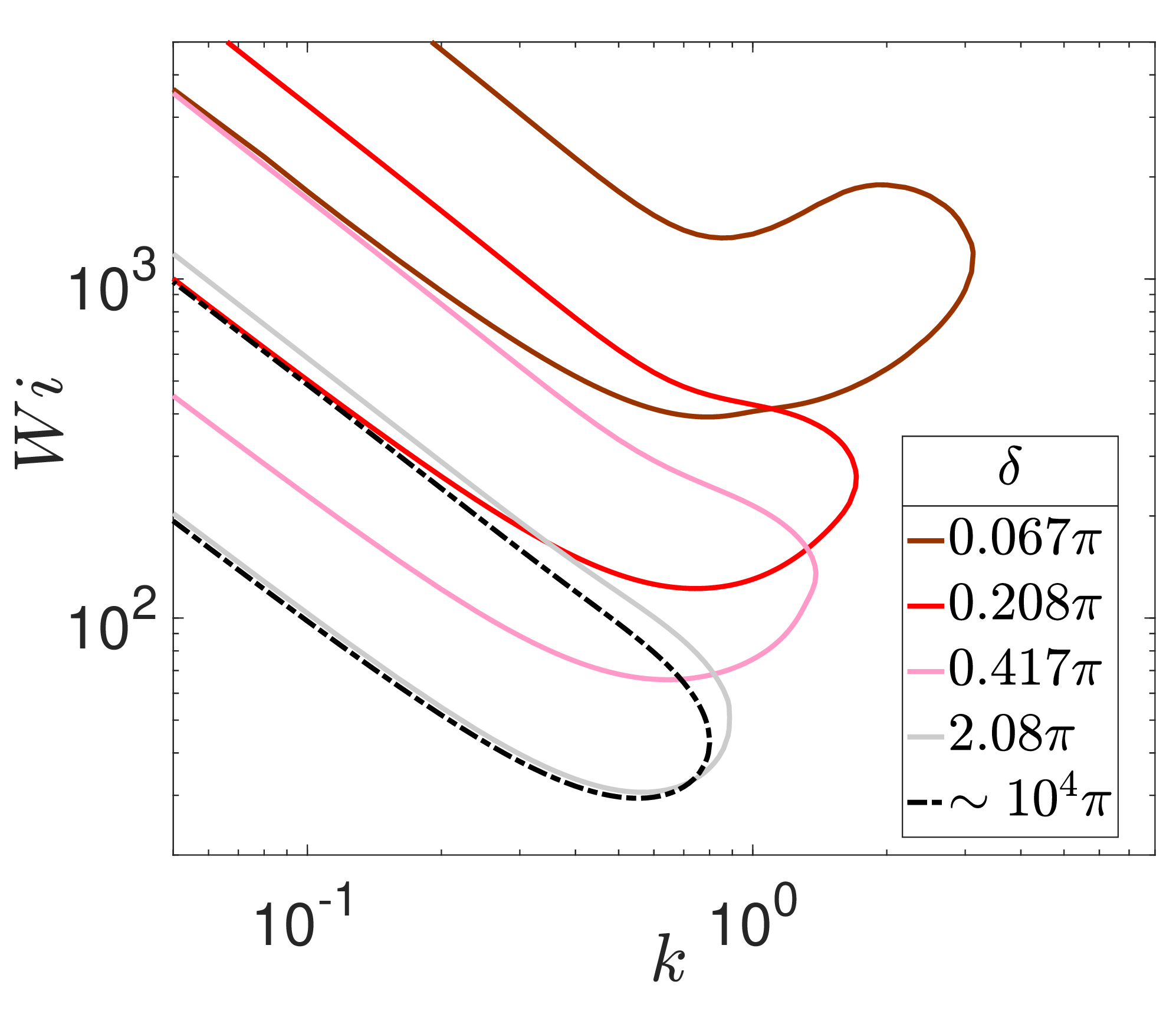}
       \put (-127,25){\footnotesize (b)}
       %       \includegraphics[width=.33\linewidth]{fig/Neutral_curve_W_k_plane_various_wd_B0pt9_ht16pi240_with_inset_of_theta_profiles.eps}
       % \put (-105,25){\footnotesize (c)}
      
\caption{Neutral curves in the $\Wi-k$ plane for Kolmogorov flow with different profiles of the nonuniform base-state polymer concentration (see the legend for the values of $\delta$). (a) Neutral curves when polymers are increasingly localised ($\delta$ is decreased) near the velocity maximum ($y = \pi$). (b)  Neutral curves when polymers are increasingly localised near the shear maximum ($y = \pi/2$). Here, the solution is very dilute with $\beta = 0.95$.} 
\label{fig:delta}
\end{figure}

Let us first examine how the stability of the flow changes as polymers are increasingly localised. Figure~\ref{fig:delta}(\textit{a}) shows how the neutral stability curve, in the $\Wi-k$ plane, changes as the base-state concentration is made increasingly nonuniform by decreasing $\delta$ in \eqref{eq:pulse}, so that the concentration profile transitions from being uniform  ($\delta \sim 10^4$) to being that of a distinct polymer-laden stream overlying the velocity-maximum ($\delta = 0.067 \pi$, see Fig.~\ref{fig:base}(a)). The neutral stability curves, which enclose the region of instability in the $\Wi-k$ plane, are seen to shift downward. Thus, decreasing $\delta$ reduces the critical Weissenberg number $\Wi_c$, defined as the minimum value of $\Wi$ required for the flow to be unstable to perturbations of any wavenumber $k$ ($\Wi_c$ is the smallest value of $\Wi$ sampled by the neutral curve). Decreasing $\delta$ also causes the neutral curves to enlarge, implying an increase in the range of unstable wavenumbers for any fixed value of $\Wi > \Wi_c$. Clearly, localising polymers about the velocity maximum destabilises the flow. 

Figure~\ref{fig:delta}(b) shows the effect of localising polymers about the location of maximum shear at $y = \pi/2$ (the corresponding base state is visualized in the \href{https://bighome.iitb.ac.in/index.php/s/g6wRiyEY8H3SNyR}{supplementary material}).
% ; i.e., the profile of \eqref{eq:pulse} is shifted by $-\pi/2$ before decreasing $\delta$. 
In this case, the neutral curves shift upwards significantly, showing that the flow is strongly stabilised. 
Since an increase in the concentration near the shear-maximum is accompanied by a decrease near the velocity-maximum, the stabilisation evidenced by Fig.~\ref{fig:delta}(b) may be attributed to the weakening of viscoelastic stress near the velocity-maximum. This result is consistent with
% Recognizing that the increase in the concentration near the location of maximum shear coincides with a decrease near the location of maximum velocity, we note that the stabilisation evidenced by Fig.~\ref{fig:delta}(b) is consistent 
the necessity of 
a maximum in the base velocity profile 
% being necessary 
for the centre-mode instability~\citep{Yadav_et_al_PRF2024}.
% with prior work that has shown the importance of having a maximum in the base velocity profile for the center-mode to be unstable~\citep{Yadav_et_al_PRF2024}. 

\section{Sudden destabilisation on increasing polymer loading}\label{sec:beta-effect}

We now focus on the nonuniform case wherein 
a polymer-laden stream is located at the velocity maximum
% polymers are localised about the velocity maximum 
(i.e., the case of $\delta = 0.067 \pi$ in Fig.~\ref{fig:base}(a)) and examine the influence of increasing the total polymer loading (decreasing $\beta$). 

In Kolmogorov flow with a uniform distribution of polymers, the centre-mode becomes unstable at lower values of $\Wi$ as the polymer loading is increased~\citep{Kerswell_page,Lewy_Kerswell_2025}. Figure~\ref{fig:neutral-beta}(a) shows how the neutral stability curves of the uniform system shift  downward to smaller $\Wi$ as $\beta$ is decreased. This trend holds in the nonuniform case as well, as evidenced by Fig.~\ref{fig:neutral-beta}(b). However, we see that the neutral curves of the nonuniform system undergo a qualitative change as $\beta$ is decreased. The curve, which has a single-lobed U-shape for $\beta$ near unity, develops a second lobe for moderate $\beta$ (see the curves for  $\beta = 0.8$ and 0.7), and then becomes single-lobed again at small $\beta$ (curves for $\beta \lessapprox 0.5$). 

\begin{figure}
       \centering
        % \begin{subfigure}[b]{0.32\textwidth}
       % \includegraphics[width=.23\linewidth]{fig/neutral_curves_Wi_k_plane_various_Beta_uniform_vKf_updated.eps}
       % \put (-25,25){\footnotesize (a)}
       % \hspace{1.5 em}
       %  \includegraphics[width=.4\linewidth]{fig/Neutral_curve_nonuniform_split_into_two_B095to8_0pt7to0pt3.eps}
       %  \put (-20,25){ \footnotesize (b)}\\
        \includegraphics[width=.333\linewidth]{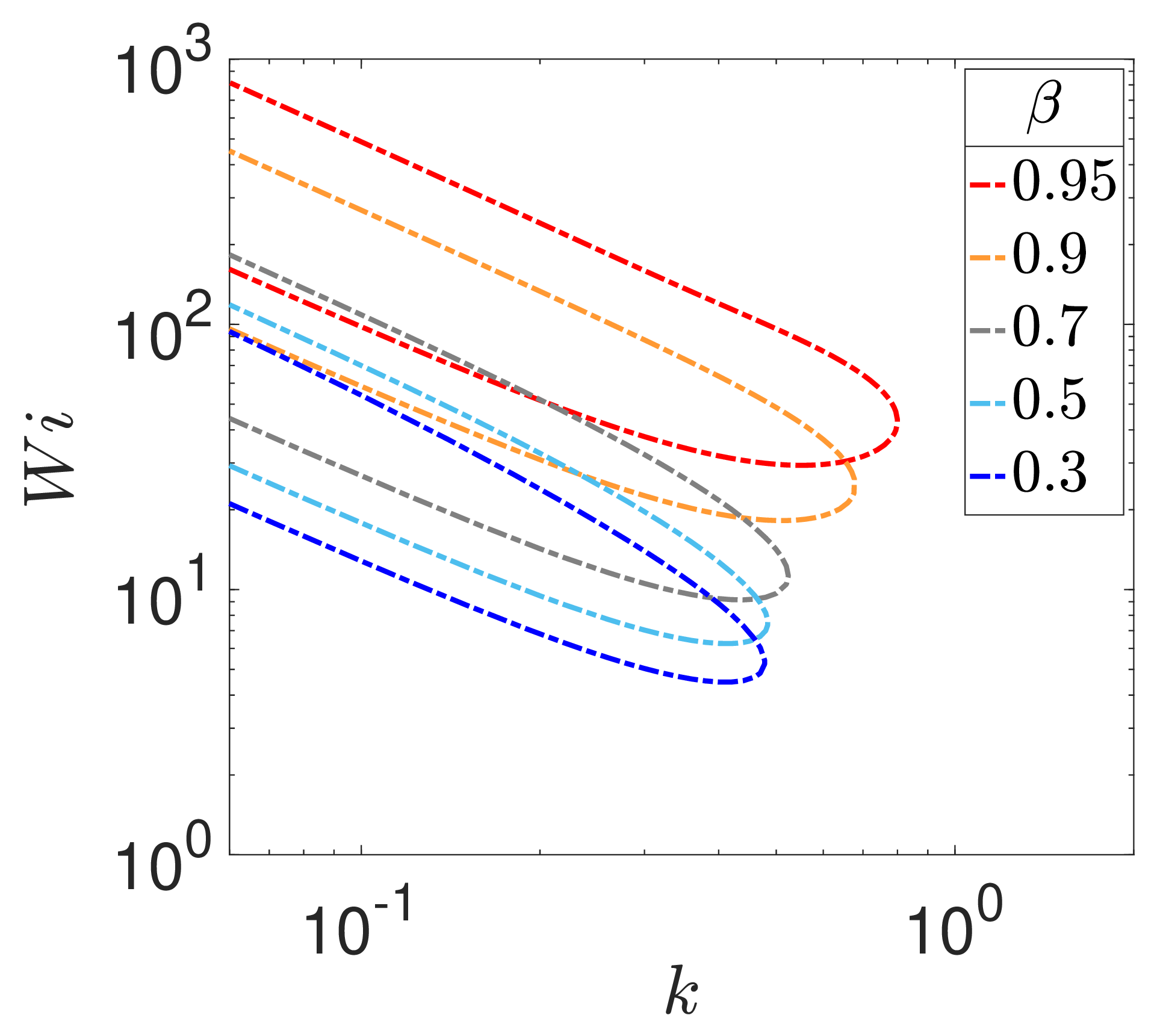}
        % neutral_curves_Wi_k_plane_various_Beta_uniform_vKf_updated_squared.eps}
       \put (-99,24){\footnotesize (a)}
       % \hspace{0.25 em}
        \includegraphics[width=.63\linewidth]{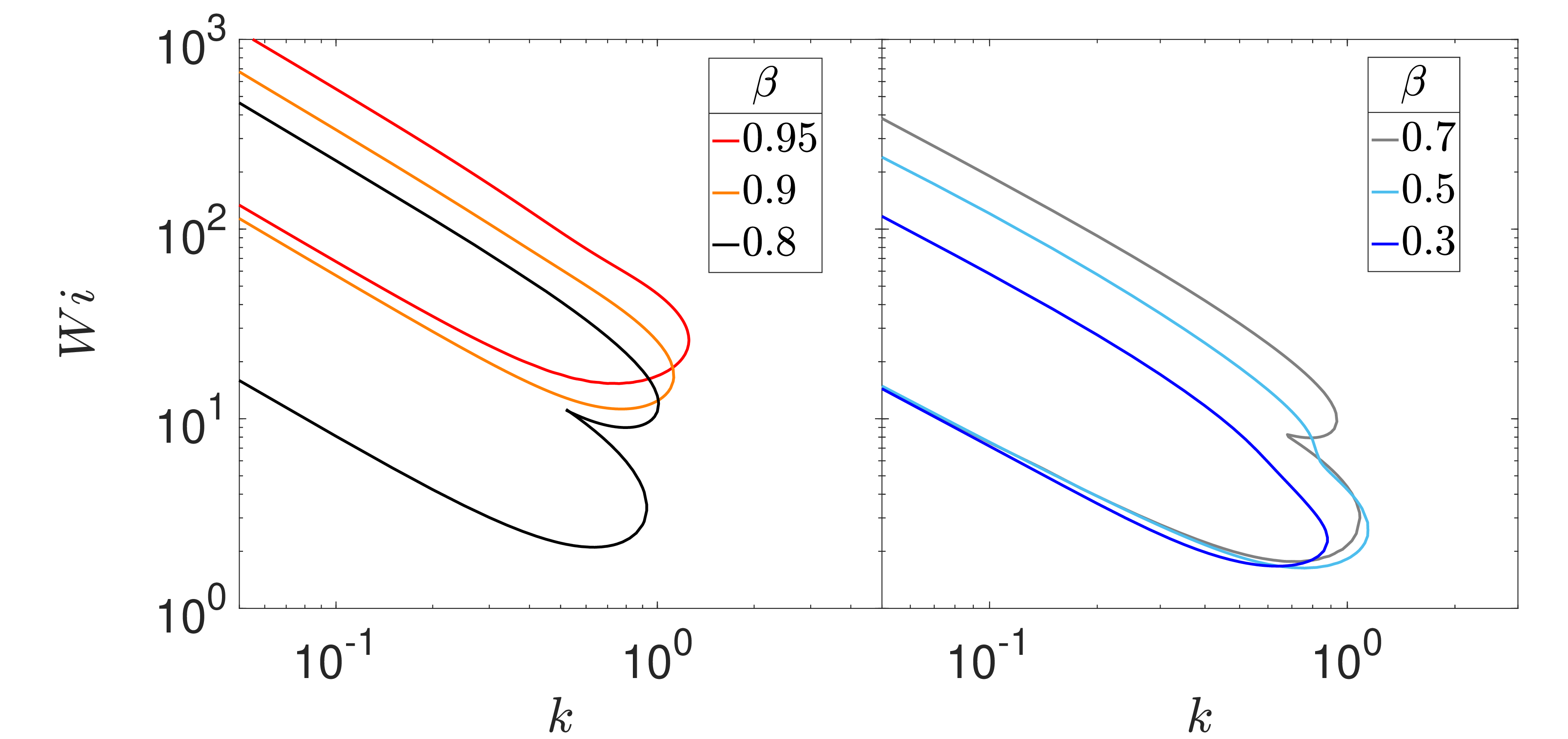}
        % Neutral_curve_nonuniform_split_into_two_B095to8_0pt7to0pt3_squared.eps}
        \put (-204,24){ \footnotesize (b)}\\
	\caption{Effect of increasing the total polymer loading (decreasing $\beta$) on the neutral curves in the $Wi-k$ plane for Kolmogorov flow with a base-state having (a) uniformly distributed polymers and (b) polymers localised about the velocity maximum ($\bar{\theta}$ corresponds to $\delta = 0.067\pi$ in Fig.~\ref{fig:base}(a)). }
	\label{fig:neutral-beta}
\end{figure}
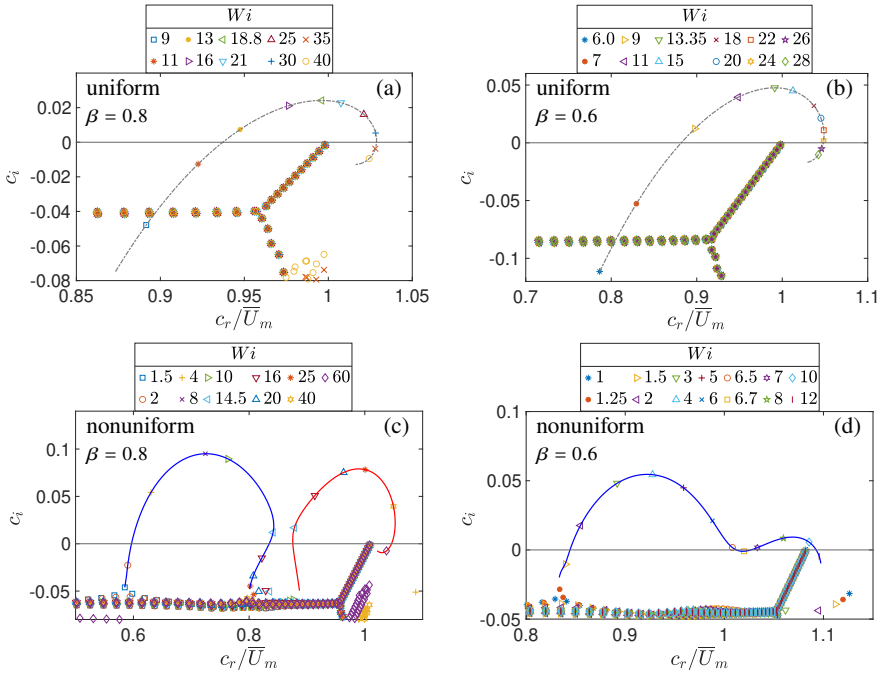
\begin{figure}
\centering
        \includegraphics[width=0.42\linewidth] {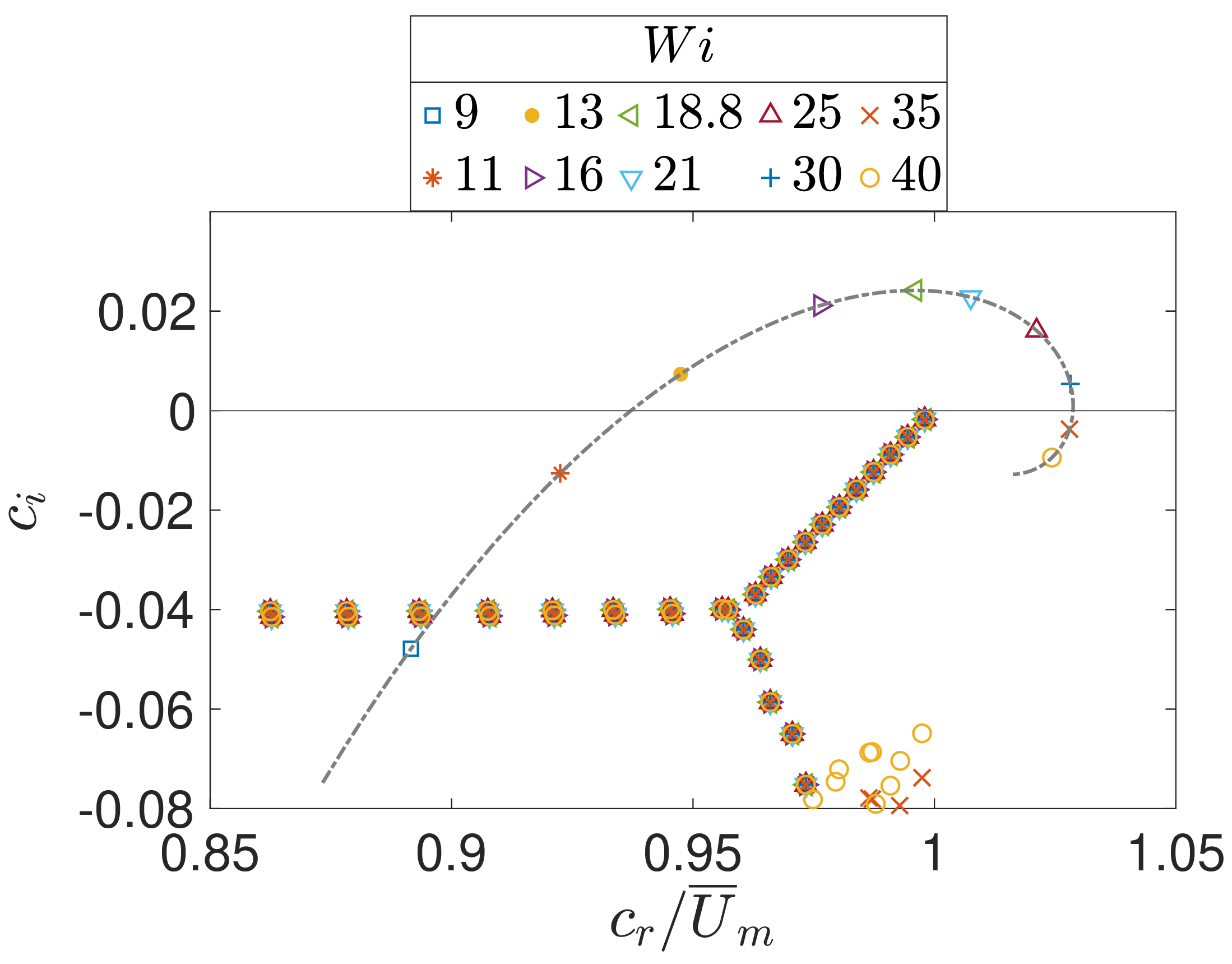}
        \put (-21,90){\footnotesize (a)}
        \put (-130,90){\footnotesize uniform}
        \put (-130,80){\scriptsize $\beta = 0.8$}
        \hspace{0.5 em}
         \includegraphics[width=0.42\linewidth] {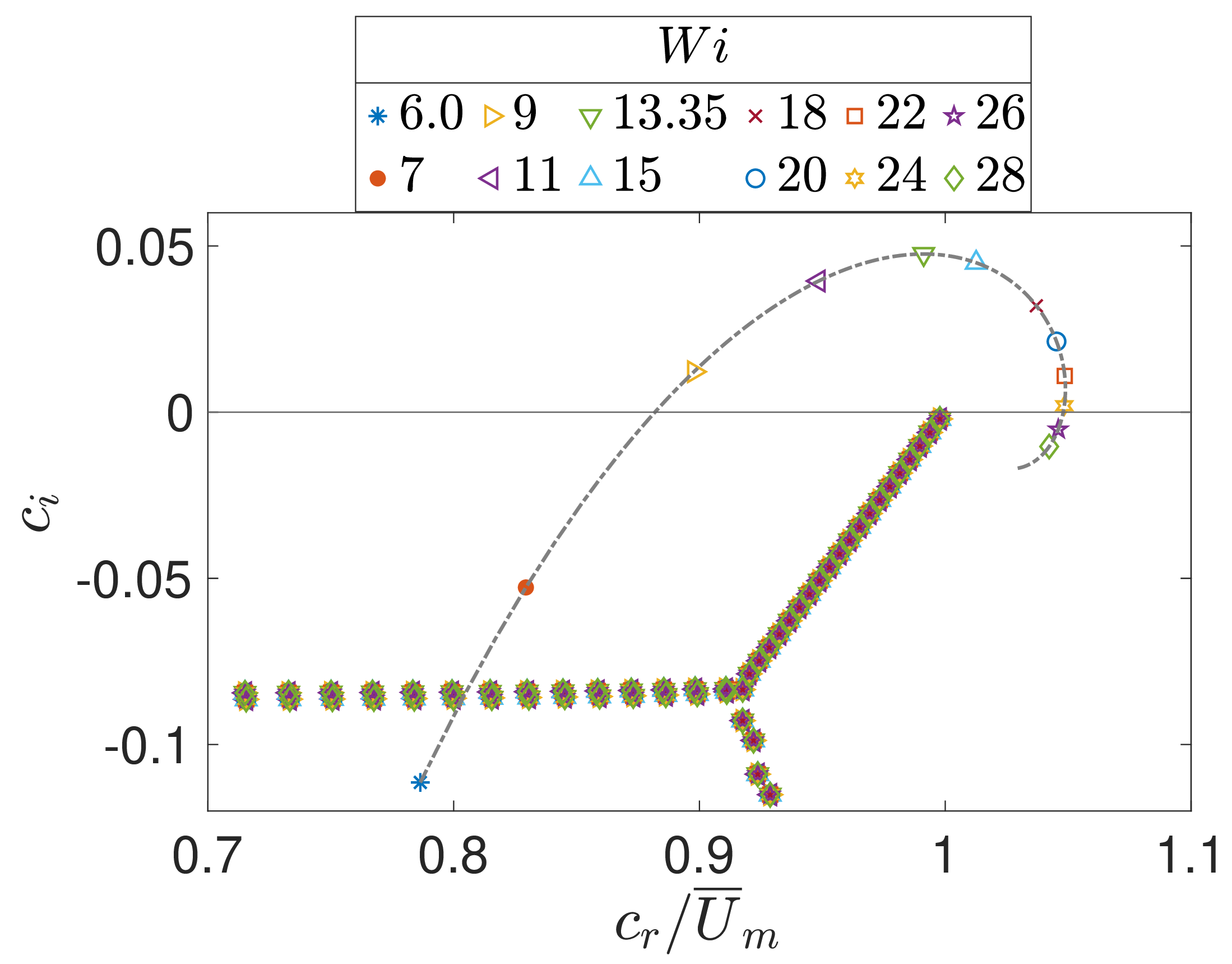}
        \put (-21,90){\footnotesize (b)}
        \put (-130,90){\footnotesize uniform} 
        \put (-130,80){\scriptsize $\beta = 0.6$} \\
            \includegraphics[width=0.42\linewidth] {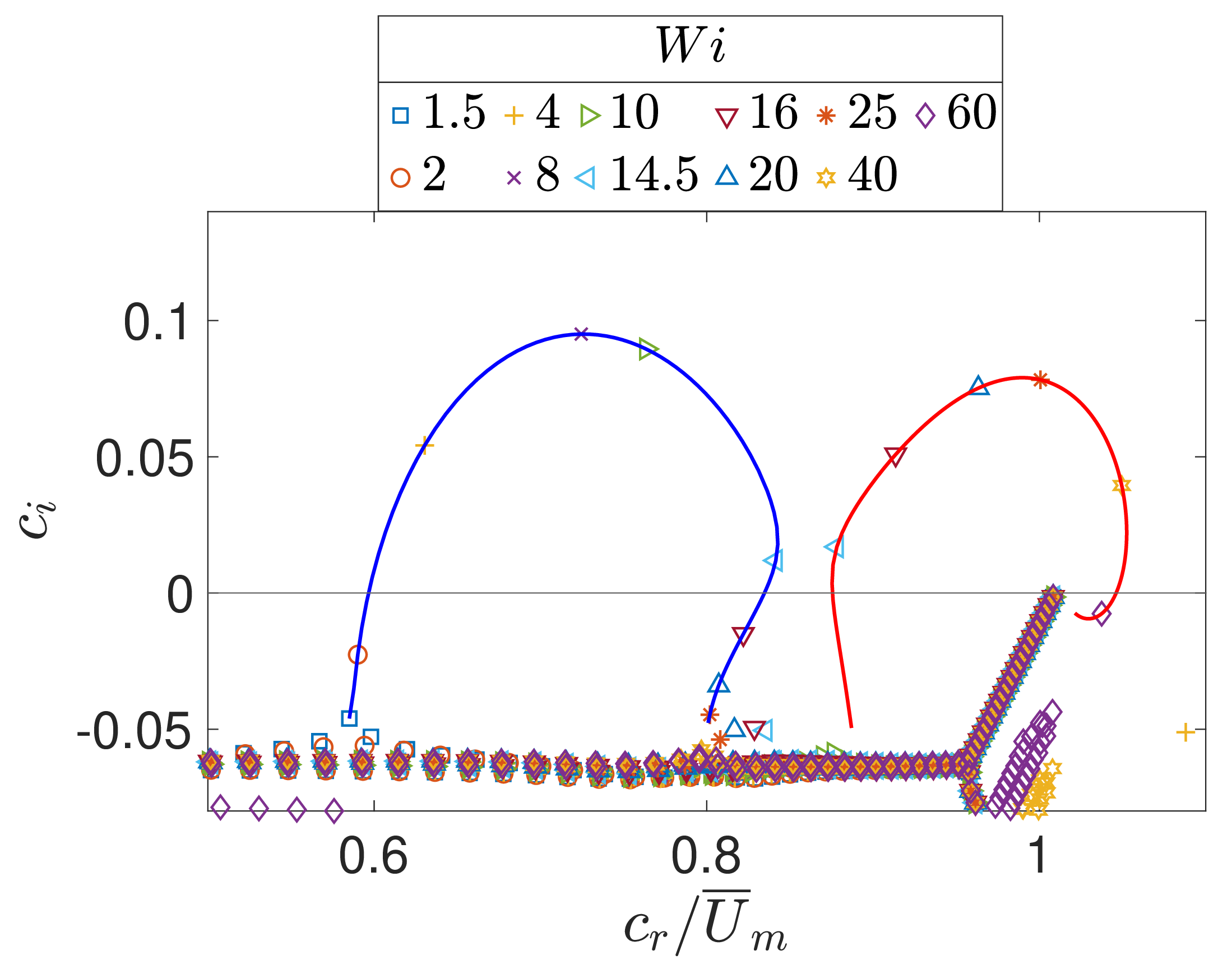}
        \put (-18,90){\footnotesize (c)}
        \put (-130,90){\footnotesize nonuniform}
        \put (-130,80){\scriptsize $\beta = 0.8$}
        \hspace{0.5 em}
    \includegraphics[width=0.42\linewidth] {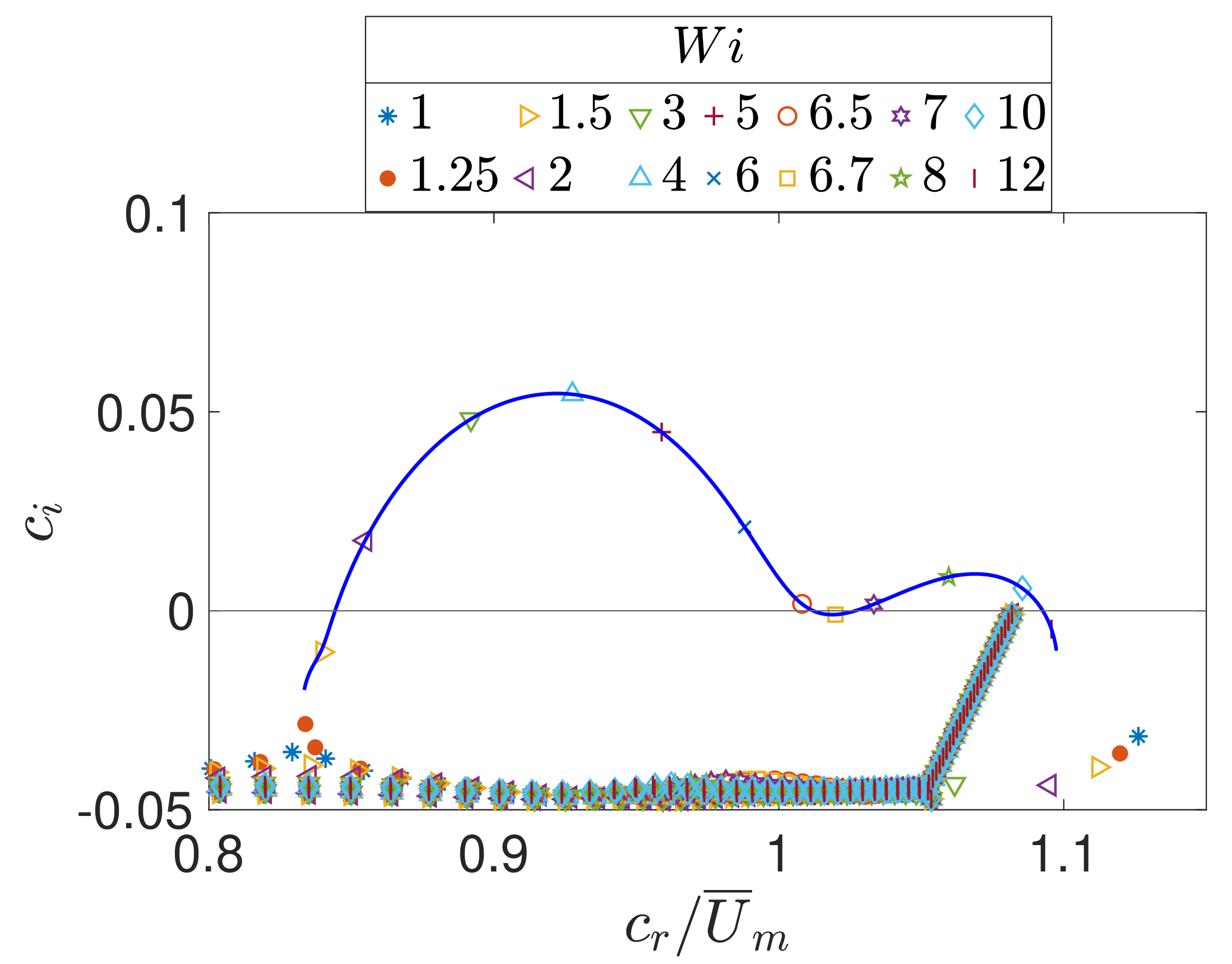}
        \put (-18,90){\footnotesize (d)}
        \put (-130,90){\footnotesize nonuniform}
        \put (-130,80){\scriptsize $\beta = 0.6$}

    \caption{Overlay of eigenspectra for Kolmogorov flow, showing how the centre-mode becomes unstable as $\Wi$ is varied (see the legends). (a,b) Uniform base-state polymer concentration with $\beta = 0.8, k = 0.4$ and $\beta =0.6, k = 0.3$, respectively; (c,d) Nonuniform base-state polymer concentration ($\bar{\theta}$ corresponds to $\delta = 0.067 \pi$ in Fig.~\ref{fig:base}(a)) with $\beta = 0.8, k = 0.4$ and $\beta = 0.6, k = 0.815$, respectively. The continuous curves shown in red and blue in panel (c) trace the motion of the two distinct eigenvalues as $\Wi$ increases. To show the evolution of the centre mode clearly, we stabilize the band of neutral modes of the continuous spectrum of the $\hat\theta$-advection equation by adding diffusion with $\Pen^{-1} = 5\times10^{-6}$.
    \label{fig:eigen-double}
    }
\end{figure}

Figure~\ref{fig:eigen-double} visualizes the variation of the eigenspectra with $\Wi$, for the uniform and nonuniform cases (first and second rows, respectively) with moderate values of $\beta = 0.8$ and $0.6$ (first and second columns, respectively). In each panel, the eigenspectra for a fixed value of $k$ and different values of $\Wi$ (see the legends) are overlaid on the same plot; while the stable eigenvalues nearly overlap, the unstable ones trace out curves that reveal how their growth rates $c_i$ vary with $\Wi$. These curves of $c_i$ correspond to constant-$k$ vertical lines in Fig.~\ref{fig:neutral-beta} and so $c_i$ will become positive for values of $\Wi$ that are enclosed by the neutral curves. Since the neutral curves are single-lobed in the uniform case (Fig.~\ref{fig:neutral-beta}(a)), the corresponding $c_i$ becomes positive over a single window of $\Wi$ values (Figs.~\ref{fig:eigen-double}(a,b)). However, in the non-uniform case, $c_i$ is positive for two distinct ranges of $\Wi$ (Figs.~\ref{fig:eigen-double}(c,d)), provided $k$ is chosen from the double-lobe region of the corresponding neutral curve (Fig.~\ref{fig:neutral-beta}(b)). For $\beta = 0.8$ (Fig.~\ref{fig:eigen-double}(c)), two distinct eigenvalues becomes positive, one for relatively large $\Wi$ (red trace) and the other for relatively small $\Wi$ (blue trace); the former has a higher wavespeed $c_r$ than the latter. The two ranges of $\Wi$ overlap slightly as demonstrated by the fact that both eigenvalues are unstable at $\Wi = 14.5$ in Fig.~\ref{fig:eigen-double}(c). 
% The eigenvalue that is unstable at smaller $\Wi$ corresponds to the lower second lobe in the corresponding neutral curve (black curve for $\beta = 0.8$ in Fig.~\ref{fig:neutral-beta}(b)).  
The behaviour of the eigenspectra is different for $\beta = 0.6$ (Figs.~\ref{fig:eigen-double}(d)); here, the eigenvalue that is unstable at large $\Wi$ becomes unstable again at smaller $\Wi$. 

%----------------------------------
\begin{figure}
\centering
          \includegraphics[width=0.33\linewidth] {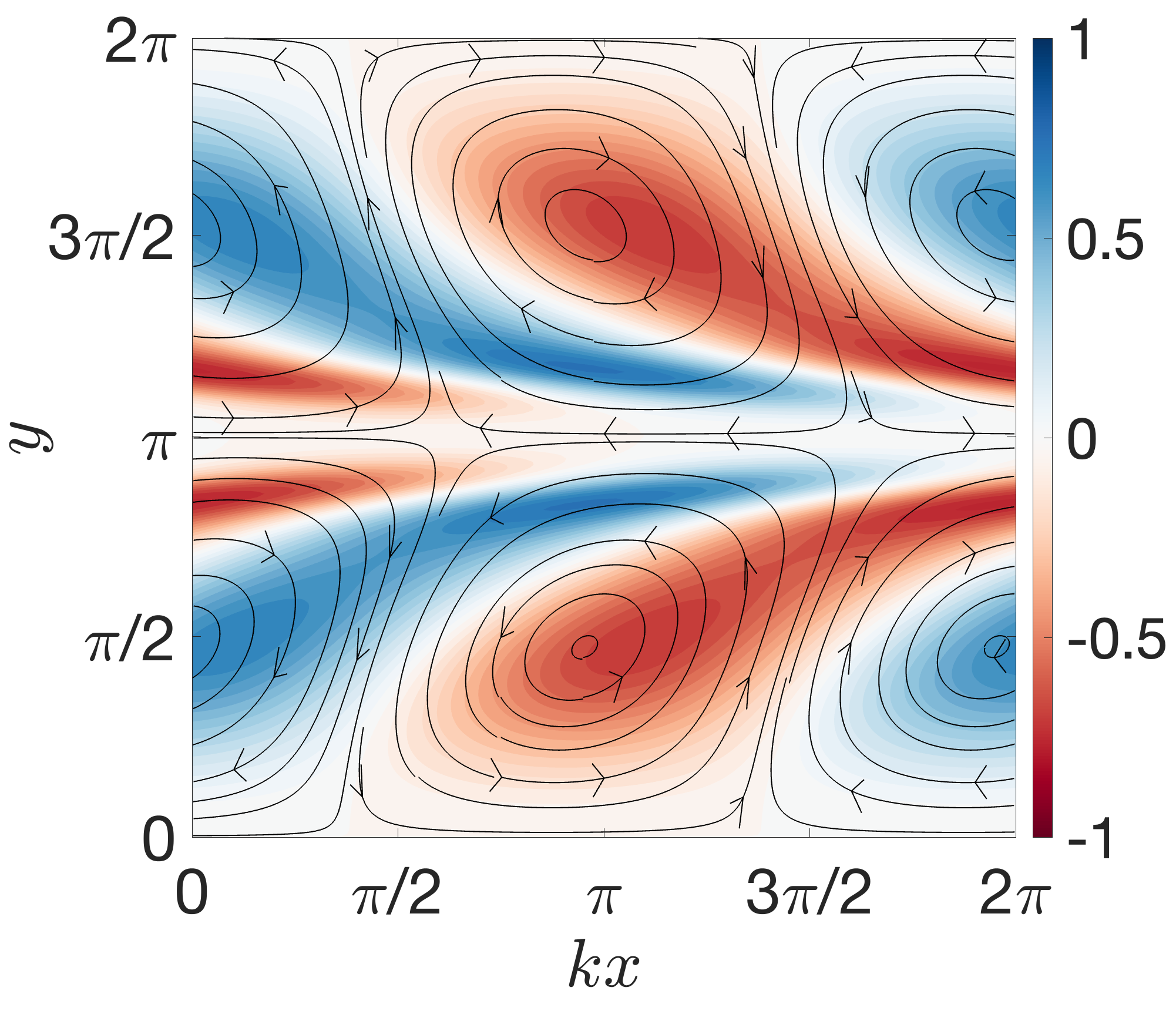}
        \put (-104,107){\footnotesize (a) uniform}
        % \put (-130,125){\footnotesize uniform}
        % \put (-130,115){\scriptsize  $\Wi = 18.8, k = 0.4$}
        % \put (-130,115){\scriptsize $\beta = 0.8$}
         % \includegraphics[width=0.33\linewidth] {fig/contour_real_trace(C)_nonuniform_second_peak_beta0pt8_Wi23pt25_k0pt4_kap5e_neg6.eps}
        \includegraphics[width=0.33\linewidth] {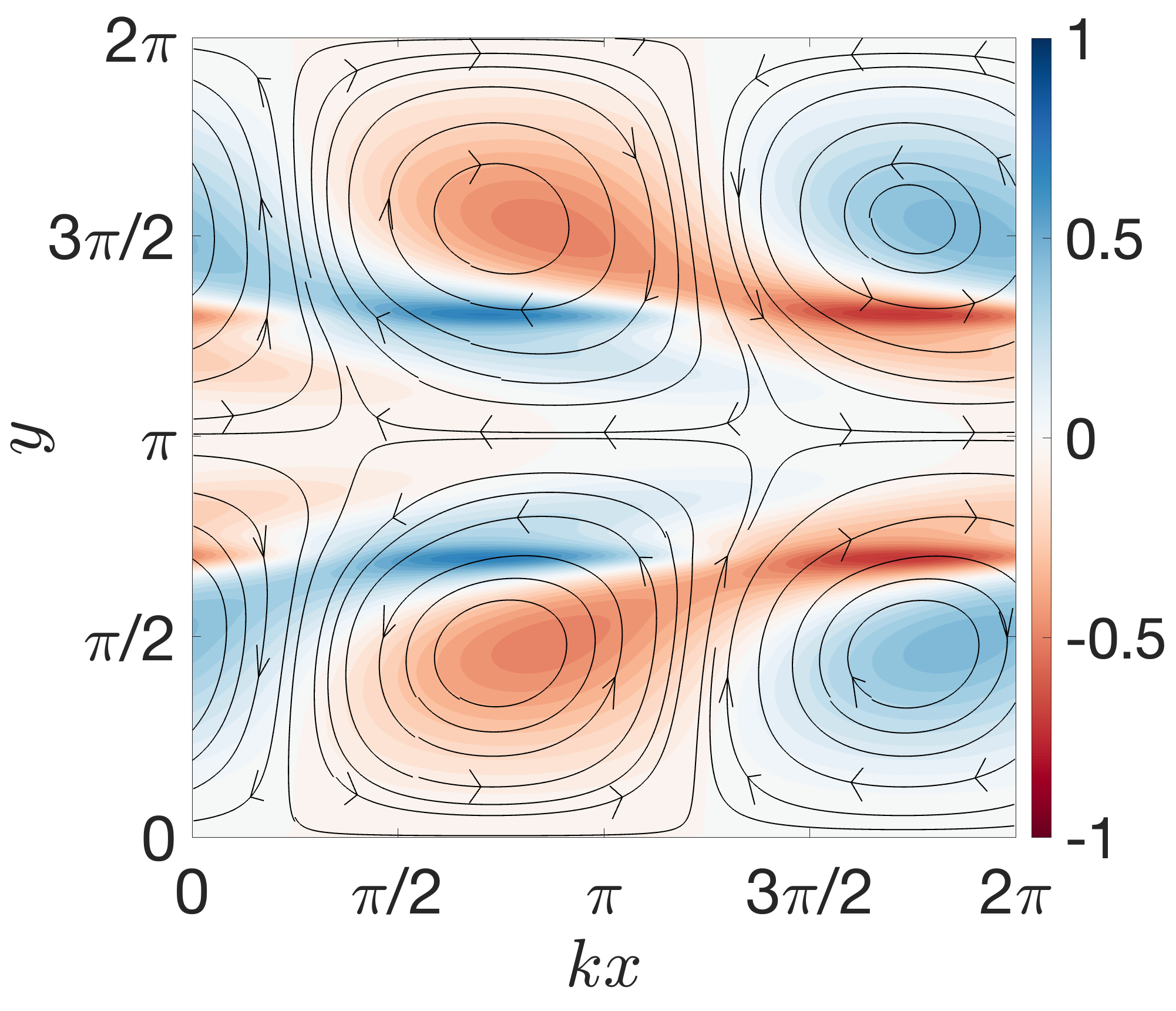}
         \put (-104,107){\footnotesize (b) nonuniform,  higher $c_r$}
        % \put (-130,125){\footnotesize nonuniform,  higher $c_i$}
        % \put (-130,115){\scriptsize  $\Wi = 23.25, k = 0.4$}
        % \put (-130,115){\scriptsize $\beta = 0.8$}
        % \hspace{0.5 em}
         % \includegraphics[width=0.33\linewidth] {fig/contour_real_trace(C)_nonuniform_first_peak_beta0pt8_Wi8_k0pt4_kap5e_neg6.eps}
         \includegraphics[width=0.33\linewidth] {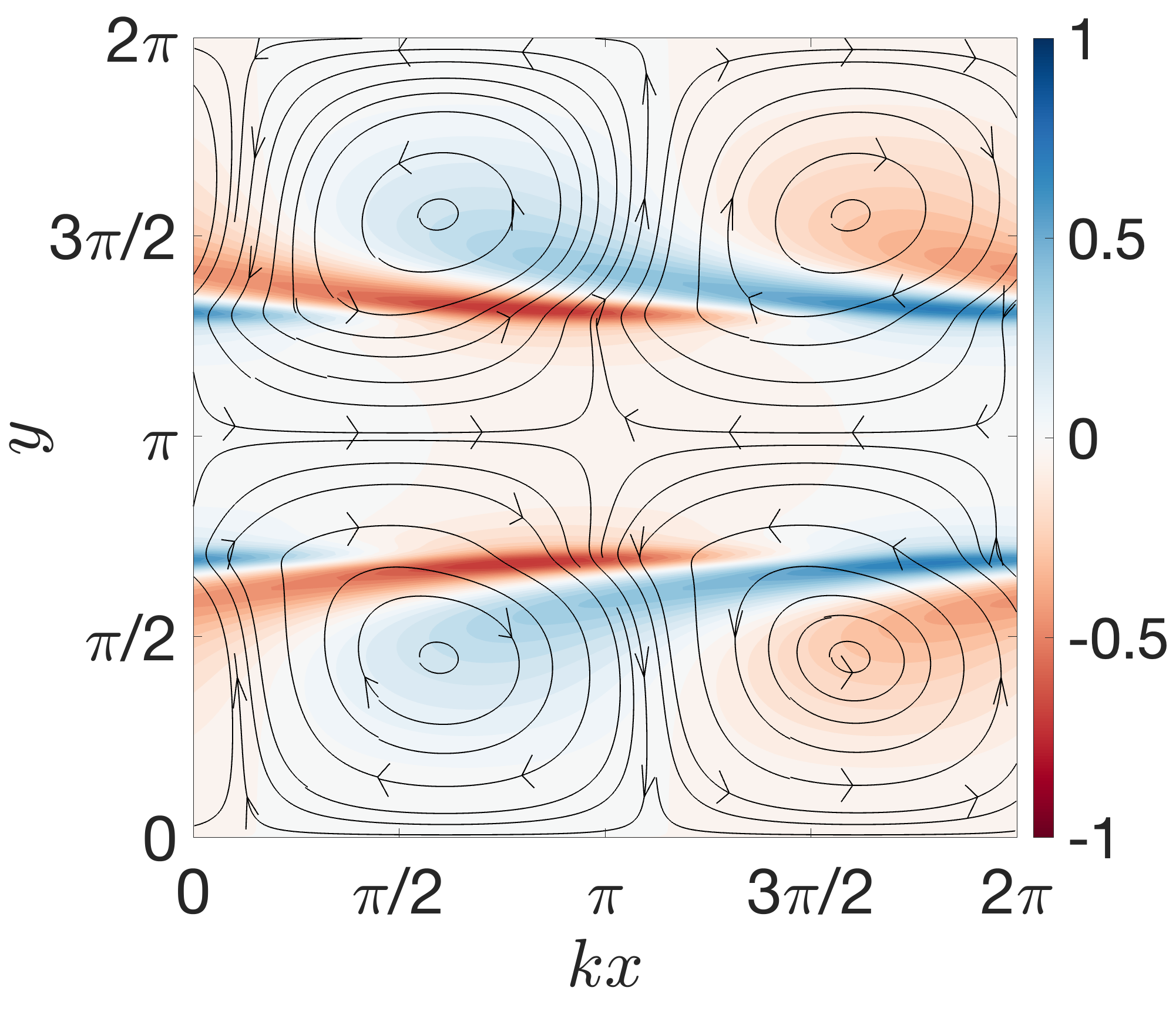}
        \put (-104,107){\footnotesize (c) nonuniform,  lower $c_r$}
        % \put (-130,125){\footnotesize nonuniform,  lower $c_i$} 
        % \put (-130,115){\scriptsize  $\Wi = 8.0, k = 0.4$} 
        % \put (-130,115){\scriptsize $\beta = 0.8$} 
    \caption{Visualisation of unstable eigenfunctions in Kolmogorov flow, showing contours of the perturbed polymer squared-extension $\mathrm{tr}(\hat{\C})$ overlaid with the streamlines of the disturbance velocity field. 
    (a) Uniform base-state concentration with $\Wi = 18.8, k = 0.4$, (b) Nonuniform base-state concentration ($\bar{\theta}$ corresponds to $\delta  = 0.067 \pi$ in Fig.~\ref{fig:base}(a)) with $\Wi = 23.25, k = 0.4$, (c) Nonuniform base-state concentration  with  $\Wi = 8.0, k = 0.4$. These eigenfunctions correspond to the fastest-growing modes (peaks of the traces) in Figs.~\ref{fig:eigen-double}(a,c) where $\beta = 0.8$. Analogous plots for $\beta = 0.6$ are presented in the \href{https://bighome.iitb.ac.in/index.php/s/g6wRiyEY8H3SNyR}{supplementary material}.
    % Before plotting normalize the field by the max(abs(min),abs(max)) so that the fields are bounded between (-1,1) and then plot; this will allow a fair comparison of spatial structure across the plots---all 
    \label{fig:eigenfunc}
    }
\end{figure}

% \begin{figure}
% \centering
%         \includegraphics[width=0.33\linewidth] {fig/Real_part_of_contour_traceC_and_streamlines_nonuniform_low_phase_speed_beta0pt8_Wi14pt5_k0pt4_kap5e_neg6.eps}
%         \put (-104,107){\footnotesize (a) low $c_r/\bar{U}_m$ mode}
%         \includegraphics[width=0.33\linewidth] {fig/Real_part_of_contour_traceC_and_streamlines_nonuniform_high_phase_speed_beta0pt8_Wi14pt5_k0pt4_kap5e_neg6.eps}
%         \put (-104,107){\footnotesize (b)  high $c_r/\bar{U}_m$ mode}
%     \caption{Contours of trace of the perturbed conformation tensor, with overlaid streamlines of perturbed velocity for a non uniform base state. Data are shown for the  two  unstable modes shown in fig.\,5c ; with $Re = 0$, $\beta = 0.8, Wi = 14.5$ and $k = 0.4$. Panel (a) shows data for low phase speed mode, while (b)  for high phase speed mode. 
%    % \label{fig:eigenfunc}
%     }
% \end{figure}
%----------------------------

The structure of the eigenfunctions of the unstable modes, for $\beta = 0.8$, are illustrated in Fig.~\ref{fig:eigenfunc}. Here, we visualize the field of $\mathrm{tr}(\hat\C)$, the perturbed squared-extension of polymers,
% the trace of the perturbed conformation tensor 
using a pseudocolour plot, over which we overlay streamlines of the perturbation velocity field. A typical eigenfunction of the centre-mode for the uniform case, corresponding to the fastest growing mode in Fig.~\ref{fig:eigen-double}(a), is shown in Fig.~\ref{fig:eigenfunc}(a). Typical eigenfunctions of the two types of nonuniform unstable modes are shown in Fig.~\ref{fig:eigenfunc}(b,c); the fastest-growing among the higher speed modes---peak of the red trace in Fig.~\ref{fig:eigen-double}(c)---is visualized in Fig.~\ref{fig:eigenfunc}(b), while the fastest-growing among the lower speed modes---peak of the blue trace in Fig.~\ref{fig:eigen-double}(c)---is visualized in Fig.~\ref{fig:eigenfunc}(c).
 % \hl{both these modes are simultaneously unstable at $\Wi - 14.5$ in} 
% The slower mode with smaller $c_r$ is shown in Fig.~\ref{fig:eigenfunc}(b) while the faster mode with larger $c_r$ is shown in Fig.~\ref{fig:eigenfunc}(c). 
The two nonuniform modes have a similar spatial structure, though the activity is slightly more spread out in the higher speed mode. Compared to the uniform case, the structure of the modes in the nonuniform case is more spatially restricted; indeed, the activity is focused near regions with large base-state polymer-concentration gradients, i.e., where the polymer-laden stream meets the polymer-free streams (compare Figs.~\ref{fig:eigenfunc}(b,c) with the $\bar \theta$ profile for $\delta = 0.067 \pi$ in Fig.~\ref{fig:base}(a)). This feature of the nonuniform eigenfunctions is even more pronounced at higher total polymer-loadings (the analogue of Fig.~\ref{fig:eigenfunc} for $\beta = 0.6$ is presented in the \href{https://bighome.iitb.ac.in/index.php/s/g6wRiyEY8H3SNyR}{supplementary material}).

\begin{figure}
       \centering
        % \begin{subfigure}[b]{0.32\textwidth}
        \includegraphics[width=0.4\linewidth]{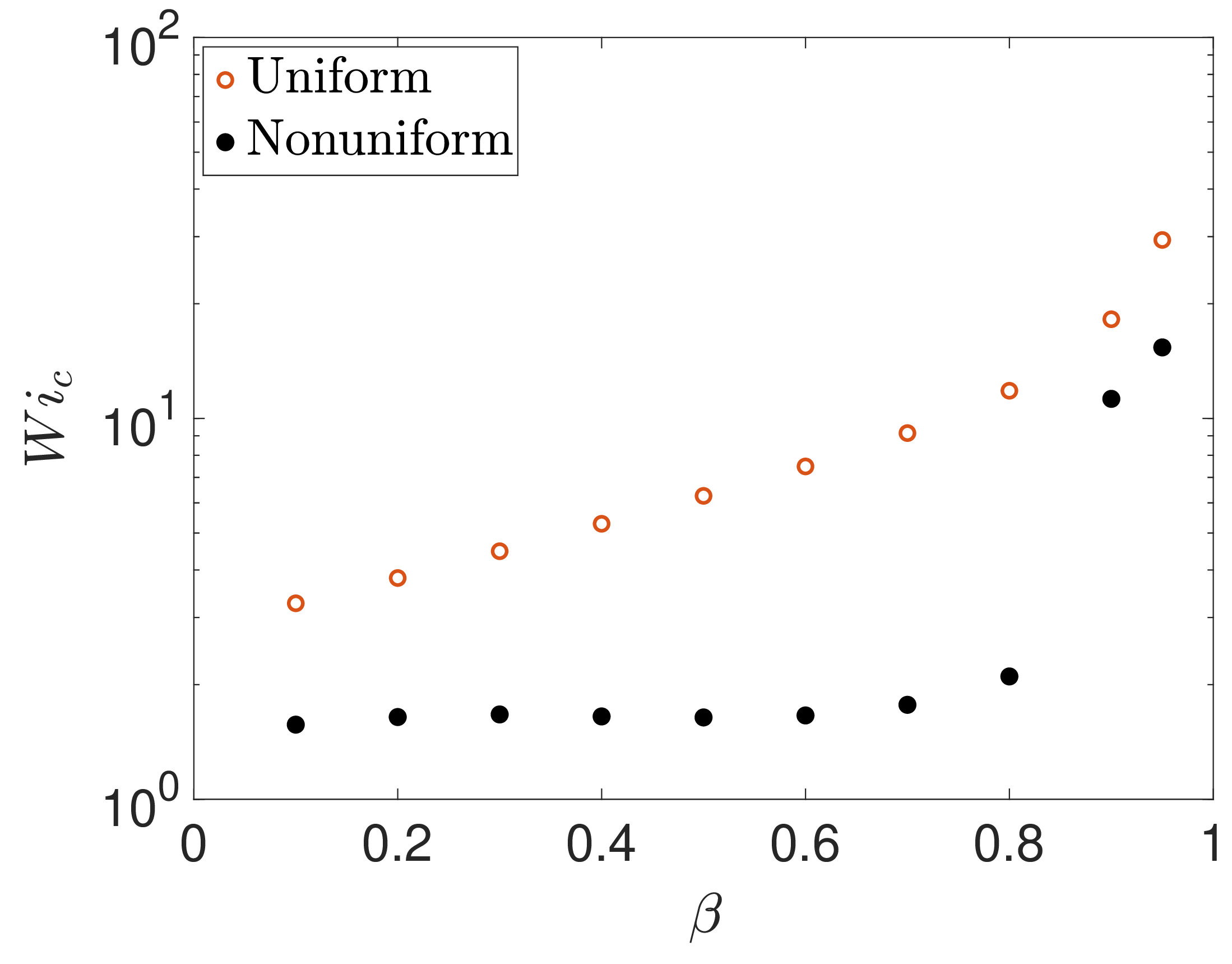}
        \put (-16,24){\footnotesize (a)}
        \hspace{1 em}
                \includegraphics[width=0.4\linewidth]{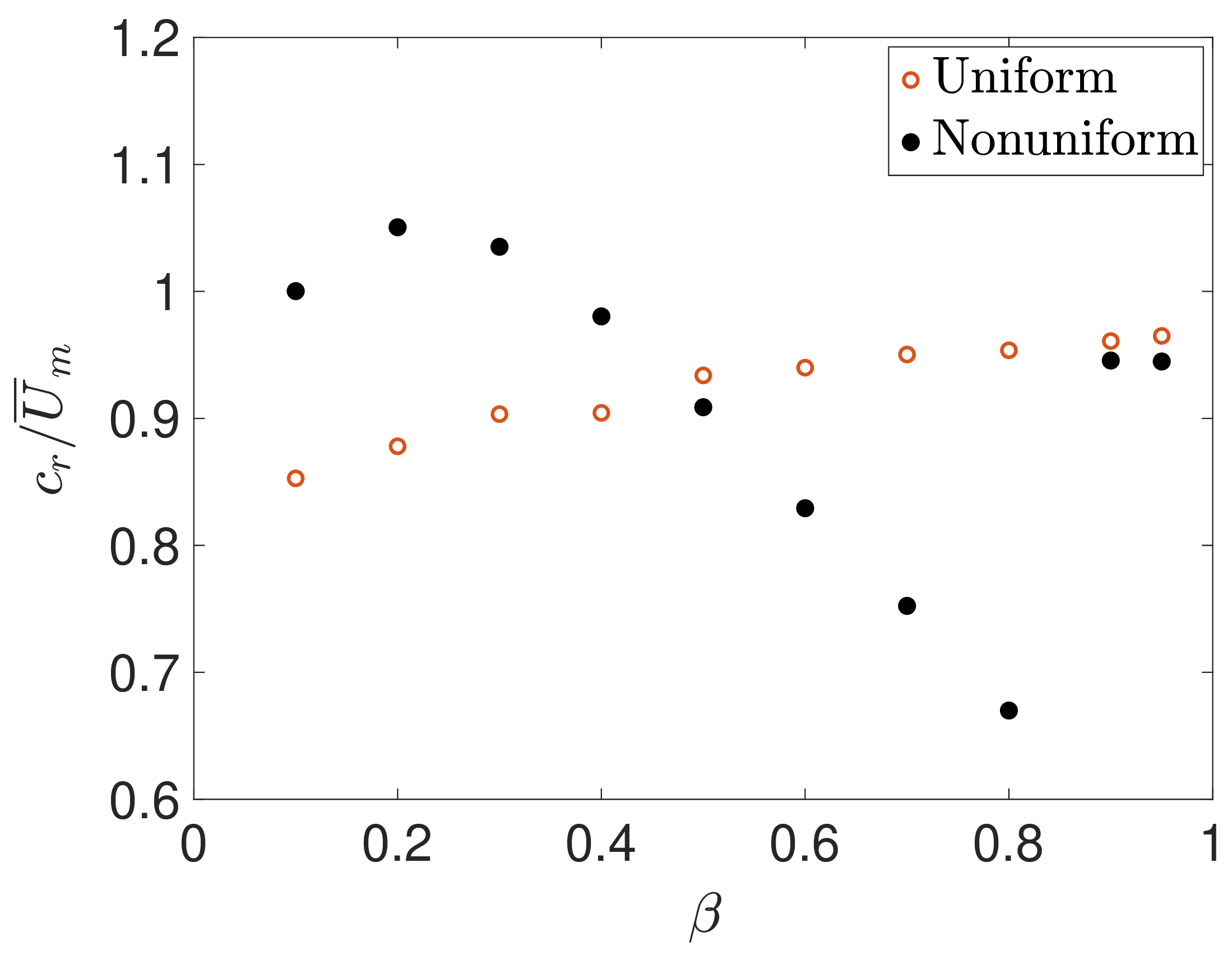}
        \put (-16,24){\footnotesize (b)}

        % \includegraphics[width=.43\linewidth]{fig/Neutral_curve_Wi_vs_k_vKf_nonunifiorm_ht16pi240_updtaed_with_beta0pt8_two_modes_updated_black.eps}
        % \put (-20,85){(d)}
       % \hspace{0.005\textwidth}
        % \caption{$Wi_c $ vs. $\beta$.}
         % \label{figch3:11c}
        % \end{subfigure}
	\caption{(a) Variation of the critical Weissenberg number $Wi_c$ as the total polymer loading is increased ($\beta$ is decreased) in Kolmogorov flow with a uniform and a nonuniform (see the legend) base-state polymer concentration. (b) Wave speed of the critical mode. }
	\label{fig:critical-beta}
\end{figure}

Selecting the minimum values of $\Wi$ sampled by the neutral curves in \ref{fig:neutral-beta}, we plot the variation of $\Wi_c$ with $\beta$ in Fig.~\ref{fig:critical-beta}(a). The wave-speed of the corresponding critical modes are plotted in Fig.~\ref{fig:critical-beta}(b). Results for both the uniform and nonuniform cases are presented. First, we see that $\Wi_c$ is always smaller in the nonuniform case, which demonstrates that localising a given amount of polymer near the velocity maximum destabilises the flow. Second, increasing the polymer loading (decreasing $\beta$) reduces $\Wi_c$, gradually in the uniform case and sharply in the nonuniform case. The emergence of the second lobe in the nonuniform neutral curve (Fig.~\ref{fig:neutral-beta}(b)) produces a dramatic decrease in $\Wi_c$ as $\beta$ is decreased past $\beta \approx 0.8$. As the critical mode switches to the new lobe, which arises from slower-speed modes, the critical wave speed drops from $\approx\bar{U}_m$ to much smaller values (Fig.~\ref{fig:critical-beta}(b)). However, the critical wave speed increases as $\beta$ decreases, so that once the neutral curve recovers a single-lobed shape, for $\beta \lesssim 0.5$, the critical wave speed is again~$\approx\bar{U}_m$.

It is interesting to note that neutral curves with multiple lobes or loops have been observed for the centre-mode previously, but in the elasto-inertial context. Specifically, in pipe flow, \citet{chaudhary_pipe_jfm_2021} show that the neutral curve in the $\Rey-k$ plane takes the form of two U-shaped loops, for a limited range of $\beta$ and $\Wi/\Rey$ (the elasticity number); these loops can be overlapping (as is the case in Fig.~\ref{fig:neutral-beta}(b)) or non-overlapping. Such double-lobed neutral curves have also been found in elasto-inertial Kolmogorov flow \citep{Lewy_Kerswell_2025}.
% also find  imilar behaviour of the neutral curve Kolmogorov flow \citet{Lewy_Kerswell_2025} show that the neutral curve in the $\Rey-k$ plane takes the form of two nonoverlapping U-shaped loops, for a limited range of $\Wi/\Rey$ (the elasticity number). It is worth noting that similar neutral curves, again in the context of elasto-inertial centre-modes, have also been observed in viscoelastic pipe flow  \citet{chaudhary_pipe_jfm_2021}. The double-lobed neutral curves in \ref{fig:neutral-beta}(b) are similar since they may be viewed as 
% the result of two overlapping or intersecting loops. 

The fact that $\Wi_c>1$ for all $\beta$, in Fig.~\ref{fig:critical-beta}(a), demonstrates the essential role of elasticity in destabilizing the nonuniform polymer solution. In the limit of $\Wi \ll 1$, an Oldroyd-B solution simplifies to a Newtonian solution with a viscosity $\mu = \mu_s/\beta$ that exceeds that of the pure solvent $\mu_s$ in proportion to the polymer concentration. Thus, the $\Wi \ll 1$ limit of our problem with a nonuniform base-state concentration corresponds to a high viscosity stream sandwiched between low viscosity streams. Fig.~\ref{fig:critical-beta}(a) shows that this viscosity stratification alone cannot destabilise the flow at small $\Rey$, regardless of the degree of stratification (value of $1-\beta)$ since the flow is stable when $\Wi<1$. This is expected to change when inertial effects become prominent, given the presence of instabilities due to viscosity stratification in high-$\Rey$ Newtonian shear flows~\citep{Govindarajan2014}. 

% In either situation, the implication of such neutral curves is that, for a range of wavenumbers, a perturbation can become unstable a second time as the control parameter ($\Rey$ or $\Wi$) is increased.
% the result of two overlapping loops, each associated with a distinct eigenvalue, as demonstrated in Fig.\ref{fig:eigen-double}(c). However, this is not the case for the double-lobed curve for $\beta = 0.6$, which is associated with a single eigenvalue (Fig.\ref{fig:eigen-double}(d)). though the two overlapping loops, each associated our double lobed curves can be seen as overlapping loops. 

\section{Energy analysis underscores the role of concentration gradients} \label{sec:energy}

To gain insight into the destabilisation induced by localising polymers, we analyse the perturbation kinetic energy budget~\citep{zhang_zakiJFM}. In the context of the centre-mode in channel flow, the kinetic energy analysis has been used recently to show that the instability is driven by elastic stresses even in the presence of inertia, despite having $\Rey \sim 10^3$~\citep{Buza_Page_Kerswell_2022}. 

A complete perturbation energy analysis would include budget equations for the rate of change of kinetic energy, polymer-conformation energy, and concentration variance (obtained from the momentum balance, the constitutive conformation-tensor equation, and the advection equation for concentration, respectively). A polymer-energy measure that is a norm, in the manner of the kinetic energy, has been proposed by \citet{Hameduddin2019} and used in an energy analysis of the centre-mode by \citet{Buza_Page_Kerswell_2022}. In the polymer energy equation, dissipation via elastic relaxation is countered by flow-induced stretching 
% due to the base or perturbation flow
~\citep{Buza_Page_Kerswell_2022}. The concentration of polymers, being absent from the conformation-tensor equation, will not appear in the polymer energy equation, and so its analysis will not directly help us understand why localising polymers destabilises the flow; we therefore do not examine the polymer energy equation here.  
% localisedthe reason has contributions whose signs are independent of the parameters of the system---a positive contribution from flow-stretching and a negative contribution from elastic relaxation. Furthermore, near the onset of instability, these two terms will always balance, and so an analysis of the polymer energy equation will not help us understand why $\Wi_c$ reduces when polymers are localised. 
The concentration variance equation is trivial, since advection will simply redistribute the concentration in space and not modify the variance integrated over the domain.
% the destabilisation due to polymer localisation.
% Furthermore, the conformation tensor equation, and therefore the polymer energy equation, does not contain the polymer concentration $\theta$. 
 We therefore focus on just the kinetic energy equation.
 
Following the procedure in \citet{zhang_zakiJFM}, we take the inner product of the velocity perturbation ${ \hat{\bm{u}}}$ with its linearised evolution equation to obtain  
% \eqref{eq:tilde-vx}-\eqref{eq:tilde-vy} 
the disturbance kinetic energy budget
% To obtain the kinetic energy budget equation, we follow the procedure described in \cite{zhang_zaki_jfm} (see  \S 2.3.2). The steps involved are as follows: first, we multiply the linearised momentum balance equation by the complex conjugate of the velocity perturbation. The resulting equation is then complex-conjugated, further these two equations are added together to obtain the kinetic energy budget equation. After further simplifications, as explained in \cite{zhang_zaki_jfm}, the kinetic energy equation for the Kolmogorov flow is given by:
\begin{gather}
% \begin{aligned}
 Re\; \dd_t \int_{0}^{2\pi} \frac{\hat{u}_i \hat{u}_i^*}{2} \, dy =  -Re \int_{0}^{2\pi} \frac{1}{2}( \hat{u}^{*}_i \hat{u}_j + \hat{u}_i \hat{u}^{*}_j)\frac{\partial \bar{U}_i}{\partial x_j} \, dy   - \int_{0}^{2\pi} \beta \frac{\partial {\hat{u}}^{*}_i}{\partial x_j} \frac{\partial {\hat{u}}_i}{\partial x_j} \, dy \nonumber \\
 - \int_{0}^{2\pi} \frac{1}{2}\left(\hat{\Tc}_{ij}\frac{\partial {\hat{u}}^{*}_i}{\partial x_j} + \hat{\Tc}^{*}_{ij}\frac{\partial {\hat{u}}_i}{\partial x_j} \right)\, dy.  \label{eq:kinetic-prime}
 % \end{aligned}
 \end{gather}
 where $*$ denotes the complex conjugate, $\langle f \rangle$ = $\int_0^{2\pi} \int_0^{2\pi} f dx dy$, and $\hat{\Tc}_{ij} $ is the linearised elastic stress tensor of the perturbation field (see \eqref{eq:stress}):
 \begin{equation}\label{eq:linstress}
 \hat{\Tc}_{ij} =  \bar{\theta}(y)\frac{1-\beta}{Wi}\hat{\Cc}_{ij}  + \hat{\theta}\frac{1-\beta}{Wi}(\bar{\Cc}_{ij} - \delta_{ij})  
\end{equation}
The two terms of \eqref{eq:linstress} have distinct physical origins, which should be borne in mind while interpreting the kinetic energy budget. The first term ($\propto \bar\theta$) represents the polymer stress due to conformation perturbations which is modulated by base-state concentration variations, while the second term ($\propto \hat\theta$) represents the stress that arises due to concentration perturbations. Only the former survives when the concentration is held constant ($\theta = 1$). Moreover, even when normal mode disturbances to the concentration are permitted ($\hat \theta = \tilde\theta e^{ik(x-ct)}$), the first term alone will survive when the base-state concentration is uniform ($\bar\theta = 1$) because $\tilde\theta = 0$ when $\bar{\theta}^\prime=\dd_y \bar{\theta} = 0$ for any nonsingular normal mode (see \eqref{eq:theta-tilde}). 

\begin{figure}
\centering
    % \includegraphics[width=0.33\linewidth] {fig-SI/theta_streaml_vKf__uni_B0pt8_Wi18pt8_k0pt4.eps}
    %     \put (-104,107){\footnotesize (a) uniform}
    \includegraphics[width=0.33\linewidth] {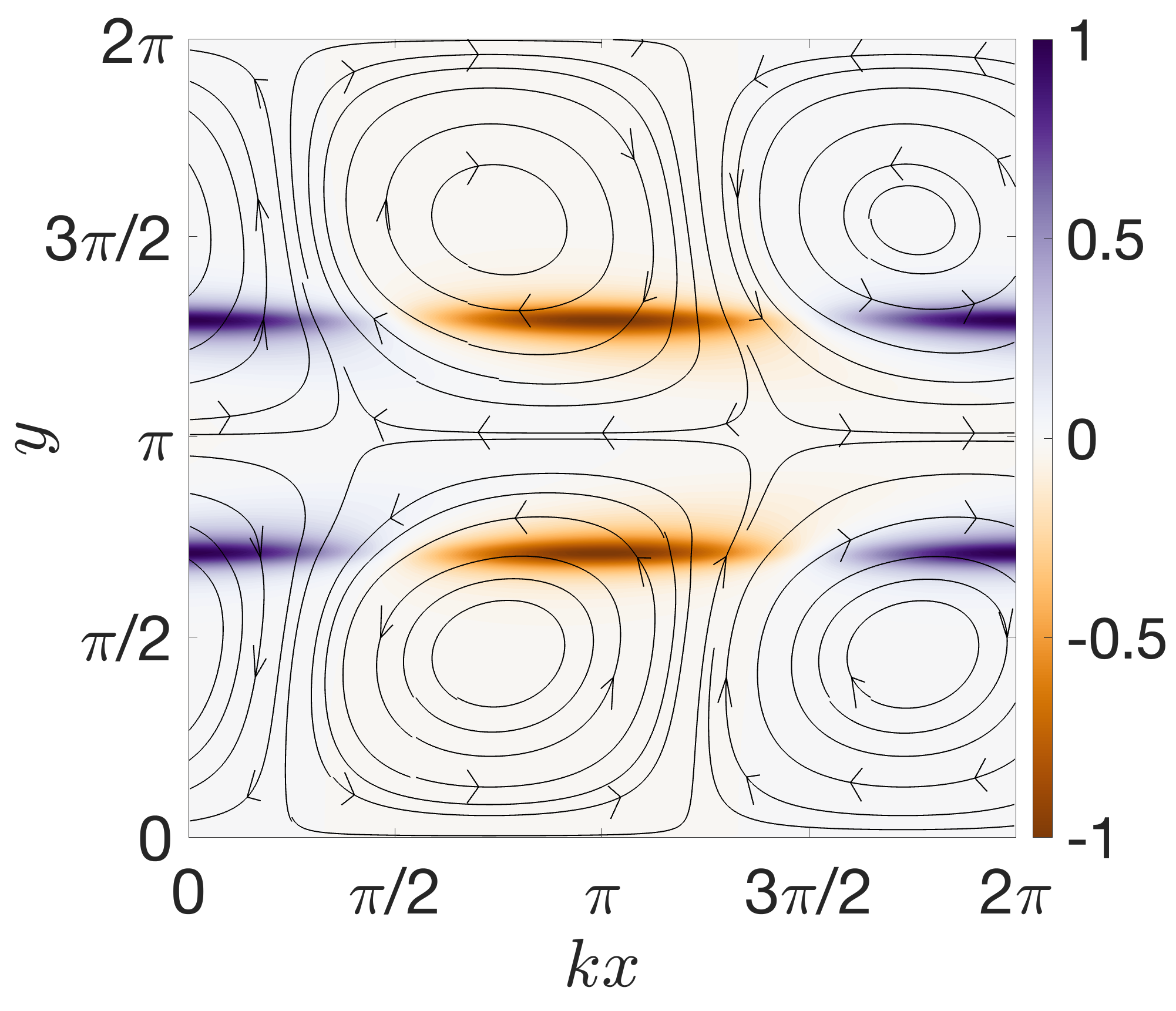}
        \put (-104,107){\footnotesize (a) nonuniform,  higher $c_r$}
        \hspace{2 em}
    \includegraphics[width=0.33\linewidth] {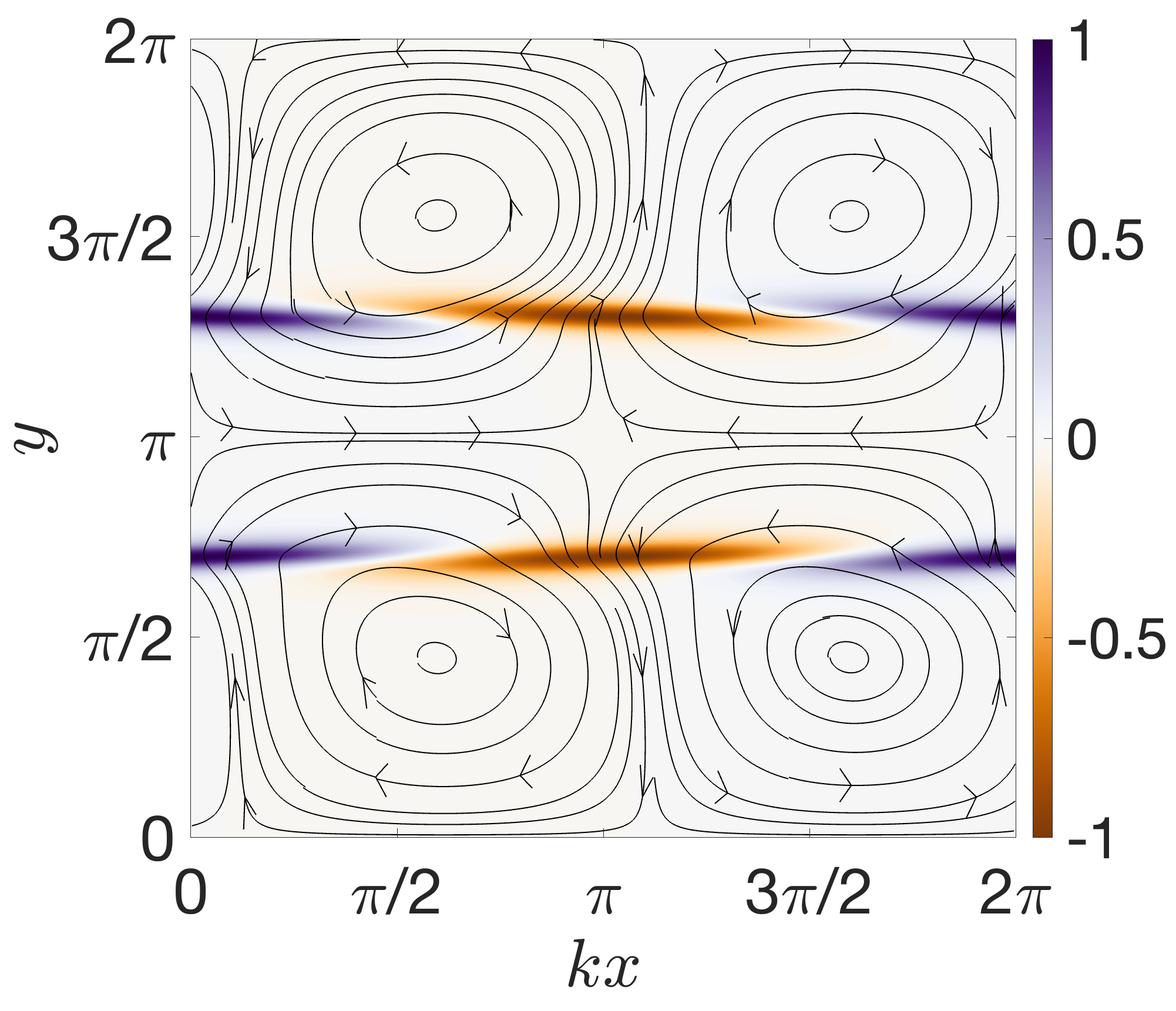}
        \put (-104,107){\footnotesize (b) nonuniform,  lower $c_r$}
    \caption{Visualisation of the concentration perturbation field $\hat{\theta}$ of unstable eigenfunctions in Kolmogorov flow with a nonuniform base-state concentration.
    % Uniform base-state concentration with $\Wi = 18.8, k = 0.4$, 
    (a) $\hat{\theta}$ of the eigenfunction in Fig.~\ref{fig:eigenfunc}(b) 
    (b)  $\hat{\theta}$ of the eigenfunction in Fig.~\ref{fig:eigenfunc}(c) 
    % Nonuniform base-state concentration ($\bar{\theta}$ corresponds to $\delta  = 0.067 \pi$ in Fig.~\ref{fig:base}(a)) with $\Wi = 23.25, k = 0.4$, (c) Nonuniform base-state concentration  with  $\Wi = 8.0, k = 0.4$. 
    These eigenfunctions correspond to the fastest-growing modes (peaks of the traces) in Fig.~\ref{fig:eigen-double}(c) where $\beta = 0.8$. 
     \label{fig:theta-eigenfunc}
    }
\end{figure}

Now, in the base flow with a polymer-rich stream sandwiched between polymer-free streams ($\bar{\theta}$ corresponding to $\delta = 0.067\pi$ in Fig.~\ref{fig:base}(a)), gradients in the base concentration occur only at the interface between the streams and so we expect the concentration perturbations $\hat{\theta}$ to be focused at the interface. This is indeed what we observe, as demonstrated in Figs.~\ref{fig:theta-eigenfunc}(a,b) which present $\hat{\theta}$ for the eigenfunctions of Fig.~\ref{fig:eigenfunc}(b,c). Transverse velocity perturbations $\hat{u}_y$ deflect the interface and induce longitudinal concentration variations that are simultaneously advected by the base flow. In the absence of advection, transverse flow perturbations directed across the interface and towards the centre of the polymer-laden stream will reduce the concentration near the interface (induce a minima in $\hat\theta$); the opposite will occur for flow perturbation directed away from the centre of the polymer-rich stream (this is evident on considering the action of the term $\hat{u}_y\dd_y \bar{\theta}$ in \eqref{eq:theta-tilde} near the interface where $\dd_y \bar{\theta} \neq 0$). However, a phase shift is produced between $\hat{u}_y$ and $\hat\theta$ by the advection that is caused by the difference between the base flow velocity near the interface and the speed $c_r$ of the normal mode (see \eqref{eq:theta-tilde}); this advection is the reason why the minima and maxima of $\hat{\theta}$ in Fig.~\ref{fig:theta-eigenfunc} are shifted relative to the locations where the perturbed flow across the interface is toward and away from the centre of the polymer-laden stream, respectively (see the streamlines in Fig.~\ref{fig:theta-eigenfunc}).

Returning to the derivation of the energy budget, we simplify \eqref{eq:kinetic-prime} for normal mode disturbances \eqref{eq:normal-mode} to obtain
 \begin{equation} \label{eq:kinetic}
 Re\,\dd_t{E} =   Re\, W  +D+P,
\end{equation}
where the kinetic energy $E$, the rate of work done by Reynolds stresses $W$, and the viscous dissipation $D$ are given by
\begin{gather}
E=   \left\langle{(|\tilde{u}_x|^2 +|\tilde{u}_y|^2)}/{2} \right\rangle,
W = -\left\langle{\bar{U}^\prime (\tilde{u}^{*}_x \tilde{u}_y +\tilde{u}_x \tilde{u}^{*}_y )}/{2} \right \rangle,\nonumber \\
	 D = -{\beta}\left\langle k^2({|\tilde{u}_x|^2 +|\tilde{u}_y|^2}) + {d}_y\tilde{u}^*_x {d}_y\tilde{u}_x + {d}_y\tilde{u}^*_y {d}_y\tilde{u}_y   \right\rangle,
\end{gather}
and the rate of work done by polymers is split into two contributions, 
\begin{equation}
    P = P_{\theta} + P_{\nabla\theta},
\end{equation}
where
\begin{align}
	   P_{\theta}  &= -\bar{\theta}\frac{1-\beta}{2Wi} \Big\langle \tilde{\Cc}_{xx} (-ik) \tilde{u}^{*}_x + \tilde{\Cc}^{*}_{xx} ik \tilde{u}_x + \tilde{\Cc}_{xy} d_y \tilde{u}^{*}_x + \tilde{\Cc}^{*}_{xy} d_y \tilde{u}_x  \\&\quad +  \tilde{\Cc}_{xy} (-ik) \tilde{u}^{*}_y + \tilde{\Cc}^{*}_{xy} ik \tilde{u}_y + \tilde{\Cc}_{yy} d_y \tilde{u}^{*}_y + \tilde{\Cc}^{*}_{yy} d_y \tilde{u}_y  \Big\rangle,\\
    P_{\nabla{\theta}}  &= -\frac{1-\beta}{2Wi} \Big\langle ik (\bar{\Cc}_{xx} -1)  (-\tilde{\theta} \tilde{u}^{*}_x  + \tilde{\theta}^{*} \tilde{u}_x) + \bar{\Cc}_{xy} (\tilde{\theta} d_y  \tilde{u}^{*}_x  +  \tilde{\theta}^{*} d_y \tilde{u}_x) \\
    &\quad + ik \bar{\Cc}_{xy} (-\tilde{\theta} \tilde{u}^{*}_y  + \tilde{\theta}^{*} \tilde{u}_y ) 
    % - (\bar{\Cc}_{yy}-1) (\tilde{\theta} d_y  \tilde{u}^{*}_y  +   \tilde{\theta}^{*} d_y \tilde{u}_y)   
    \Big\rangle.
\end{align}
$P_\theta$ is the usual polymer rate-of-work (or power) term (which would obtain for a constant polymer concentration) now modulated by the base-state concentration $\bar\theta$. $P_{\nabla{\theta}}$ is a new polymer power term that arises only when the polymer concentration is nonuniform; specifically, it is zero unless $\dd_y\bar\theta$ is nonzero since $\tilde\theta$ would be zero otherwise, as discussed above. Thus, $P_{\nabla{\theta}}$ may be attributed to gradients in the base-state concentration. In contrast, $P_\theta$ is merely modulated by variations in the base-state concentration and is present even in a uniform solution.

The interpretation of \eqref{eq:kinetic} is straightforward when $\Rey$ is nonzero and the growth rate of the disturbance $c_i$ is positive. Then, the kinetic energy grows, and the cause of the instability can be inferred from the 
% ordering of the magnitudes of the
dominant positive power term on the right hand side of \eqref{eq:kinetic} (the dissipation $D$ is always negative and must be overcome by one of the power terms). However, even when considering a neutral mode for which $c_i=0$, \eqref{eq:kinetic} can still provide causal information, since a suitable small perturbation to $\Wi$ will render $c_i>0$ 
while leaving the values of the dominant terms in \eqref{eq:kinetic} nearly unchanged.
% without changing the signs of the dominant terms in \eqref{eq:kinetic}. 
The same reasoning allows us to apply \eqref{eq:kinetic} even when $\Rey = 0$. Since the limit $\Rey \rightarrow 0$ is regular, a small increase in $\Rey$ will modify a zero-$\Rey$ mode into a weakly-inertial mode, with a non-zero evolving kinetic energy, but with nearly the same values for the dissipation and polymer power terms. With this understanding, we now apply \eqref{eq:kinetic} to the critical zero-$\Rey$ modes of Fig.~\ref{fig:critical-beta} for both the uniform and nonuniform cases. We have tested the effect of increasing $\Rey$ to a small value and found that these modes attain a small positive growth rate of kinetic energy without altering the identity of the dominant positive work term (see the \href{https://bighome.iitb.ac.in/index.php/s/g6wRiyEY8H3SNyR}{supplementary material}).
% ($\Rey W$ is subdominant in low $\Rey$ flows).  which will be driven by the dominant positive work term which will remain unaffected. 

 \begin{figure}
    \centering
    \includegraphics[width=0.45\linewidth] {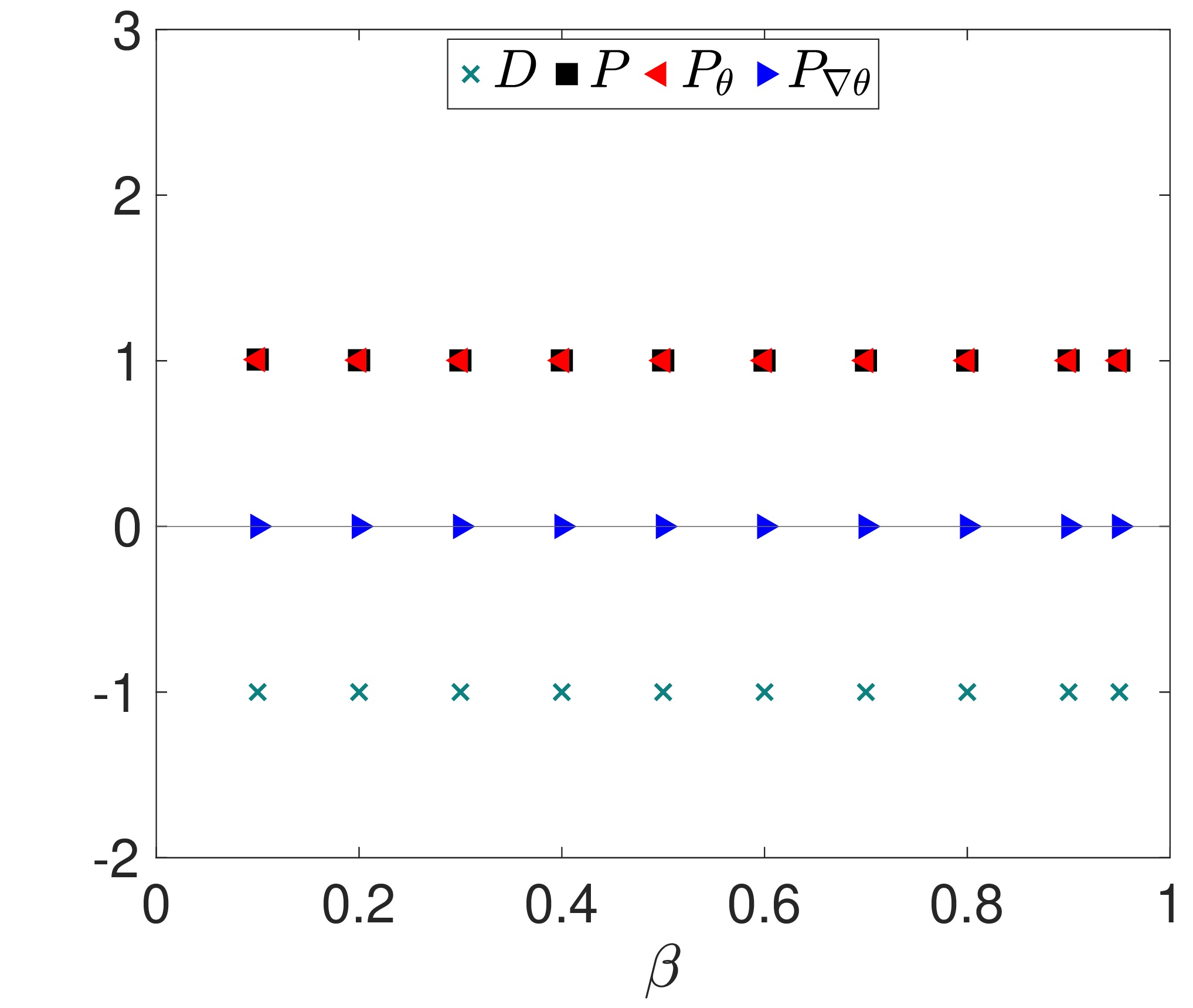}
    \put (-62,26){\footnotesize (a)}
    \put (-50,26){\footnotesize uniform}
        \hspace{1 em}
    \includegraphics[width=0.45\linewidth] {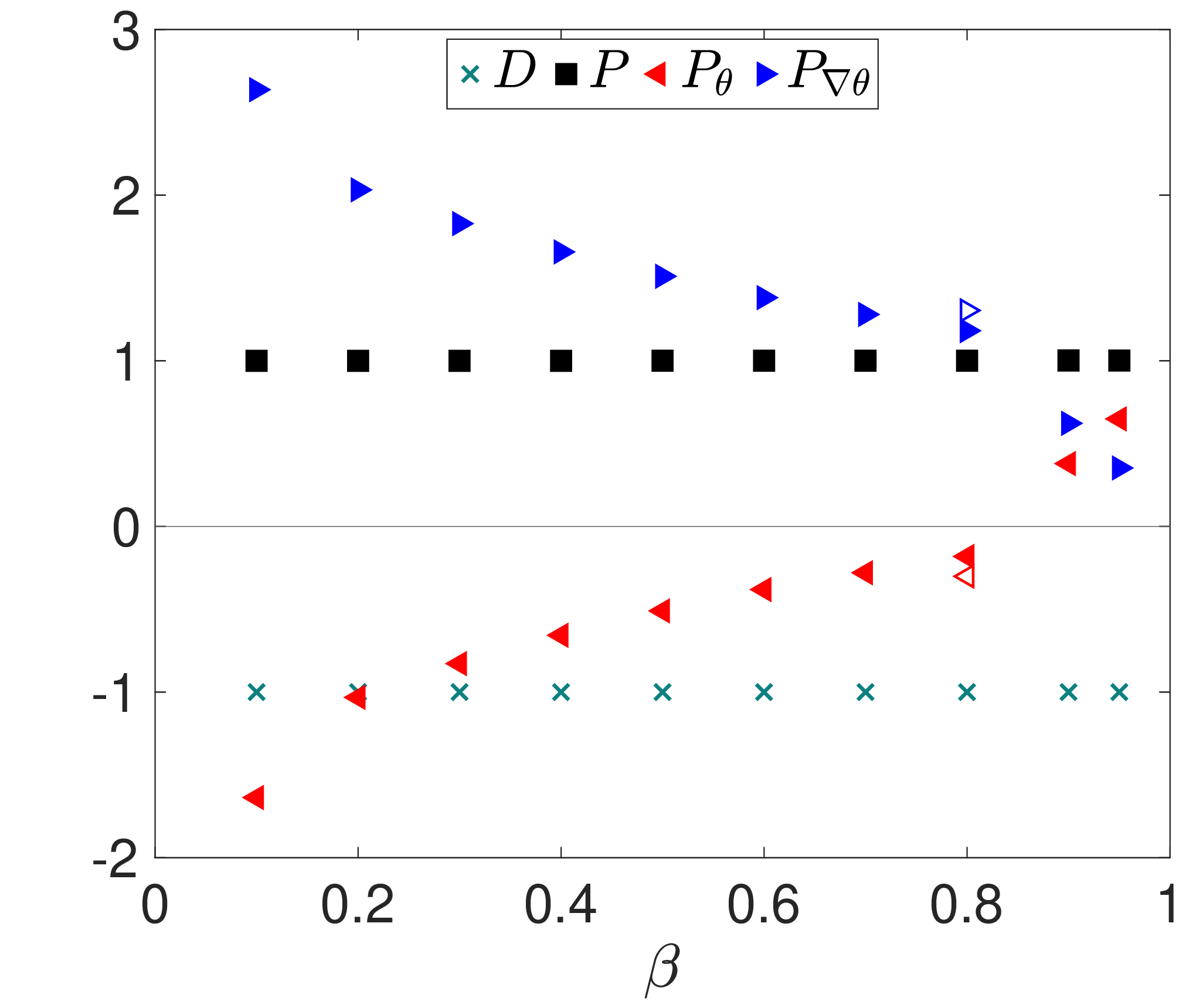}
    \put (-62,26){\footnotesize (b)}
    \put (-50,26){\footnotesize nonuniform}

    % \includegraphics[width=0.45\linewidth] {fig/Energy_plot_nonuniform_vKf_Re0pt01_neutral_mode_with_corrected_PD_terms_with_data_for_B0pt8_both_lower_upper_mode_data.eps}
    % \put (-150,113){(b)} \\
	\caption{ Rate of work contributions to the kinetic energy budget for critical modes of inertialess ($\Rey = 0$) Kolmogorov flow with (a) uniform and (b) nonuniform base state polymer concentrations. All terms are normalised by the magnitude of the viscous dissipation and so $D = -1$. The magnitudes of the power terms remain nearly the same (and their signs and relative magnitudes are unchanged) on increasing $\Rey$ to a small finite value to endow the modes with finite inertia and a small positive growth rate (see the \href{https://bighome.iitb.ac.in/index.php/s/g6wRiyEY8H3SNyR}{supplementary material} for the finite-$\Rey$ analogue of panel (b)).} \label{fig:energy}
\end{figure}

Let us first examine the power contributions in the case with a uniform base-state concentration. Figure~\ref{fig:energy}(a) presents results for the critical modes (i.e., for which $\Wi = \Wi_c(\beta)$ and $k = k_c(\beta)$) of the uniform case in Fig.~\ref{fig:critical-beta}(a). Here, all terms have been normalised by the magnitude of dissipation and so $D = -1$. As anticipated, $P_{\nabla{\theta}}=0$ and $P_\theta$ is the dominant positive power term, for all values of $\beta$. The results are strikingly different for the critical modes of the nonuniform case (Fig.~\ref{fig:critical-beta}(a)), wherein polymers are localised about the velocity-maximum: we see, in Fig.~\ref{fig:energy}(b) (filled symbols), that $P_{\nabla{\theta}}$ is nonzero and that the relative contributions of $P_{\nabla{\theta}}$ and $P_{{\theta}}$ to $P$ (which equals $D$ when $\Rey = 0$) vary with $\beta$. In the limit of very low polymer loading ($\beta \rightarrow 1$), the dominant positive term is $P_\theta$, just as it is in the uniform case (see the data for $\beta = 0.95$ in Fig.~\ref{fig:energy}(b)). However, even a modest departure from this limit results in the dominant term switching to $P_{\nabla{\theta}}$ ($\beta = 0.9$ in Fig.~\ref{fig:energy}(b)). A further increase in polymer loading (decrease in $\beta$) produces a striking change in the work contributions: $P_{{\theta}}$ becomes negative, so that the elastic instability can be entirely attributed to $P_{\nabla{\theta}}$, i.e., to work done by polymer stresses arising from gradients in the polymer concentration. 

This transition in the energetic characteristics of the instability, which occurs at $\beta \approx 0.8$ (Fig.~\ref{fig:energy}(b)), coincides with the appearance of two-lobes in the neutral curves of the nonuniform case (Fig.~\ref{fig:neutral-beta}(b)) and therefore with the dramatic decrease of $\Wi_c$ with increasing polymer loading (Fig.~\ref{fig:critical-beta}(a)). These results indicate that in a nonuniform solution, where polymers are localised about the maximum velocity, the enhancement of the centre-mode instability is caused by gradients in the polymer concentration. 

To appreciate how concentration gradients can affect the flow, note that the polymer feedback-force on the fluid can be written as $\nabla\cdot(\theta \T_0)$ where $\T_0 = (\C-\mathsfbi{I}) (1-\beta)/\Wi$ is the polymer stress associated with the reference concentration $c_\mathrm{ref}$ (i.e., the stress that would be exerted by polymers if $\theta = 1$). Expanding this expression yields $\theta \nabla\cdot \T_0 + \nabla \theta \cdot\T_0$, which can be interpreted as a sum of the usual polymeric force modulated by the concentration and an additional polymeric force produced by gradients in the concentration. The latter would arise when the concentration is nonuniform even if the polymer is stretched in the exact same way (i.e., $\T_0$ is the same) everywhere in space. It is this additional elastic force that is responsible for the destabilisation of the flow when polymers are localised.

When the solution is very dilute ($\beta \rightarrow 1$), the work done by the polymer force due to concentration gradients $P_{\nabla \theta}$ adds to the usual polymer power $P_\theta$ to produce a weak destabilisation (see Fig.~\ref{fig:energy}(b) and Fig.~\ref{fig:critical-beta}(a)). As the polymer loading is increased ($\beta$ is decreased below unity), the flow is rendered increasingly unstable by the action of the polymer force induced by concentration gradients, with a dramatic destabilisation occurring just as $P_{\nabla \theta}$ becomes the sole driver of the instability. 

These findings are consistent with the nature of the eigenfunctions in the nonuniform case, visualized in Figs.~\ref{fig:eigenfunc}(b,c)---the perturbed polymer extension is focused in the interface region between the polymer-free stream and the polymer-rich stream, where the base-state concentration gradients are strongest and thus where concentration perturbations appear (Fig.~\ref{fig:theta-eigenfunc}). Since this feature is shared by the two eigenfunctions in Figs.~\ref{fig:eigenfunc}(b,c) which correspond to the upper and lower lobes of the neutral curve at $\beta = 0.8$, it is interesting to compare the energy signatures at the local minima of the two lobes. The minimum of the lower lobe is the critical mode whose energy signatures have just been discussed (filled symbols in Fig.~\ref{fig:energy}(b)). The power contributions for the minimum of the upper lobe are computed for $\beta = 0.8$ and presented as open symbols in Fig.~\ref{fig:energy}(b). $P_{\nabla \theta}$ is seen to be the sole positive work term for both lobes, confirming that polymer forces arising from concentration gradients drive the instability once $\beta \lesssim 0.8$.
% for both the the higher-$\Wi$, higher-$c_i$ mode and the lower-$\Wi$, lower-$c_i$, critical mode. 
% for both upper lobe as well. The results for theare presented as filled symbols fishare the feature of being  the most unstable modes of these two lobes This is true not only for Fig.~\ref{fig:eigenfunc}(c), which corresponds to the neutral curve's lower lobe of the neutral curve at contains the critical mode but also for the upper lobe (Fig.~\ref{fig:eigenfunc}(c)) which$c_i$ mode which corresponds to the lower lobe of the neutral curve but also for the higher $c_i$ mode which corresponds to the upper lobe. The energy analysis ... 

\begin{figure}
\centering
	\includegraphics[width=0.33\textwidth]{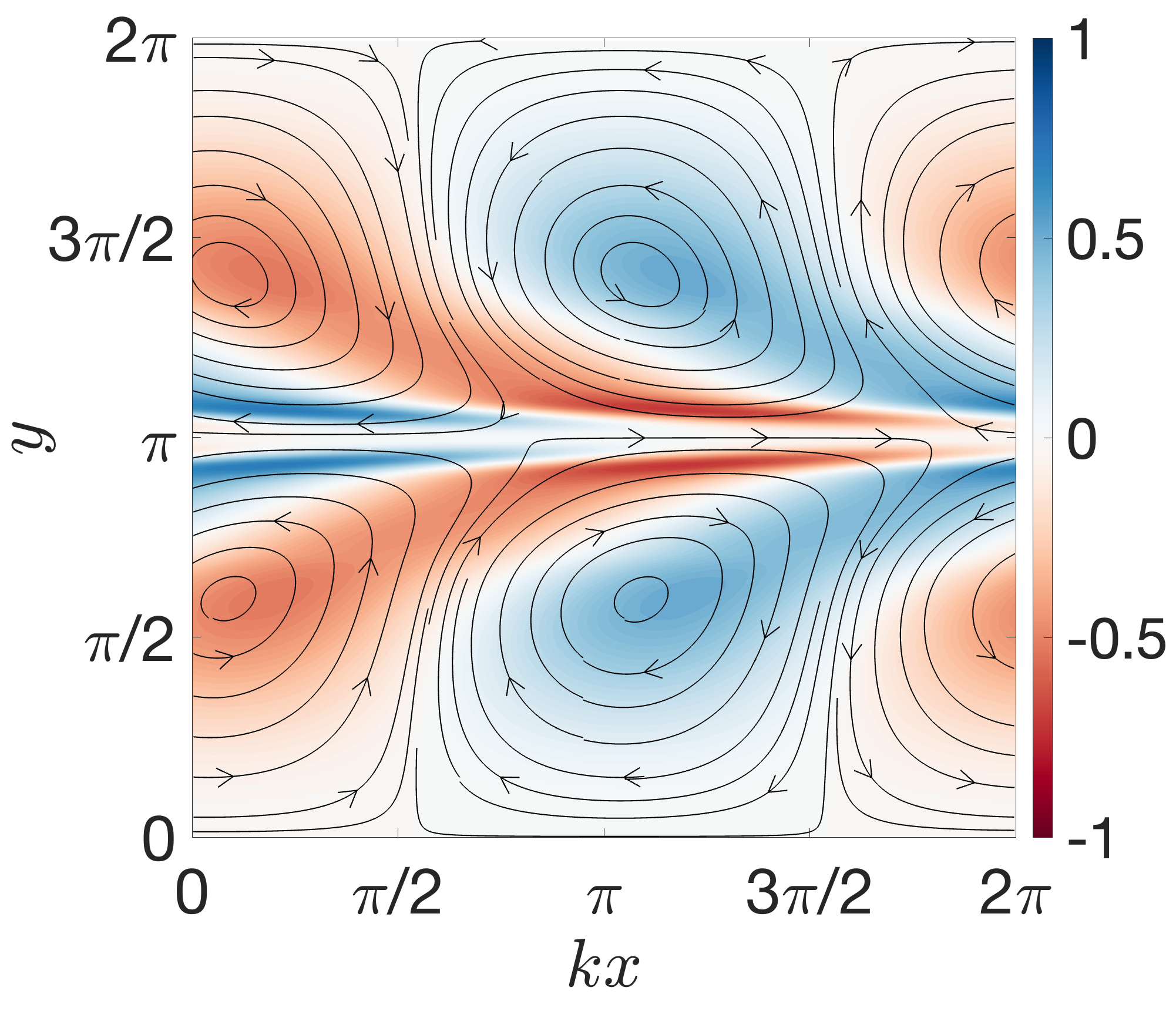}
    \put (-106,107){ \footnotesize (a) nonuniform \scriptsize $\beta = 0.99$}
     \hspace{1 em}
\includegraphics[width=0.33\textwidth]{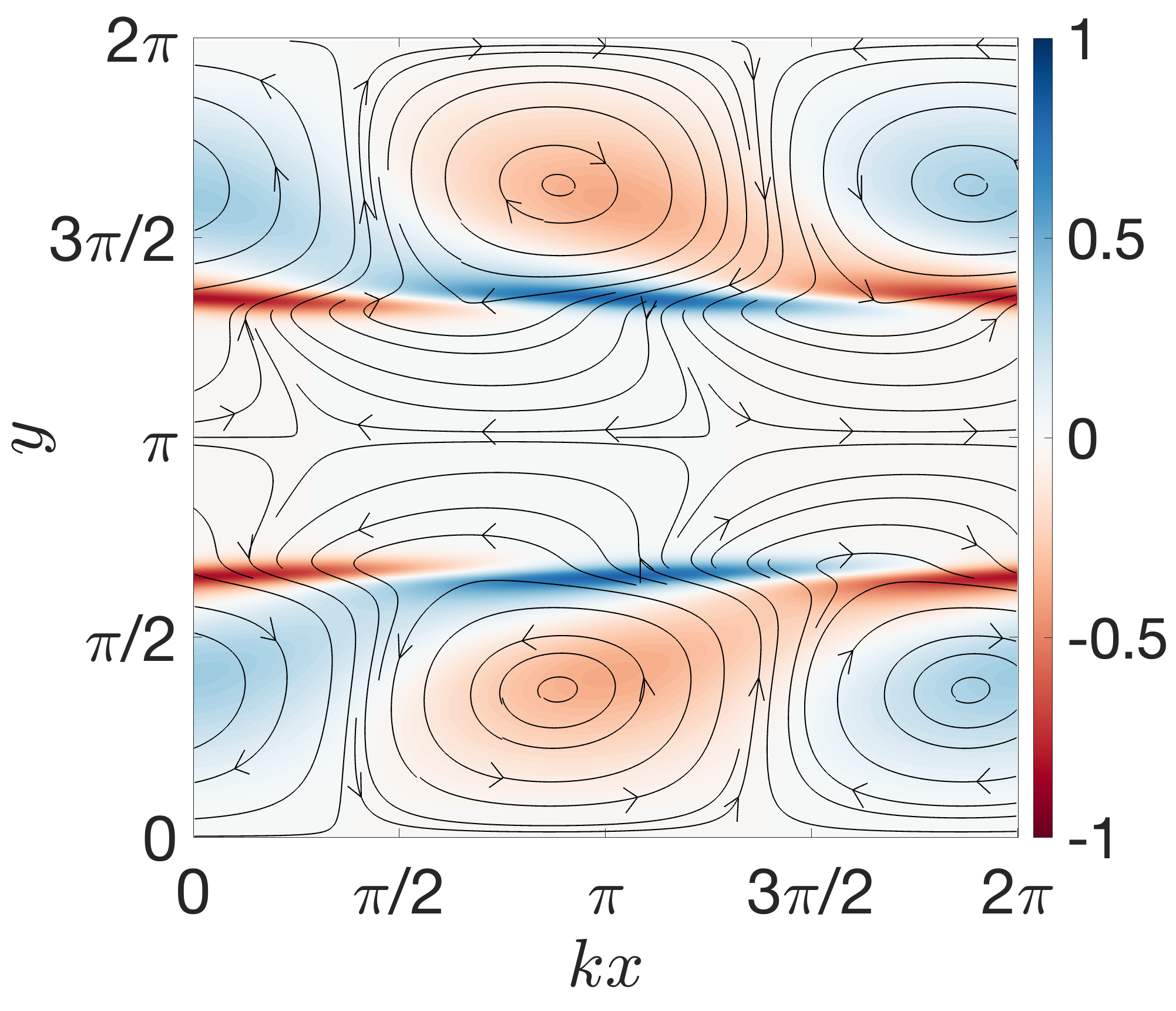}
    \put (-104,107){\footnotesize (b) nonuniform \scriptsize $\beta = 0.30$}
        % \hspace{1 em}
% \includegraphics[width=0.33\textwidth]{fig-SI/Neutral_curve_B0pt9_nonuniform_ht16pi240_&_B_star0pt75_uniform.eps}
    % \put (-100,20){(c)}
    % \includegraphics[width=0.67\linewidth]
    % {fig/neutral_curve_Wi_k_plane_for_ht16pi240_comparison_of_nonuni_beta0pt9_and_0pt5_and_betastar0pt75_0pt25.eps}
    % \put (-90,30){(b)}
	\caption{Visualisation of unstable eigenfunctions in Kolmogorov flow with a nonuniform base-state polymer concentration ($\bar{\theta}$ corresponds to $\delta  = 0.067 \pi$ in Fig.~\ref{fig:base}(a)) and (a) very low and (b) relatively-high levels of total polymer loading. The corresponding perturbed concentration fields are presented in the \href{https://bighome.iitb.ac.in/index.php/s/g6wRiyEY8H3SNyR}{supplementary material}. Parameter values: (a) $\beta = 0.99$ (with $\Wi = 100, k = 1$, for which $c_i = 0.0117076$) and (b) $\beta = 0.3$ (with $\Wi = 3, k = 0.6$, for which $c_i = 0.1007025$).} 
	\label{fig:eigen-Kol-lowbeta}
\end{figure}

The correspondence between the energy analysis and the features of the eigenfunction holds even in the very dilute limit $\beta \rightarrow 1$. Fig.~\ref{fig:energy}(b) shows that $P_{\nabla \theta}$ becomes subdominant in this limit, rendering the energy signature of the nonuniform case similar to that in the uniform case. One may then expect the eigenfunctions of the two cases to be similar as well, in very dilute solutions. This is indeed what we observe. For example, the eigenfunction in the nonuniform case for $\beta = 0.99$, presented in Fig.~\ref{fig:eigen-Kol-lowbeta}(a),
% (see the \href{https://bighome.iitb.ac.in/index.php/s/g6wRiyEY8H3SNyR}{supplementary material})
is \textit{not} focused at the interface between the polymer-free and polymer-rich streams and resembles the eigenfunction in the uniform case (Fig.~\ref{fig:eigenfunc}(a)). The perturbed polymer extension field of the eigenfunction becomes focused in the region with a high concentration-gradient only once the polymer-loading is sufficiently high ($\beta \lessapprox 0.8$) for $P_{\nabla \theta}$ to be the dominant polymer power term (Figs.~\ref{fig:eigenfunc}(b,c)); it then retains this feature as the polymer loading is further increased (see the eigenfunction for $\beta = 0.3$ in Fig.~\ref{fig:eigen-Kol-lowbeta}(b)). 

\section{Localising polymers near the centreline destabilises channel flow}\label{sec:channel}

The preceding sections have shown that localising polymers about the velocity-maximum strongly destabilises Kolmogorov flow. We now check whether this result extends to channel flow when polymers are localised about the centreline, which is the location of the velocity maximum.

\begin{figure}
\includegraphics[width=0.33\linewidth] {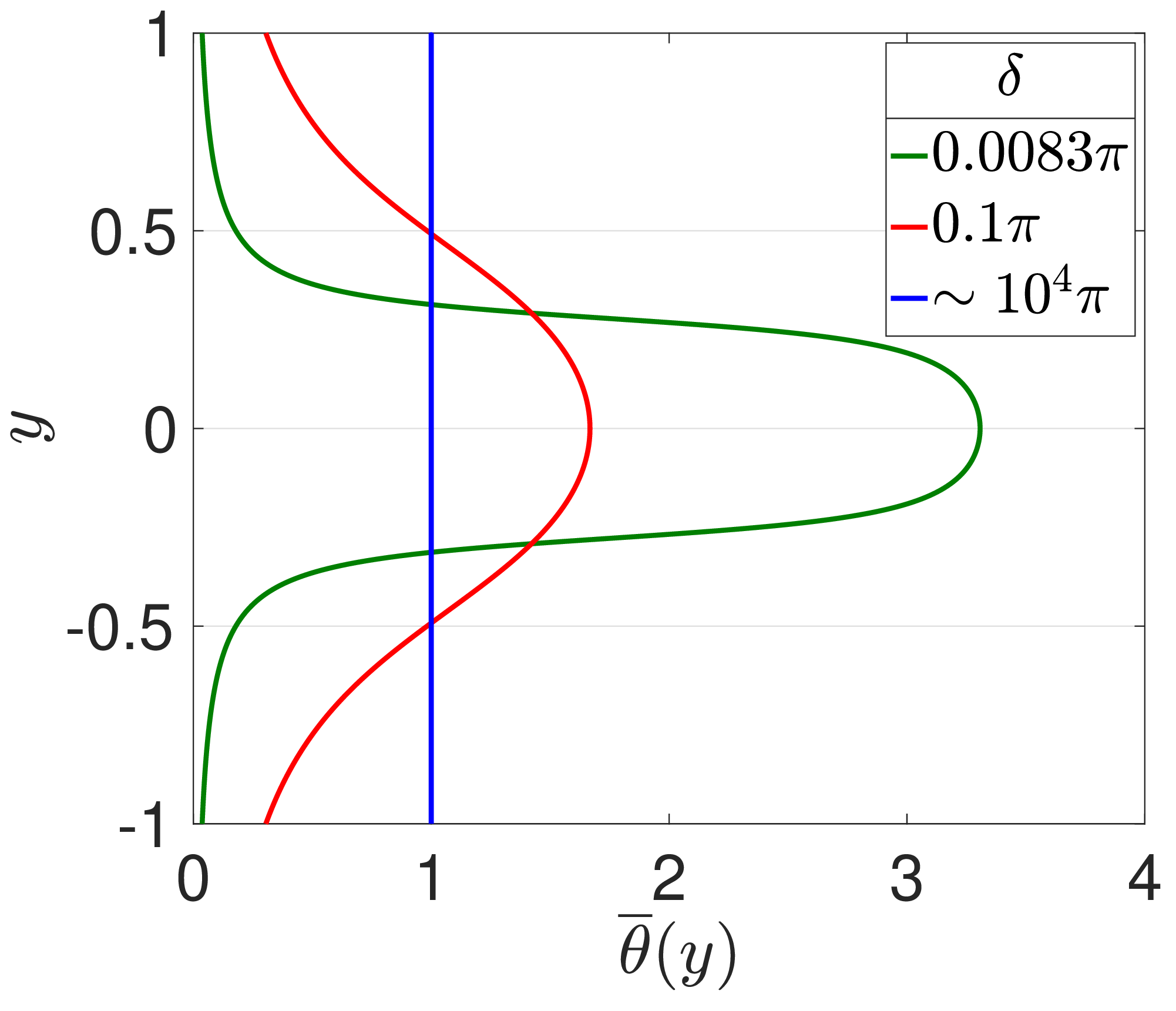}
    \put (-18,25){\footnotesize (a)}
    % \hspace{1 em}
\includegraphics[width=0.33\textwidth]{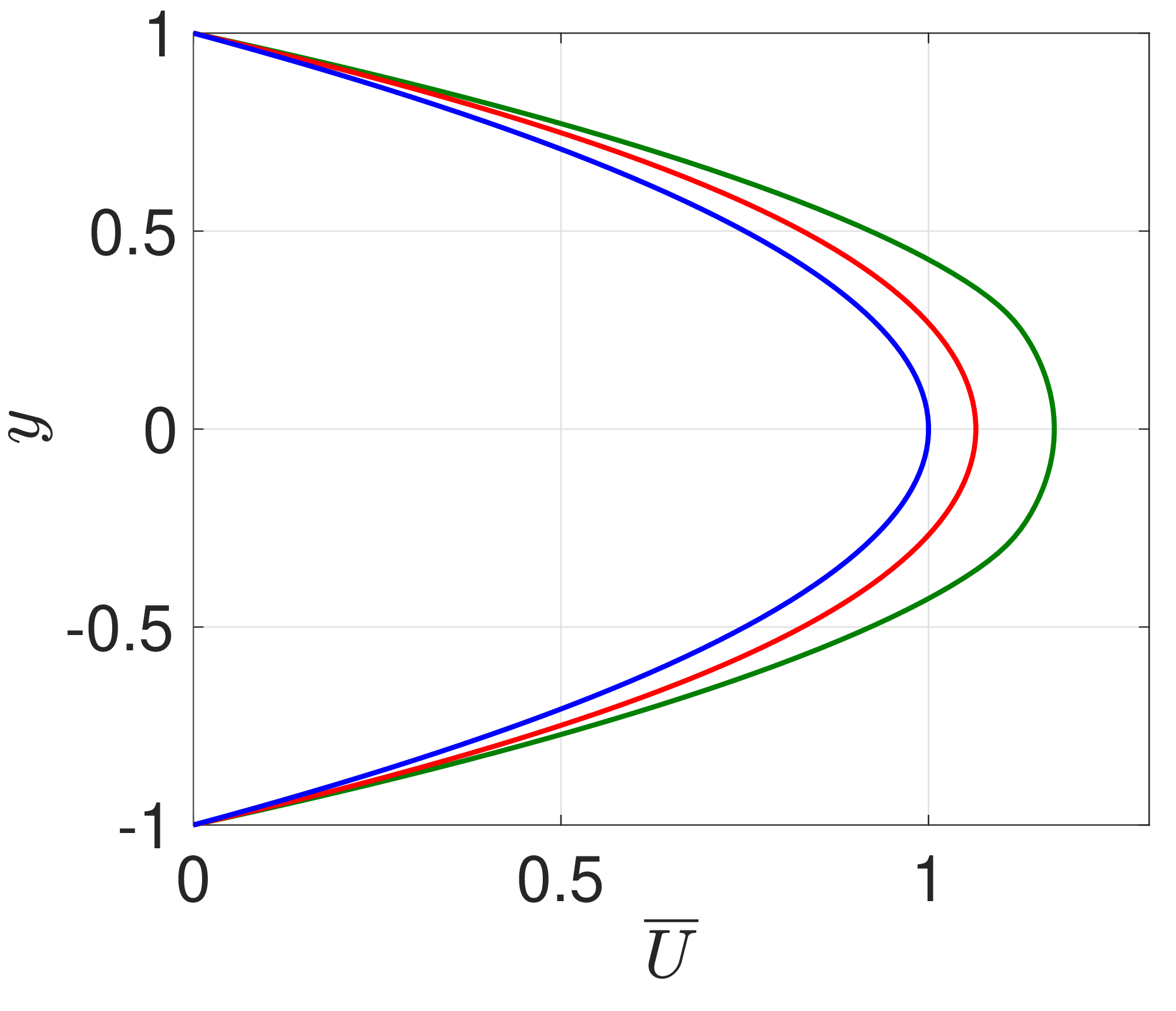}
    \put (-18,25){\footnotesize(b)}
        % \hspace{1 em}
\includegraphics[width=0.33\textwidth]{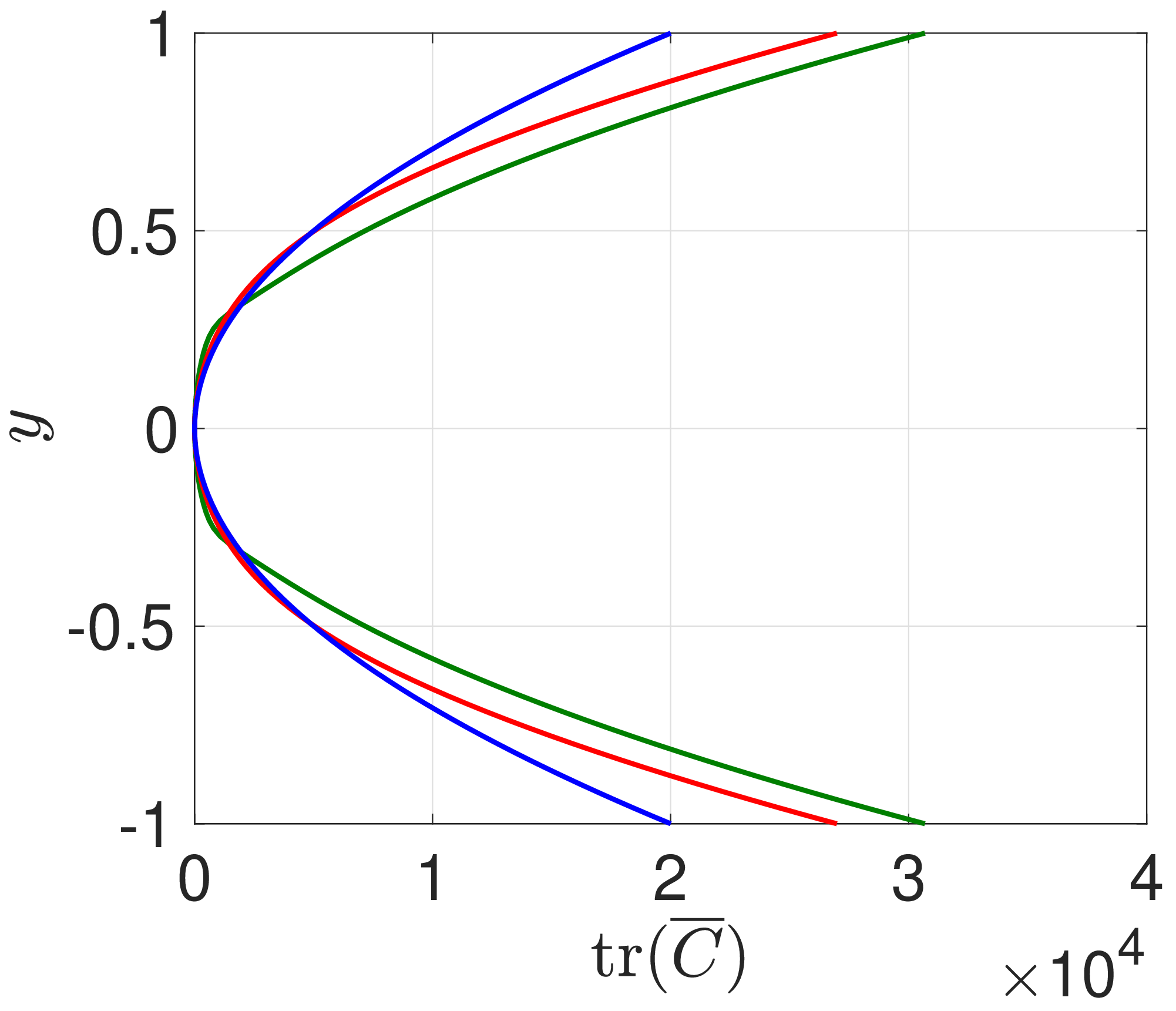}
    \put (-18,25){\footnotesize(c)}
    \caption{ \label{fig:base-channel} 
    Channel flow base-states with uniform and nonuniform polymer concentrations (localised about the centreline, i.e., the velocity maximum). (a) Concentration profiles for increasing extents of localisation (decreasing $\delta$); (b) base-state velocity; (c) base-state polymer squared extension. Here, $\Wi = 50$ and $\beta = 0.8$; results for $\beta = 0.5$ are presented in the \href{https://bighome.iitb.ac.in/index.php/s/g6wRiyEY8H3SNyR}{supplementary material}.}
 \end{figure}

Figure~\ref{fig:base-channel}(a) shows three different base-state polymer concentration profiles in a channel with walls at $y = \pm1$. We rescale the argument of the function \eqref{eq:pulse} so that $y$ ranges from -1 to +1 and then vary $\delta$ to realise a uniform profile (large $\delta$, blue line), a profile that mimics a polymer-laden stream sandwiched between polymer-free streams (smallest $\delta$, green line), and an intermediate profile (red line). The corresponding base-state velocity $\bar U$ and polymer square extension $\trCb$ fields are presented in Figs.~\ref{fig:base-channel}(b,c), respectively, for $\beta = 0.8$ (results for $\beta = 0.5$ are provided in the \href{https://bighome.iitb.ac.in/index.php/s/g6wRiyEY8H3SNyR}{supplementary material}).

To analyse the stability of these channel flows, we solve the eigenvalue problem 
\,(\ref{eq:tilde-cont}-\ref{eq:theta-tilde}) with no-slip and no penetration conditions at the walls, i.e., $\tilde{\bm u}=0$ at $y = \pm1$, using the Chebyshev spectral collocation method \citep{boyd,Ranga-spectral}. The code has been benchmarked against previous studies of viscoelastic channel flow with a uniform base-state concentration \citep{khalid_jfm, khalid_prl,Yadav_et_al_PRF2024}. As in the case of Kolmogorov flow, we facilitate the tracking of the neutral mode by adding very weak diffusion to the perturbation equations, so as to stabilise the continuous modes of the $\theta$-advection equation while leaving the centre-mode unchanged; homogeneous Neumann boundary conditions on $\tilde{\C}$ and $\tilde{\theta}$ are applied. We have verified that our neutral curves are unaltered by diffusion (e.g., via the polymer diffusive instability) by solving the eigenvalue problem without diffusion for select ($\Wi,k$) below and above the neutral curves.

\begin{figure}
\centering
% \includegraphics[width=0.33\linewidth] {fig/channel_Base_theta_profiles_uniform_nonuniform.eps}
%     \put (-18,25){\footnotesize (a)}
  % \includegraphics[width=0.4\linewidth] {fig/channel_base_velocity_profiles_B0pt994.eps}
  %   \put (-20,30){(b)}
  %   \\
    \includegraphics[width=0.4\linewidth] {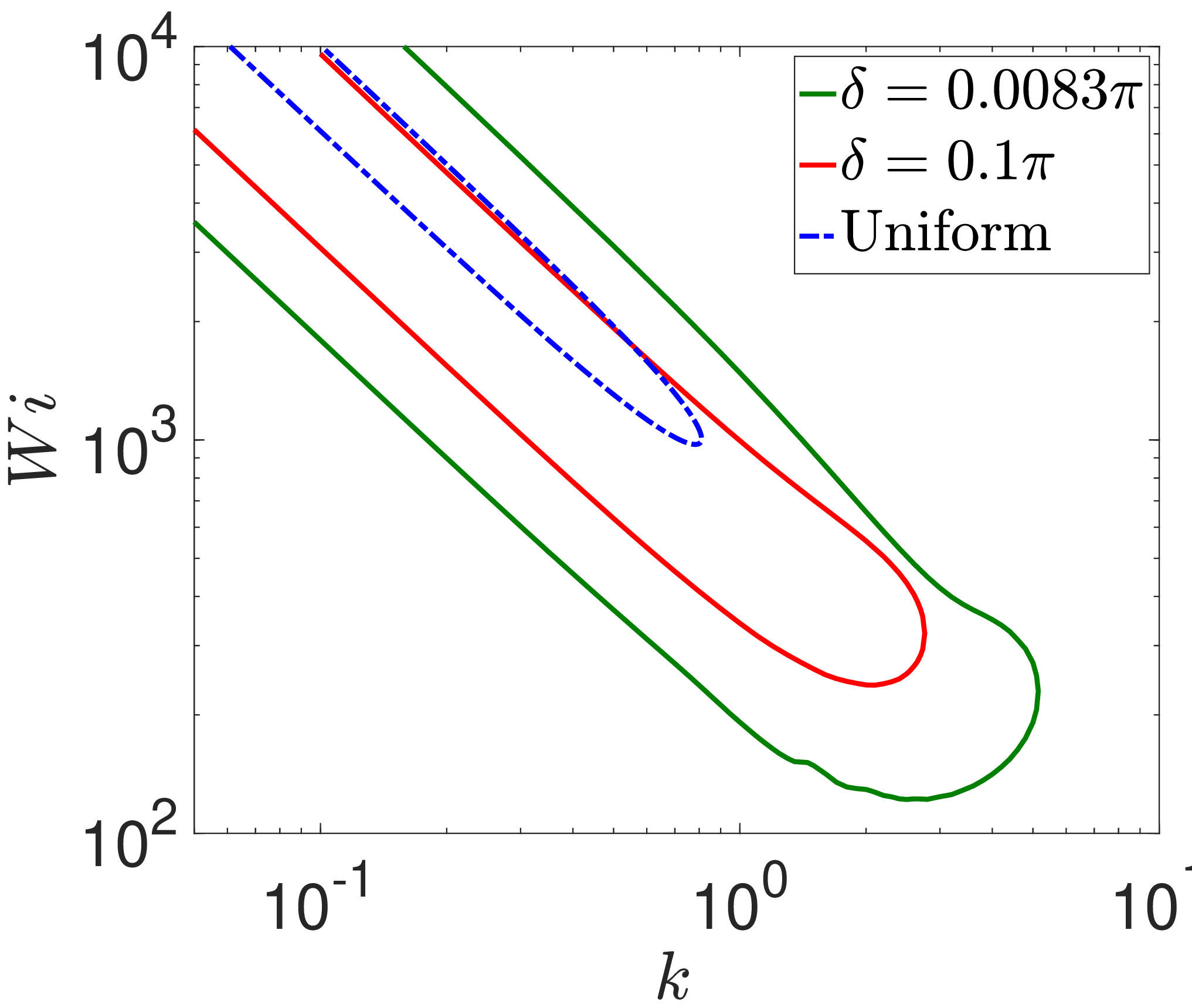}
    \put (-125,26){\footnotesize (a)}
    % \includegraphics[width=0.4\linewidth] {fig/Channel_neutral_curve_W_k_plane_strip_atWall_both_uni_nonuni_beta0pt994_ht4pi240_8pi240.eps}
    % \put (-21,30){(d)}\\
    \hspace{ 1 em}
     \includegraphics[width=0.4\linewidth] {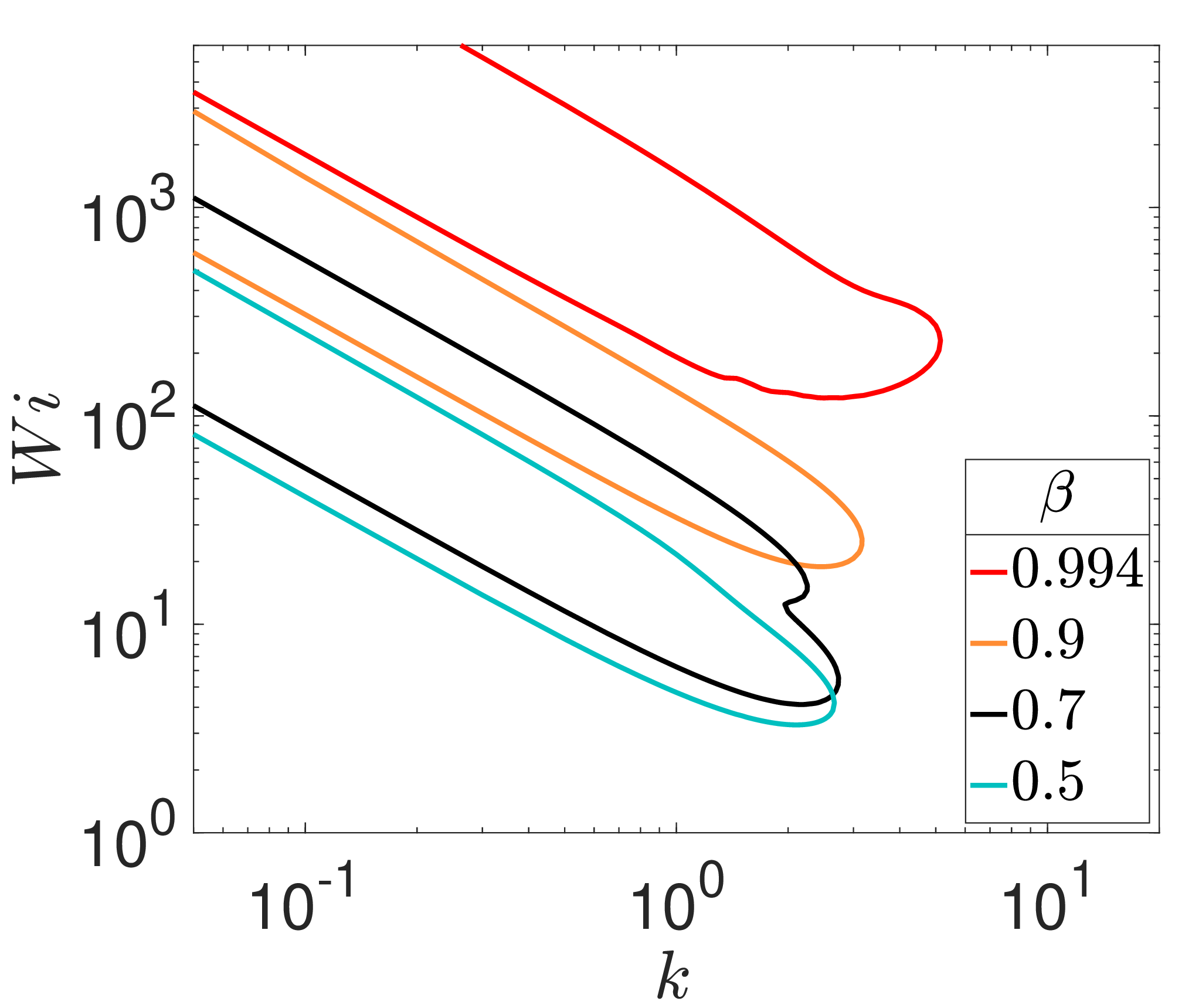}
        \put (-125,26){\footnotesize (b)}
    %  \includegraphics[width=0.4\linewidth] {fig/dispersion_curves_Uni_and_nonuni_W500_B0pt994_ht2pi240_kap1e_neg8.eps}
    % \put (-20,137){(f)}
	\caption{
    Neutral curves in the $\Wi-k$ plane for channel flow. (a) Effect of localising polymers near the centreline (decreasing $\delta$); the solution is sufficiently dilute ($\beta = 0.994$) for the uniform base state to be unstable. (b) Effect of increasing the total polymer loading (decreasing $\beta$) on the neutral curve for the case of a nonuniform base-state concentration ($\bar{\theta}$ corresponds to $\delta = 0.0083\pi$ in Fig.~\ref{fig:base-channel}(a)). } 
    \label{fig:channel}
\end{figure}

Figure~\ref{fig:channel}(a) presents neutral curves in the $\Wi-k$ plane, for the three base-state concentration profiles in Fig.~\ref{fig:base-channel}(a), considering a very dilute solution with $\beta = 0.994$. We see that localising polymers about the centreline destabilises the flow, reducing $\Wi_c$ from $\sim 10^3$ to $\sim10^2$. These results are analogous to those for Kolmogorov flow in Fig.~\ref{fig:delta}(a). 

Next, we consider the effect of increasing the polymer loading (decreasing $\beta$). 
With a uniform polymer concentration, the purely-elastic centre-mode in the channel flow of an Oldroyd-B fluid is unstable only in the very dilute regime where $\beta \gtrapprox 0.99$~\citep{khalid_prl}. With a nonuniform base-state concentration that is localised about the centreline, the centre-mode not only remains unstable but destabilises at lower $\Wi$ as the polymer loading is increased. Figure~\ref{fig:channel}(b) presents neutral curves for the nonuniform case ($\delta = 0.0083\pi$ in Fig.~\ref{fig:base-channel}(a)) with various values of $\beta$. As in Kolmogorov flow (Fig.~\ref{fig:neutral-beta}(b)), the neutral curves shift downward with decreasing $\beta$ and take on a double-lobed shape for intermediate $\beta$ (see the black curve for $\beta = 0.7$ in Fig.~\ref{fig:channel}(b)). Remarkably, with sufficient polymer-loading, localising the polymers about the centreline leads to an instability at $\Wi \sim 10$---a reduction of two orders of magnitude relative to the uniform case (compare the uniform result in Fig.~\ref{fig:channel}(a) with the nonuniform results for $\beta \leq 0.7$ in Fig.~\ref{fig:channel}(b)).

 \begin{figure}
    \centering
    \includegraphics[width=0.43\linewidth] {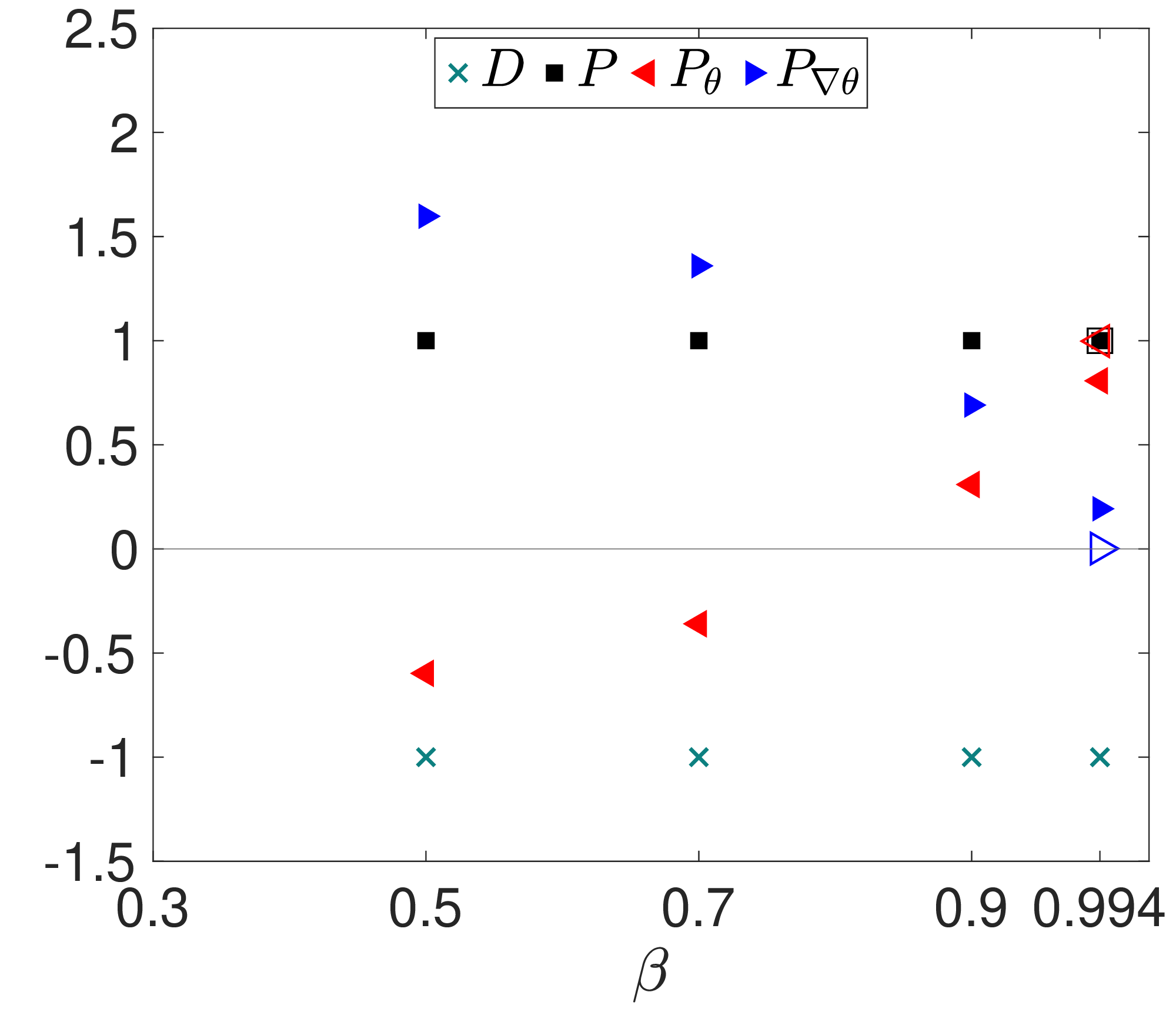}
    \put (-140,26){\footnotesize (a)}
    % \put (-135,26){\footnotesize $\Rey = 0$}
    %     \hspace{1 em}
    % \includegraphics[width=0.43\linewidth] {fig/Energy_Uni_NonU_PPF_Re0pt05_unstable_mode.eps}
    % \put (-140,26){\footnotesize (b)}
    % \put (-135,26){\footnotesize $\Rey = 0.05$}
	\caption{ Budget of rate of work contributions for critical modes of inertialess ($\Rey = 0$) channel flow with a nonuniform base state polymer concentration. Results for the uniform base state polymer concentration at $\beta = 0.994$ are shown with open symbols. All terms are normalised by the magnitude of the viscous dissipation and so $D = -1$. The magnitudes of the power terms remain nearly the same (and their signs and relative magnitudes are unchanged) on increasing $\Rey$ to a small finite value to endow the modes with finite inertia and a small positive growth rate (see the \href{https://bighome.iitb.ac.in/index.php/s/g6wRiyEY8H3SNyR}{supplementary material} for the finite-$\Rey$ analogue of this figure).
    \label{fig:energy-channel}}
    %The magnitudes of the power terms remain nearly the same (and their signs and relative magnitudes are unchanged) on increasing $\Rey$ to a small finite value to endow the modes with finite inertia and a small positive growth rate (see the \href{https://bighome.iitb.ac.in/index.php/s/g6wRiyEY8H3SNyR}{supplementary material} for the finite-$\Rey$ analogue of panel (b)).} \label{fig:energy}
\end{figure}

The disturbance kinetic energy budget for the nonuniform critical modes (minimum of the neutral curves) of Fig.~\ref{fig:channel}(b) is presented in Fig.~\ref{fig:energy-channel} (filled symbols). The results for the uniform critical mode at $\beta = 0.994$ (minimum of the neutral curve for the uniform solution in Fig.~\ref{fig:channel}(a)) is also shown for comparison (open symbols). The results are entirely analogous to those in Kolmogorov flow (Fig.~\ref{fig:energy}(a,b)). In the uniform solution, $P_{\nabla\theta} = 0$ and $P_\theta$ is the sole positive power term (see the open blue and red triangles at $\beta = 0.994$). Being very dilute, the corresponding nonuniform mode (filled triangles) has $P_\theta$ as the dominant power term, though $P_{\nabla\theta}$ now makes a positive contribution to the power input. As the polymer loading is increased ($\beta$ is decreased), $P_{\nabla\theta}$ first becomes the dominant positive power term (at $\beta = 0.9$) and then becomes the sole positive power term (for $\beta \leq 0.7$). As in Kolmogorov flow, a second lobe appears in the neutral curve when $P_\theta$ first becomes negative as $\beta$ is decreased (compare the results for $\beta = 0.7$ in Fig.~\ref{fig:energy-channel} and Fig.~\ref{fig:channel}(b)).

\begin{figure}
\centering
        % \includegraphics[width=0.33\linewidth] {fig/Real_part_for_Channelflow_of_contour_traceC_and_streamlines_uniform_beta0pt994_Wi973pt98_k0pt7828_kap1e_neg8.eps}
        % \put (-104,107){\footnotesize (a) uniform, $\beta = 0.994$}
        % \includegraphics[width=0.33\linewidth] {fig/Real_part_for_Channelflow_of_contour_traceC_and_streamlines_of_critical_mode_nonuniform_ht2pi240_beta0pt994_Wi121pt91_k2pt5_kap1e_neg8.eps}
        % \put (-104,107){\footnotesize (b) nonuniform, $\beta = 0.994$}
        % \includegraphics[width=0.33\linewidth] {fig/Real_part_for_Channelflow_of_contour_traceC_and_streamlines_of_critical_mode_nonuniform_ht2pi240_beta0pt7_Wi4pt123_k2pt16_kap5e_neg7.eps}
        % \put (-104,107){\footnotesize (c) nonuniform, $\beta = 0.7$}
         \includegraphics[width=0.33\linewidth] {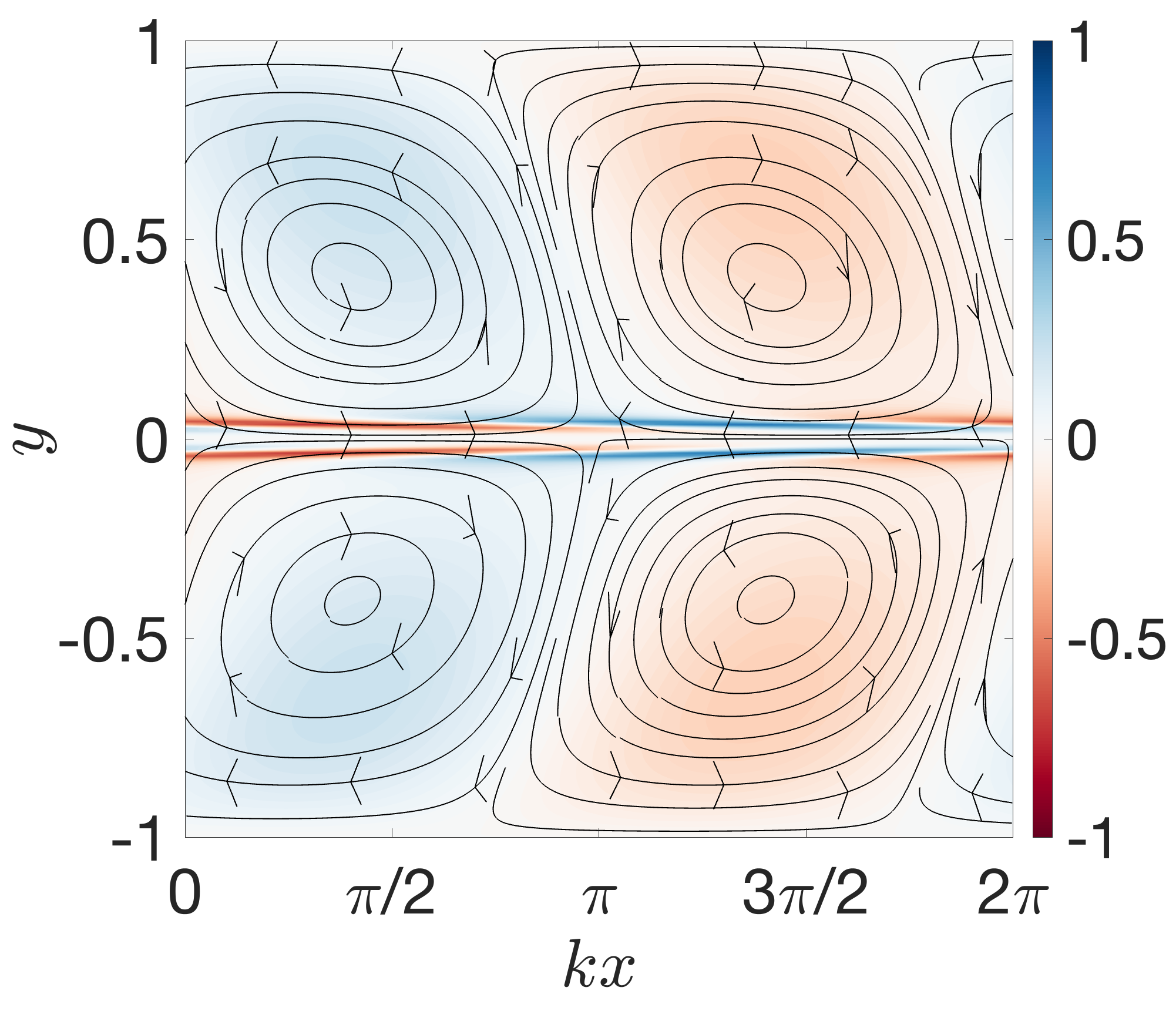}
        \put (-106,107){\footnotesize (a) uniform, \scriptsize  $\beta = 0.994$}
        % \hspace{1.5 em}
         \includegraphics[width=0.33\linewidth] {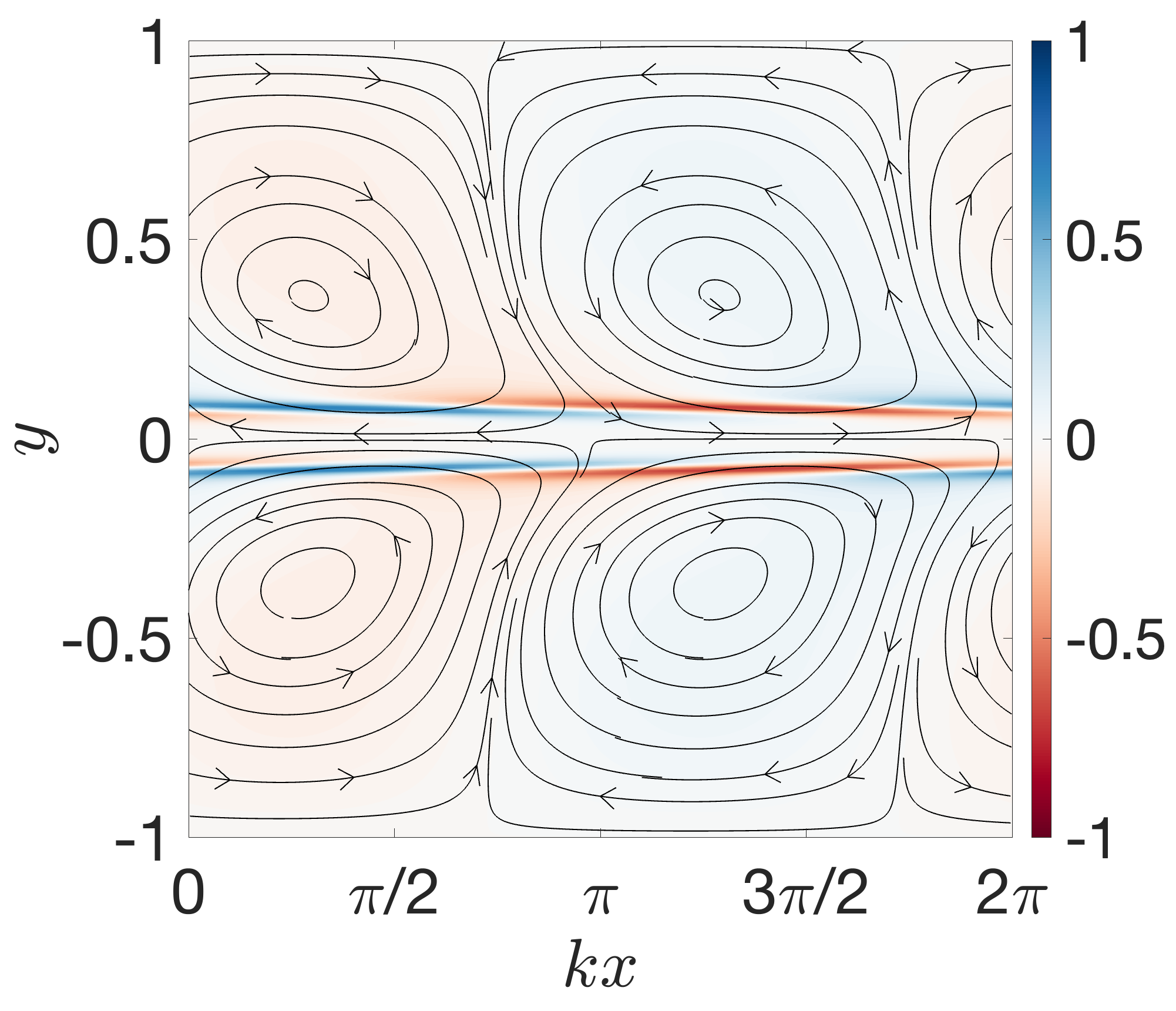}
         \put (-106,107){\footnotesize (b) nonuniform, \scriptsize $\beta = 0.994$}
        \includegraphics[width=0.33\linewidth] {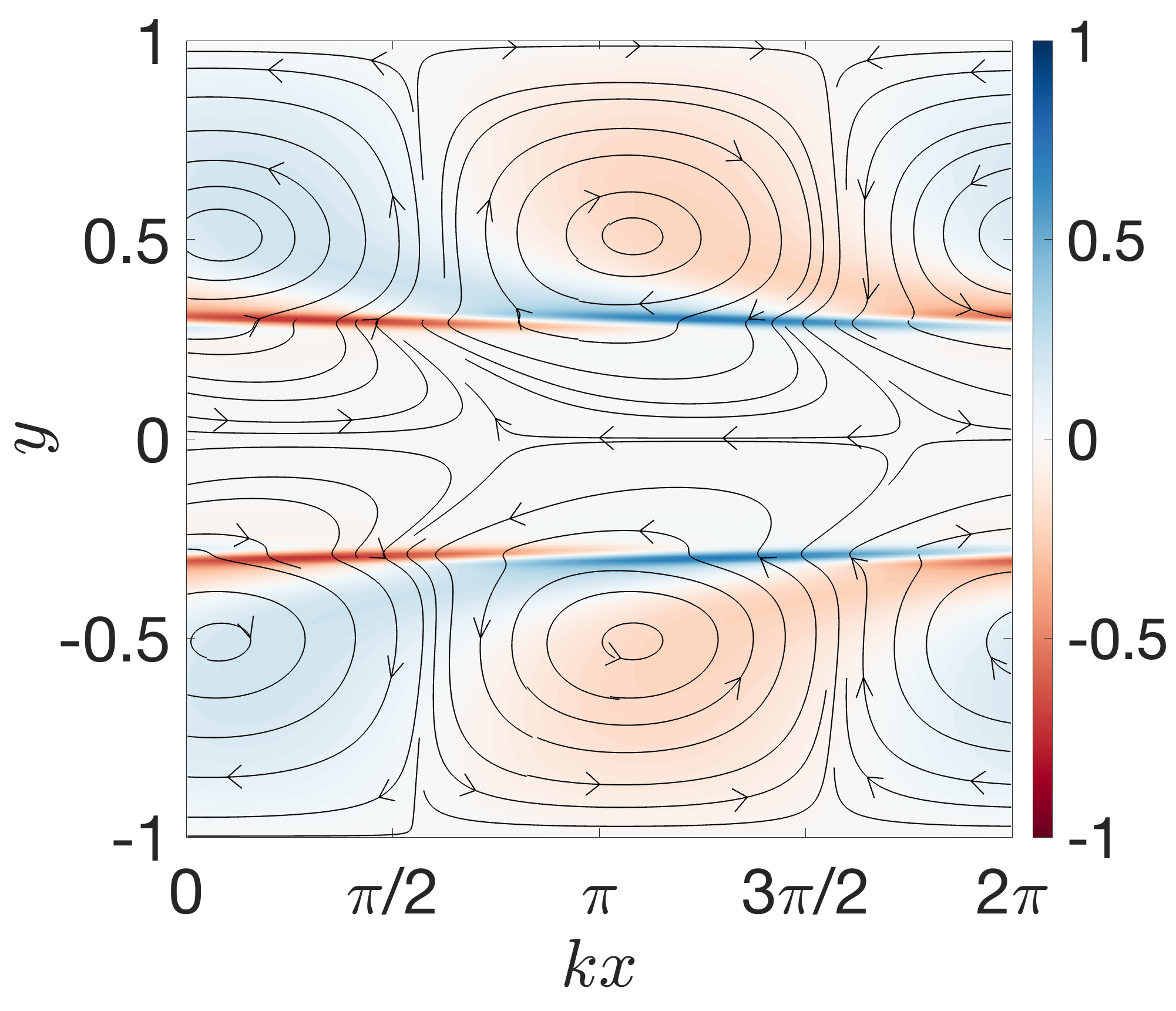}
        \put (-106,107){\footnotesize (c) nonuniform, \scriptsize  $\beta = 0.7$}
    \caption{Visualisation of unstable eigenfunctions in channel flow, showing contours of the perturbed polymer squared-extension $\mathrm{tr}(\hat{\C})$ overlaid with the streamlines of the disturbance velocity field. These modes lie just beyond the onset of instability marked by the neutral curves for $\beta = 0.994$ and $\beta = 0.7$ in Figs.~\ref{fig:channel}(a,b). 
    (a)~Uniform base-state concentration with $\beta = 0.994, \Wi = 1050, k = 0.78$ and $ c_i = 1.084\times 10^{-5}$; 
    (b)~Nonuniform base-state concentration ($\bar{\theta}$ is localised about the centreline and corresponds to $\delta  = 0.0083 \pi$ in Fig.~\ref{fig:base-channel}(a)) with 
    a very low level of total polymer loading, $\beta = 0.994$, and $\Wi = 130, k = 2.5$, $ c_i = 4.93\times 10^{-4}$;
    % (b) Nonuniform base-state concentration with $\beta = 0.994, \Wi = 130, k = 2.5$ and $ c_i = 4.93\times 10^{-4}$; 
    (c)~Nonuniform base-state concentration, as in panel (b), but with a higher polymer-loading, $\beta = 0.7$, and $\Wi = 6, k = 2.16$, $c_i = 4.976\times10^{-3}$. The perturbed concentration fields for the eigenfunctions of the nonuniform cases in panels (b,c) are presented in the \href{https://bighome.iitb.ac.in/index.php/s/g6wRiyEY8H3SNyR}{supplementary material}.
    \label{fig:channel-eigen}
    }
\end{figure}

Eigenfunctions for the uniform and nonuniform cases are compared in Fig.~\ref{fig:channel-eigen}. The field of $\mathrm{tr}(\hat{\C})$ in the uniform case, for $\beta = 0.994$, exhibits thin layers of large perturbed-extension located at a small distance on either side of the centreline (Fig.~\ref{fig:channel-eigen}(a)), as has been shown and explained in past work \citep{khalid_prl,Kerswell_page}. The eigenfunction in the nonuniform case, at the same low level of polymer loading ($\beta = 0.994$), has a similar spatial structure though the thin layers of large perturbed-extension are not as close to the centreline (Fig.~\ref{fig:channel-eigen}(b))
% , though the thin layers of large extension are slightly further away from the centreline. 
When the polymer loading is increased to $\beta = 0.7$, however, the eigenfunction of the nonuniform case becomes focused in the region with a large base-state concentration gradient, i.e., the thin layers of large perturbed-extension shift to the interface between the polymer-rich and polymer-free streams
(Fig.~\ref{fig:channel-eigen}(c)). 
% This distinctive feature of the eigenfunctions in nonuniform solutions with moderate polymer loading, namely that of being focused in regions of large base-state concentration gradients, is shared by the nonuniform eigenfunctions in Kolmogorov flow (Fig.~\ref{fig:eigen-double}(b,c)). 
% again similar to the in Kolmogorov flow. $\beta = 0.7$ qualitative difference between the eigenfunctions of the uniform and nonuniform flowThis feature of the eigenfunction in the nonuniform solution at $\beta = 0.7$ is consistent with the instability being driven by is akin to that in Kolmogorov flow. 
Thus, just as in Kolmogorov flow (Fig.~\ref{fig:eigen-Kol-lowbeta} and the associated discussion), the eigenfunctions at the onset of instability change their nature on increasing polymer loading (decreasing $\beta$).
The eigenfunction of the nonuniform case in channel flow resembles that in the uniform case when the solution is very dilute, such that $P_\theta$ is the dominant positive power term in the disturbance kinetic energy budget (e.g., when $\beta = 0.994$, see Fig.~\ref{fig:energy-channel} and Figs.~\ref{fig:channel-eigen}(a,b)). However, once the total polymer loading is increased sufficiently for $P_{\nabla\theta}$ to dominate, the eigenfunction changes form and becomes focused in regions of large-concentration gradients (e.g., when $\beta = 0.7$, see Fig.~\ref{fig:energy-channel} and Figs.~\ref{fig:channel-eigen}(c)). 

The similarity of the channel flow results with those of Kolmogorov flow extends to the stabilisation of the flow when polymers are localised near the region of maximum shear, which for channel flow is the near-wall region; we find for $\beta = 0.994$ that the neutral curve shifts upward, relative to the uniform-concentration case, when polymers are localised near the walls (results not shown). 
It is interesting, and of practical relevance, to further investigate how the location and width of the polymer stream affects the instability. The exploration of different nonuniform base-state concentration profiles, including two-layer arrangements with different volume fractions,  will be taken up in future work. 
% Further exploration of the stability of viscoelastic channel flow with different base-state polymer-concentration profiles and for a wider range of parameter values is left for future work.
% The shape of the neutral curve and thus the range of unstable wavenumbers depends on where the polymers are localised in the channel. These aspects, along with questions regarding the influence of finite inertia, the energy signature, of the nonuniform channel center-mode, are left for future work. 

% \begin{figure}
%    \includegraphics[width=0.4\linewidth] {fig/Channel_neutral_curve_W_k_plane_strip_atWall_both_uni_nonuni_beta0pt994_ht4pi240_8pi240.eps}
% 	\caption{ Neutral curves for polymer strip shifted towards channel wall; data are shown for $\beta  =0.994$} 
% \end{figure}

\section{Concluding remarks}\label{sec:conclusions}

Rectilinear flows of dilute polymer solutions can become unstable via the centre-mode instability, even in the absence of inertia, provided the base flow has a velocity-maximum. Here, we have shown in Kolmogorov and channel flow that the instability is strongly promoted when the polymer concentration in the base-state is nonuniform, such that polymers are localised about the velocity-maximum.
For a fixed total amount of polymer, we have shown that a flow with a distinct polymer-rich stream overlying the velocity-maximum is destabilized at a much smaller value of $\Wi$ compared to a flow with a uniform distribution of polymers.
% In the simplified setting of periodic Kolmogorov flow, we have shown how the minimum $\Wi$ required for instability can be reduced, for a fixed total amount of polymer, by localising polymers near the velocity maximum. 
This destabilizing effect of localising polymers is found to increase strongly as the total polymer loading is increased from very low to moderate levels. In the latter case, the perturbation eigenfunction of the nonuniform-concentration instability 
% resembles that of the uniform-concentration centre-mode but with its activity 
is focused 
at the interface between the polymer-laden and polymer-free streams, i.e., where
concentration gradients are present. 
% the gradients in the base-state concentration are large. strong concentration gradients. 
An energy analysis helps explain this observation by showing that the work required for driving the instability is primarily done by elastic stresses originating from gradients in the polymer concentration.
% The final section on channel flow confirms that these results---the destabilisation of the centre-mode by a nonuniform distribution of polymers---extends to channel flow wherein polymers are localised about the centreline where the base-velocity is maximum. 

It is clear from our results that localising polymers about the velocity-maximum does not merely mimic a uniform base flow with a higher total polymer concentration. This is apparent from the altered nature of the neutral stability curves, which for the nonuniform case become double-lobed at moderate polymer loading, and from the qualitative changes in the eigenfunctions. The energy analysis directly confirms this point by revealing the role of concentration gradients. With a nonuniform polymer distribution, the total polymer loading becomes an effective means of promoting flow instability. This is especially true for channel flow, which in the uniform case is stabilised by an increase in the total polymer loading but in the nonuniform case is strongly destabilised. 

The required nonuniform polymer distribution for promoting the centre-mode instability in channel flow can be realised by introducing three streams of fluid into a channel, via a T-junction for example, with the central stream being a polymer solution and the other two streams being pure solvent. In this context, our results imply that polymer-driven flow instability and consequent mixing could be enhanced by introducing polymers into just the central stream rather than premixing polymers in all streams. It would therefore be interesting, in future work, to investigate  how the non-uniform instability grows in the non-linear regime and whether the resulting dynamics lead to efficient mixing. 

More broadly, our work suggests a new line of research into how the spatial distribution of polymers could be used to suppress or enhance viscoelastic instabilities, including the elasto-inertial centre-mode and polymer-diffusive instabilities in rectilinear flows, and hoop-stress-driven instabilities in curvilinear flows.

\begin{bmhead}[Acknowledgments.]
The authors are grateful to Dario Vincenzi (LJAD, Nice) and Mayank Chopra (IIT Bombay) for helpful discussions.
 J.R.P.~acknowledges his Associateship with the International Centre for Theoretical Sciences (ICTS), Tata Institute of Fundamental Research, Bangalore, India. 
  % The authors thank the National PARAM Supercomputing Facility \textit{PARAM SIDDHI-AI} at CDAC, Pune for computing resources; simulations were also performed on the IIT Bombay workstations \textit{Gandalf} (procured through DST-SERB grant SRG/2021/001185), and \textit{Faramir} and \textit{Aragorn} (procured through the IIT-B grant RD/0519-IRCCSH0-021).
\end{bmhead}

\begin{bmhead}[Funding information.]
This work was supported by 
the Indo-French Centre for the Promotion of Advanced Research (IFCPAR/CEFIPRA, Project No. 6704-1).
\end{bmhead}

\begin{bmhead}[Declaration of interests.]
The authors report no conflict of interest.
\end{bmhead}

\begin{bmhead}[Author ORCIDs.]
\\
Shailendra~Kumar~Yadav, https://orcid.org/0009-0001-2139-776X
\\
Jason R. Picardo, https://orcid.org/0000-0002-9227-5516.
\end{bmhead}

%\begin{bmhead}[Author ORCIDs.]
%\\
%Jason R. Picardo, https://orcid.org/0000-0002-9227-5516;
%\\
%Dario Vincenzi, https://orcid.org/0000-0003-3332-3802.
%\end{bmhead}

% \backsection[Author ORCID]{S.~Hazra, https://orcid.org/0009-0000-3728-5888; J.~R.~Picardo, https://orcid.org/0000-0002-9227-5516}

\appendix

\section{Varying the width of the polymer-laden layer}\label{app:wd}

\begin{figure}
       \centering
       \includegraphics[width=.33\linewidth]{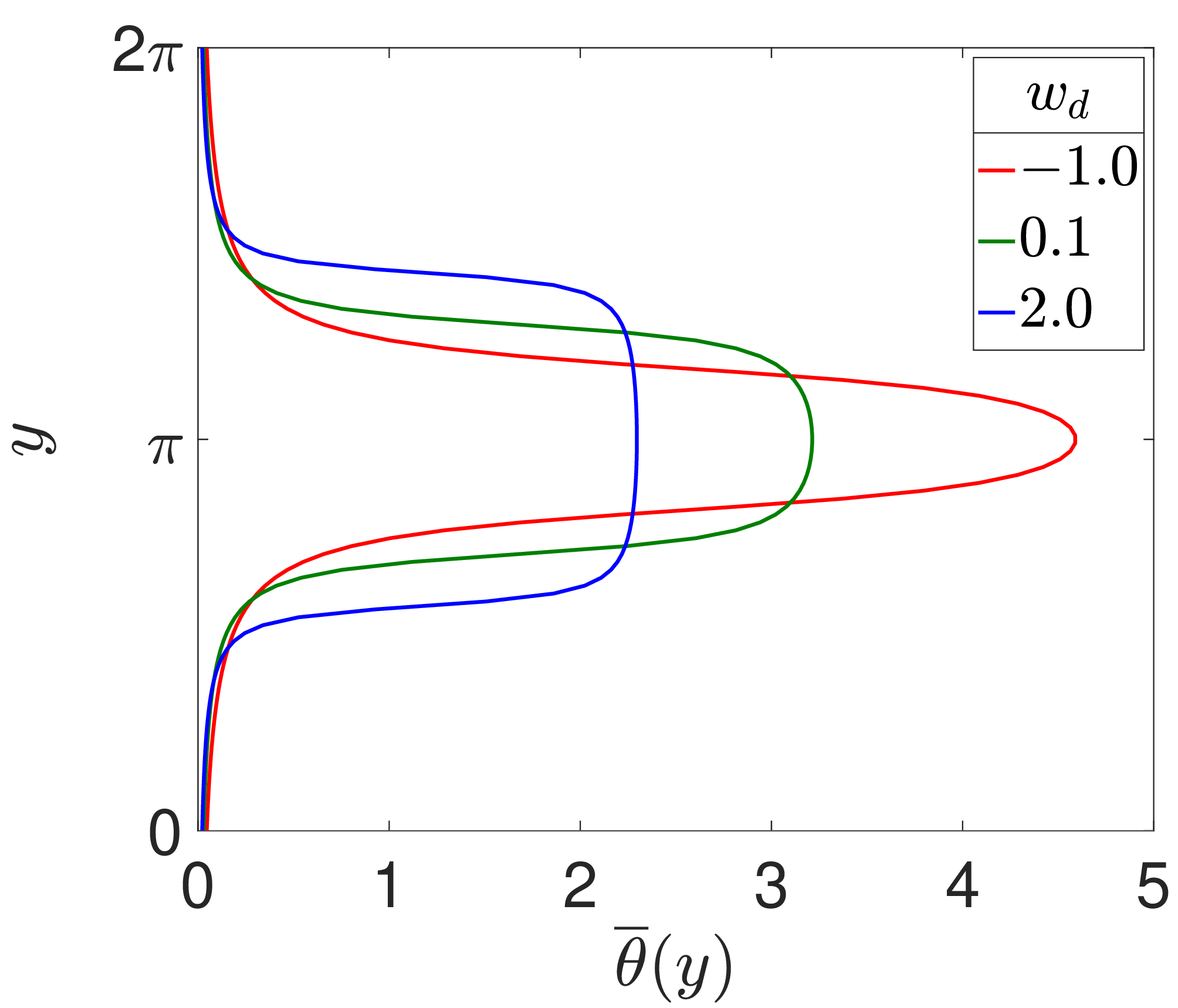}
       \put (-100,25){\footnotesize (a)}
      \includegraphics[width=.33\linewidth]{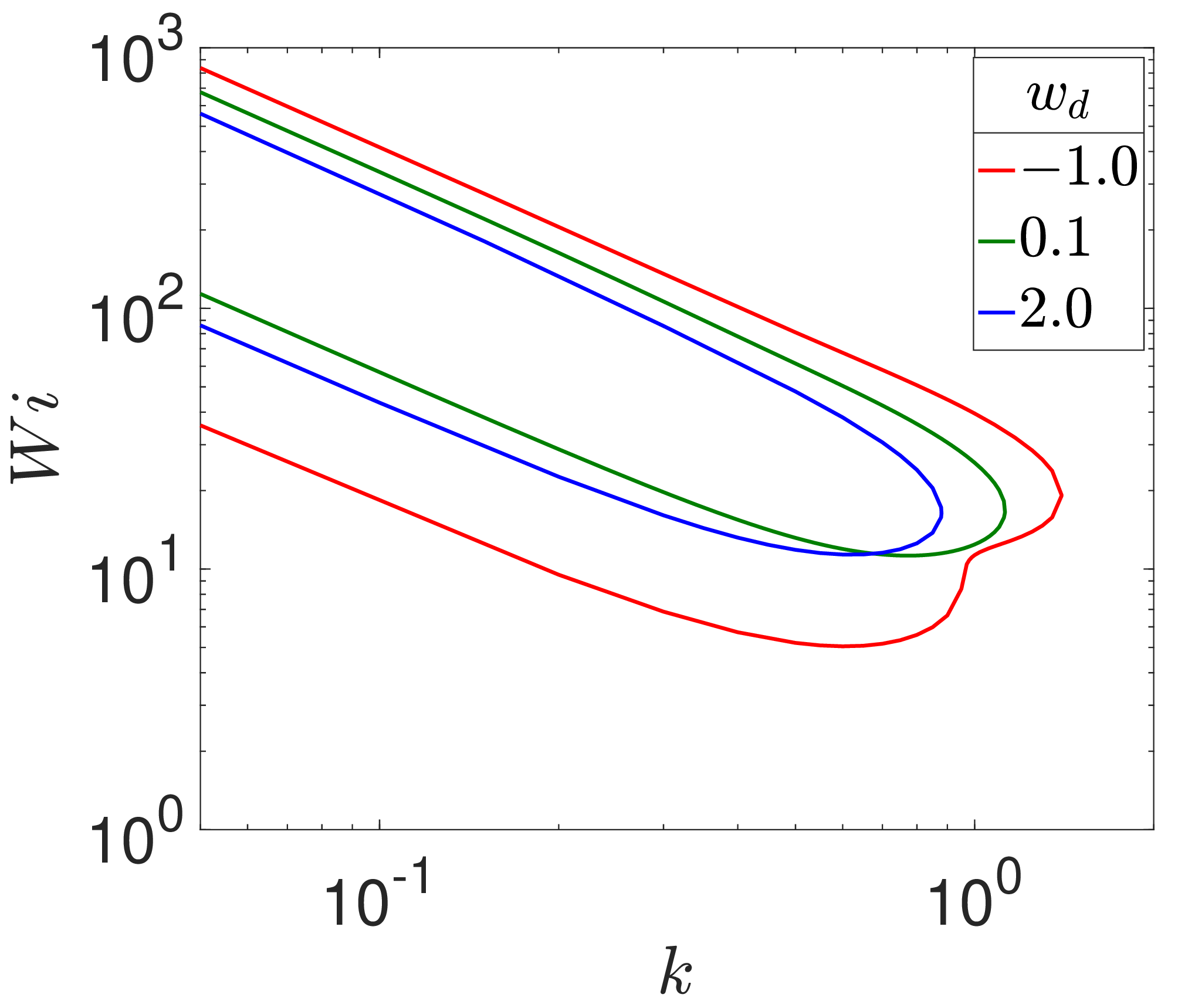}
       \put (-100,25){\footnotesize (b)}
              % \hspace{.5 em}
%       \includegraphics[width=.33\linewidth]{fig/Neutral_curve_W_k_plane_various_wd_B0pt7_ht16pi240.eps}
%        \put (-100,25){\footnotesize (c)}
% \hspace{.5 em}
       \includegraphics[width=.33\linewidth]{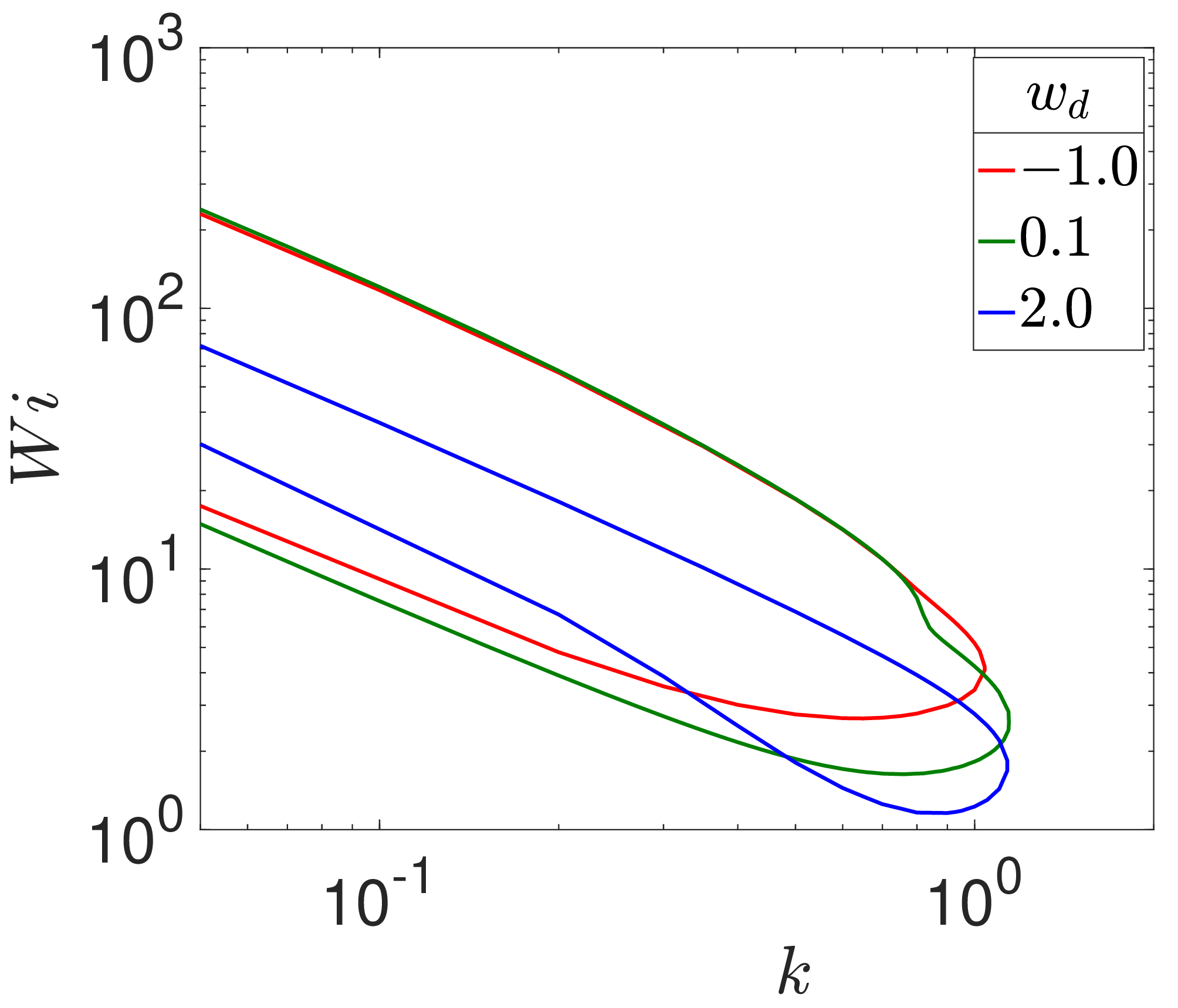}
       \put (-100,25){\footnotesize (c)}
       %       \includegraphics[width=.33\linewidth]{fig/Neutral_curve_W_k_plane_various_wd_B0pt9_ht16pi240_with_inset_of_theta_profiles.eps}
       % \put (-105,25){\footnotesize (c)}
    
\caption{Effect of the width of the polymer-laden layer, centred at the velocity maximum, on the stability of Kolmogorov flow. (a) Base concentration $\bar{\theta}$ profiles with three different widths, obtained by adjusting $w_d$ in \eqref{eq:pulse}. The corresponding neutral curves are compared for (b) $\beta = 0.9$
% (c) $\beta = 0.7$, 
and (c) $\beta = 0.5$.} 
\label{fig:wd-neutral}
\end{figure}

In the main text, we set $w_d = 0.1$ in the profile of the base-state concentration \eqref{eq:pulse} so that the polymer-laden stream (realised for $\delta = 0.067 \pi$) had a fixed width. Here, we illustrate the effect of varying $w_d$ on the stability of Kolmogorov flow. Figure~\ref{fig:wd-neutral}(a) presents the base-state concentration profiles for two additional values of $w_d$ corresponding to narrower and wider polymer-rich streams than that considered in the main text; all streams are centred about the velocity maximum. Figures~\ref{fig:wd-neutral}(b-c) present the corresponding neutral stability curves for different values of $\beta$. We see that the effect of the stream-width depends on the total polymer loading. The flow with the narrowest polymer stream is most unstable for low polymer loading (Fig.~\ref{fig:wd-neutral}(b) for $\beta = 0.9$), while the widest stream is most unstable for higher polymer loading (Fig.~\ref{fig:wd-neutral}(c) for $\beta = 0.5$). 
% The intermediate We have also found intermediate polymer loadings for which the intermediate case, examined in the main text, can also be the most unstable at intermediate polymer loading (Fig.~\ref{fig:wd-neutral}(d) for $\beta = 0.7$). 
These results show that the profile of the base-state concentration could be tuned to promote or suppress the instability.  
Further quantification of this effect in different flows is left for future work. 

\FloatBarrier

\bibliographystyle{jfm-FLM-arxiv}
\bibliography{BiBsk}

\end{document}